\documentclass[11pt,twoside,a4paper]{book}
\pdfoutput=1
\usepackage[T1]{fontenc}
\usepackage[english]{babel}
\usepackage[utf8]{inputenc}
\usepackage[pdftex]{graphicx}           % usamos arquivos pdf/png como figuras
\usepackage{setspace}                   % espaçamento flexível
\usepackage{indentfirst}                % indentação do primeiro parágrafo
\usepackage{makeidx}    
\usepackage{amsmath}
\usepackage{mathtools} 
\usepackage{scalerel}
\usepackage{bm}
\usepackage{algorithm}  
\usepackage{apacite}
\usepackage[noend]{algpseudocode}
\makeatletter
\def\BState{\State\hskip-\ALG@thistlm}
\makeatother

\usepackage{amsmath,graphicx}
\newcommand{\widesim}[2][1.5]{
  \mathrel{\overset{#2}{\scalebox{#1}[1]{$\sim$}}}}

\usepackage{float}         % índice remissivo
\usepackage[nottoc]{tocbibind}          % acrescentamos a bibliografia/indice/conteudo no Table of Contents
\usepackage{courier}                    % usa o Adobe Courier no lugar de Computer Modern Typewriter
\usepackage{type1cm}                    % fontes realmente escaláveis
\usepackage{listings}                   % para formatar código-fonte (ex. em Java)
\usepackage{titletoc}
\usepackage[fixlanguage]{babelbib}
\usepackage[font=small,format=plain,labelfont=bf,up,textfont=it,up]{caption}
\usepackage[usenames,svgnames,dvipsnames]{xcolor}
\usepackage[a4paper,top=2.54cm,bottom=2.0cm,left=2.0cm,right=2.54cm]{geometry} % margens
\usepackage[pdftex,plainpages=false,pdfpagelabels,pagebackref,colorlinks=true,citecolor=DarkGreen,linkcolor=NavyBlue,urlcolor=DarkRed,filecolor=green,bookmarksopen=true]{hyperref} % links coloridos
\usepackage[all]{hypcap}                % soluciona o problema com o hyperref e capitulos
\usepackage[square,sort,nonamebreak,comma]{natbib}  % citação bibliográfica alpha (alpha-ime.bst)
\fontsize{60}{62}\usefont{OT1}{cmr}{m}{n}{\selectfont}

\newtheorem{theorem}{Theorem}[section]

\usepackage{fancyhdr}
\renewcommand{\chaptermark}[1]{\markboth{\MakeUppercase{#1}}{}}

\graphicspath{{./figuras/}}             % caminho das figuras (recomendável)
\makeindex                              % para o índice remissivo
\fontsize{60}{62}\usefont{OT1}{cmr}{m}{n}{\selectfont}
\cleardoublepage
\normalsize

\begin{document}
\frontmatter 
% cabeçalho para as páginas das seções anteriores ao capítulo 1 (frontmatter)
\fancyhead[RO]{{\footnotesize\rightmark}\hspace{2em}\thepage}
\setcounter{tocdepth}{2}
\fancyhead[LE]{\thepage\hspace{2em}\footnotesize{\leftmark}}
\fancyhead[RE,LO]{}
\fancyhead[RO]{{\footnotesize\rightmark}\hspace{2em}\thepage}

\onehalfspacing  % espaçamento

% ---------------------------------------------------------------------------- %
% CAPA
% Nota: O título para as dissertações/teses do IME-USP devem caber em um 
% orifício de 10,7cm de largura x 6,0cm de altura que há na capa fornecida pela SPG.
\thispagestyle{empty}
\begin{center}
    \vspace*{2.3cm}
    \textbf{\Large{Full Bayesian Modeling for fMRI Group Analysis}}\\
    
    \vspace*{1.2cm}
    \Large{Johnatan Cardona Jim\' enez}
    
    \vskip 2cm
    \textsc{
    %Doctoral thesis submitted\\[-0.25cm] 
    %in the\\[-0.25cm]
    %Qualification Exam \\[-0.25cm]
    Institute of Mathematics and Statistics\\[-0.25cm]
    of the\\[-0.25cm]
    University of São Paulo\\[-0.25cm]
    %in fulfilment of the requirements\\[-0.25cm]
    %for the degree \\[-0.25cm]
    %of\\[-0.25cm]
    %Doctor of Philosophy
    }
    
    \vskip 1.5cm
    Program: Statistics\\
    Supervisor: Dr. Carlos Alberto de Bragan\c{c}a Pereira\\
    %: Dr. 

   	\vskip 1cm
    %\normalsize{During the development of this work the author received financial support from CAPES}
    
    \vskip 0.5cm
    \normalsize{São Paulo, December 7, 2017}
\end{center}

\pagenumbering{roman}     % começamos a numerar 

\chapter*{Abstract}
\noindent Jiménez, J.C. \textbf{Full Bayesian Modeling for fMRI Group Analysis}. 
%2017. 120 f.
Tese (Doutorado) - Instituto de Matemática e Estatística,
Universidade de São Paulo, São Paulo, 2017.
\\

Functional magnetic resonance imaging or functional MRI (fMRI) is a non-invasive way to assess brain activity by detecting changes associated with blood flow. In this work, we propose a full Bayesian procedure to analyze fMRI data for individual and group stages. For the individual stage we use a multivariate dynamic linear model (MDLM), where the temporal dependence is modeled through the state parameters and the spatial dependence is modeled only locally, taking the nearest neighbors of each voxel location. For the group stage we take advantage of the posterior distribution of the state parameters obtained in the individual stage and create a new posterior distribution that represents the updated beliefs for the group analysis. Since the posterior distribution for the state parameters is indexed by the time $t$, we propose an algorithm that allows on-line estimated curves of the state parameters to be drawn and posterior probabilities computed in order to assess brain activation for both individual and group analysis. We propose an alternative analysis for the group stage using a Gaussian process ANOVA model, where the on-line estimated curves obtained in the individual stage are modeled as a functional response.  Finally, we assess our proposed modeling procedure using real resting-state data and computing empirical false-positive brain activation rates. 
\\

\noindent \textbf{Keywords:} Dynamic linear model, functional MRI, Gaussian process ANOVA model.

% ---------------------------------------------------------------------------- %
% Sumário
\tableofcontents    % imprime o sumário

% ---------------------------------------------------------------------------- %
%\chapter{Lista de Abreviaturas}
%\begin{tabular}{ll}
%         CFT         & Transformada contínua de Fourier (\emph{Continuous Fourier Transform})\\
 %        DFT         & Transformada discreta de Fourier (\emph{Discrete Fourier Transform})\\
 %       EIIP         & Potencial de interação elétron-íon (\emph{Electron-Ion Interaction Potentials})\\
  %      STFT         & Tranformada de Fourier de tempo reduzido (\emph{Short-Time Fourier Transform})\\
%\end{tabular}

% ---------------------------------------------------------------------------- %
%\chapter{Lista de Símbolos}
%\begin{tabular}{ll}
%        $\omega$    & Frequência angular\\
%        $\psi$      & Função de análise \emph{wavelet}\\
%        $\Psi$      & Transformada de Fourier de $\psi$\\
%\end{tabular}

% ---------------------------------------------------------------------------- %
% Listas de figuras e tabelas criadas automaticamente
%\listoffigures            
%\listoftables            

% ---------------------------------------------------------------------------- %
% Capítulos do trabalho
\mainmatter

% cabeçalho para as páginas de todos os capítulos
\fancyhead[RE,LO]{\thesection}

\singlespacing              % espaçamento simples
%\onehalfspacing            % espaçamento um e meio

%% ------------------------------------------------------------------------- %%
\chapter{Introduction}
\label{cap:introducao}

Magnetic resonance imaging (MRI) is a non-invasive technique that is used to create elaborate anatomical images of the human body. Specifically, this technique can be used to obtain detailed brain images that can help to identify different types of tissue such as, for example, white matter and gray matter, and can also be used to diagnose aneurysms and tumors. Another important facet of this technique is that it can be used to visualize dynamic or functional activity in the brain. Functional Magnetic Resonance Imaging (fMRI) can be described as a generalization of the MRI technique, where the focus is not just one high-resolution image of the brain, but rather a sequence of low-resolution images that allows for identification, at least in an indirect way, of neuronal activity through the blood-oxygen-level dependent (BOLD) contrast.  Statistical models are very useful for analyzing the post-processed data that compose the sequence of images obtained in an fMRI experiment. As we will see in the following chapters of this manuscript, the fMRI data present a spatiotemporal feature that usually is ignored in most experiments based on this technique. For example, the most popular statistical model used to identify a brain-region reaction to an external stimulus is the normal regression linear model, usually known as GLM in the fMRI literature, which assumes spatiotemporal independence among the data, an unrealistic assumption that can lead to incorrect inference, as we will show later. It is worth mentioning that in the fMRI literature, a variety of Bayesian models has been proposed (see \cite{zhang2015bayesian} for a detailed review of the subject), where some of them account for the spatiotemporal structure present in this type of data. \\

\begin{equation} \label{equ1}
\begin{array}{lccl}
\text{Observation:}\ &\mathbf{Y}_t & =& F^{'}_t\mathbf{\Theta}_t + \mathbf{\nu}^{'}_t \\
 \text{Evolution:}\ &\mathbf{\Theta}_t& =& G_t\mathbf{\Theta}_{t-1} + \Omega_t\\
\end{array}
\end{equation}
                     
\begin{equation*}
\begin{array}{l}
\nu_t \sim N\left[\mathbf{0}, V_t \mathbf{\Sigma}\right],\\
\Omega_t \sim N\left[\mathbf{0}, \mathbf{W}_t, \mathbf{\Sigma}\right].
\end{array}
\end{equation*}                     
                       
In this thesis we propose a procedure to model fMRI experiments that is related to Block and Event-Related designs. For the voxel-wise individual analysis we use a Bayesian multivariate dynamic linear model (\ref{equ1}) to identify local brain reactions related to the design being performed. Our main purpose using this type of modeling is to take into account both the temporal structure, through the evolution equation in \ref{equ1}, and the local spatial structure, through the matrix $\mathbf{\Sigma}$. Since this type of model is mainly used to forecast, there is not a clear way -at least to our knowledge- on how to make inferences on the effects $\mathbf{\Theta}_t$. \cite{west1997bayesian} stated that one can make an inference on the state $\mathbf{\Theta}_t,$ making use of the posterior distribution $p(\mathbf{\Theta}_t|D_t)$, which is available for every $t \in \left\{1,\ldots,T\right\}$. Thus, if one wants to test, for example,  $\{\mathbf{\Theta}_t>0\}$ (which will be our case), some questions arise.  
Is it necessary to do this test for every $t$ that is observed, or is it enough to take only the last period $T,$ given the sequential update mechanism of $P(\mathbf{\Theta}_t|D_t)$? 
In case that one decides to perform the test for every $t$ is likely that the test $\left\{\mathbf{\Theta}_t>0\right\}$ can be accepted for some observations and rejected for the remaining, then how can one conclude whether the test of interest is true or false? In this sense, we take advantage of the posterior distribution $p(\mathbf{\Theta}|D_t)$ and the Monte Carlo integration technique to propose a simple way to perform the test in question, based on the joint distribution $p(\mathbf{\Theta}_k, \mathbf{\Theta}_{k+1}, \ldots, \mathbf{\Theta}_T|D_t)$, where $k\in\left\{1,\ldots,T\right\}$ must be carefully chosen. In chapter~\ref{cap:2} we explain the theoretical aspects of this model in more detail.

For the voxel-wise group analysis, we take the posterior distribution $p(\mathbf{\Theta}|D_t)$ as an input for this stage and propose three different alternatives to perform the inference related to brain activation patterns in a single group and the comparison of the brain activation patterns between two groups. One alternative is an inference procedure based on the last posterior distribution $p(\theta_T|D_T)$.  The other relies on the same idea for the individual analysis mentioned above, and the last option is based on Bayesian functional ANOVA modeling using Gaussian process prior distributions (\cite{kaufman2010bayesian}). In the latter case, \cite{kaufman2010bayesian} developed graphical displays to analyze the sources of variability and the effects on functional response, but they left the formal testing for future work. We handle this last problem proposing an inferential procedure based on posterior probabilities and Monte Carlo integration. These procedures will be implemented in an R package for free use. 

\subsection*{Outline}
In the remaining part of this chapter we introduce basic concepts on fMRI experiments, going through the explanation of what is a Magnetic Resonance Image (MRI) and the acquisition process in a functional Magnetic Resonance Image (fMRI) experiment. We also explain the important concepts of blood-oxygen-level dependet (BOLD) contrast imaging and the hemodynamic response function. We then introduce the construction of the expected BOLD response, the definition of GLM, and finally, we metion some of the most popular software packages for fMRI data analysis and the models implemeted in them. In chapter~\ref{cap:2} we give theoretical background on dynamic linear models and the Gaussian process ANOVA model. In chapter~\ref{cap:3}, we present our proposal for the voxel-wise group analysis, which includes the individual stage.

\subsection{What is an Magnetic Resonance Image (MRI)?}

\begin{figure}[H]
  \centering
  \includegraphics[width=.40\textwidth]{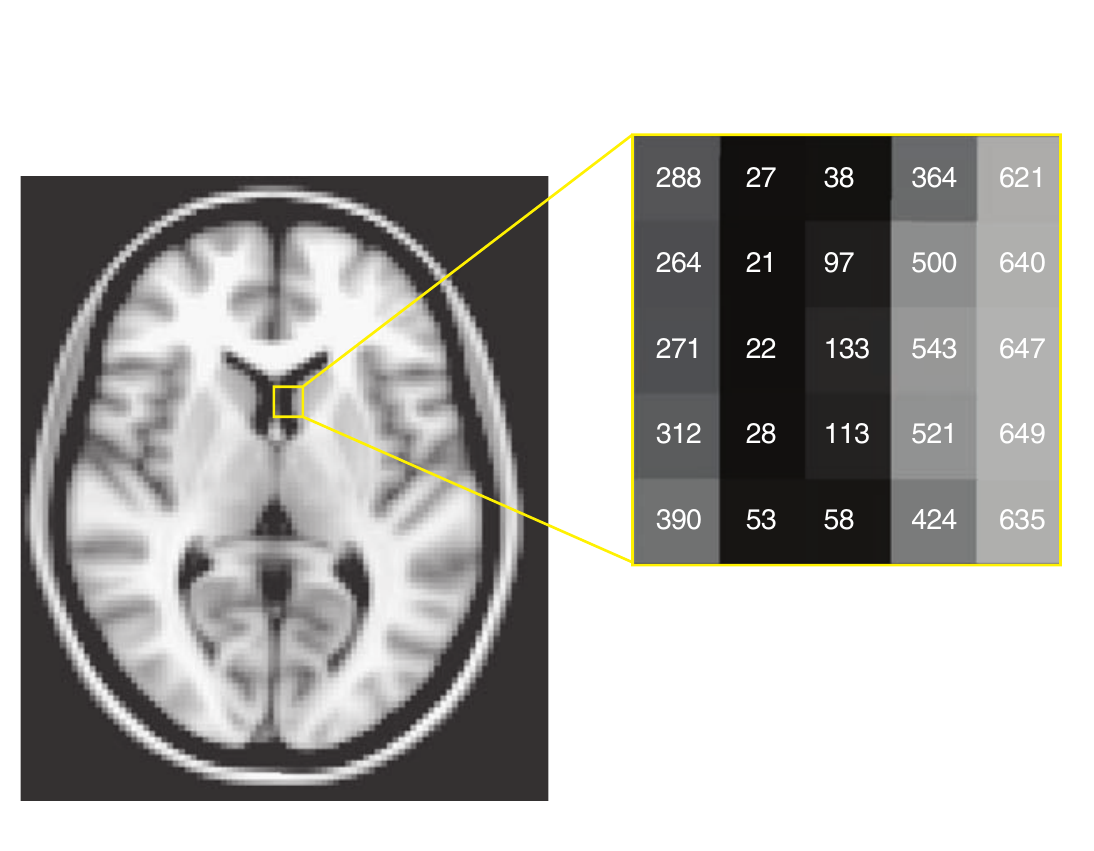} 
  \caption{An image as a graphical representation of a matrix. The grayscale values in the image on the left correspond to numbers, shown for a specific set of voxels in the closeup section on the right (\citep{poldrack2011handbook}).}
  \label{fig:11} 
\end{figure}

We now give a brief description of what a Magnetic Resonance Image is.  This can be important in order to understand some basic characteristics of the random variables we are going to use. It is worthwhile to point out that some preprocessing is necessary in this type of image in order to take raw data from the scanner and prepare them for statistical analysis. Some of the steps usually applied in the image preprocessing are motion correction, slice timing correction, spatial filtering, intensity normalization, and temporal filtering. Here we are not going to explain those in detail, the interested reader can refer to \cite{poldrack2011handbook}. In this case, an MRI can be viewed as a matrix of numbers that correspond to spatial locations. When we view an image, we do so by representing the numbers in the image in terms of grayscale values and each element in the image is called as a voxel, which is the three-dimensional analog to a pixel.  See figure \ref{fig:11} for a visual description.

\subsection{What is an fMRI experiment?}

\begin{figure}[H]
  \centering
\includegraphics[width=.40\textwidth]{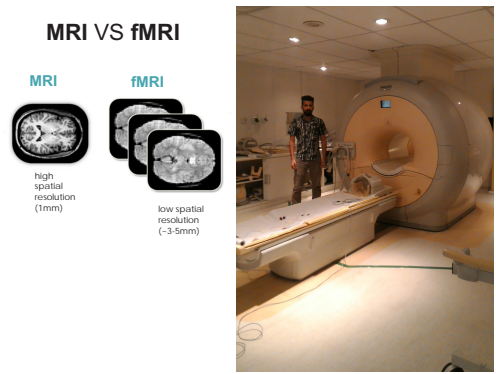}
  \caption{Left panel: Diferences between MRI and fMRI images. Right panel: The MRI scanner at the Institute of Radiology of the University of São Paulo.}
  \label{fig:dos} 
\end{figure}

An MRI is usually a high-resolution picture of the brain that is obtained using an MRI scanner.  See the right panel in figure~\ref{fig:dos}).  For that purpose, the individual has to lie in the scanner for a period of time, usually 5 minutes, in a rest state until the entire brain is scanned. This type of image can help to identify different types of tissue, such as, for example, white matter and gray matter, and can also be used to diagnose aneurysms and tumors.  On the other hand, an fMRI experiment is a process where the individual has also to stay lying in the scanner and for the same period of time --five minutes-- but up to 100 low-resolution images  can be obtained.  See the left side of figure~\ref{fig:dos}). In an fMRI experiment, the individual can receive a sequence of stimuli according to a specific design or just stay in a resting position without any external stimulation.

The types of stimuli presented to the subject in the scanner in an fMRI experiment depend on the interests of the researcher, and usually include \textit{red}{sensory}, visual and/or auditory stimuli. There are three different types of experimental designs or paradigms that are usually applied in an fMRI experiment: block design, event-related design, and a composition of the two. In a block design, a stimulus is presented in a continuous way, say, in intervals with durations of 20 to 30 seconds, with sequences of resting intervals where the stimulus is not presented.  See panel (a) of figure~\ref{fig:tres}.  On the other hand, in an event-related design, a stimulus of short duration --2 seconds or less-- is presented, followed by more long periods of rest. In this kind of paradigm, the stimuli can be presented in a random or fixed way.  See panel (c) of figure~\ref{fig:tres}. In general, block design are statistically powerful and straighforward to analyze. The disadvantage of this design is that the participant may start to predict or anticipate the task. In contrast, event-related design can be more easily randomized because of the short stimulus duration, and allows for stimulus events from different experimental conditions to be displayed in one run, something that is not possible with block designs. The main disadvantage of this type of deign is the weak signal-to-noise ratio (SNR), leading to a loss of statistical power. In general, the paradigm chosen will depend on the goal of the experiment (\cite{kashou2014practical}).

\begin{table}[H]
\begin{figure}[H]
  \centering
\begin{center}
\begin{tabular}{cc}
(a) Block design& (b) Mixed block/event related design\\
\includegraphics[width=.35\textwidth]{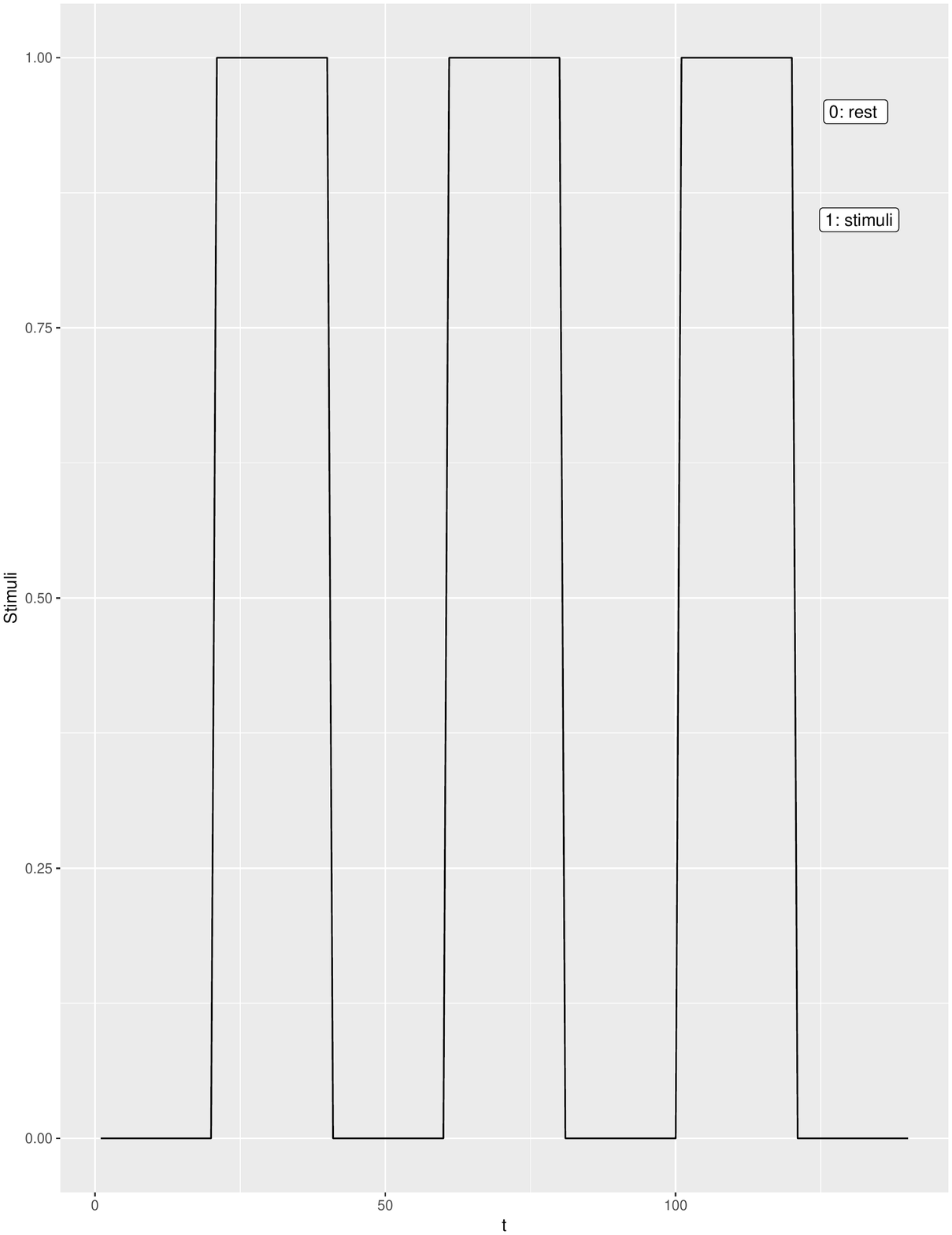}&\includegraphics[width=.35\textwidth]{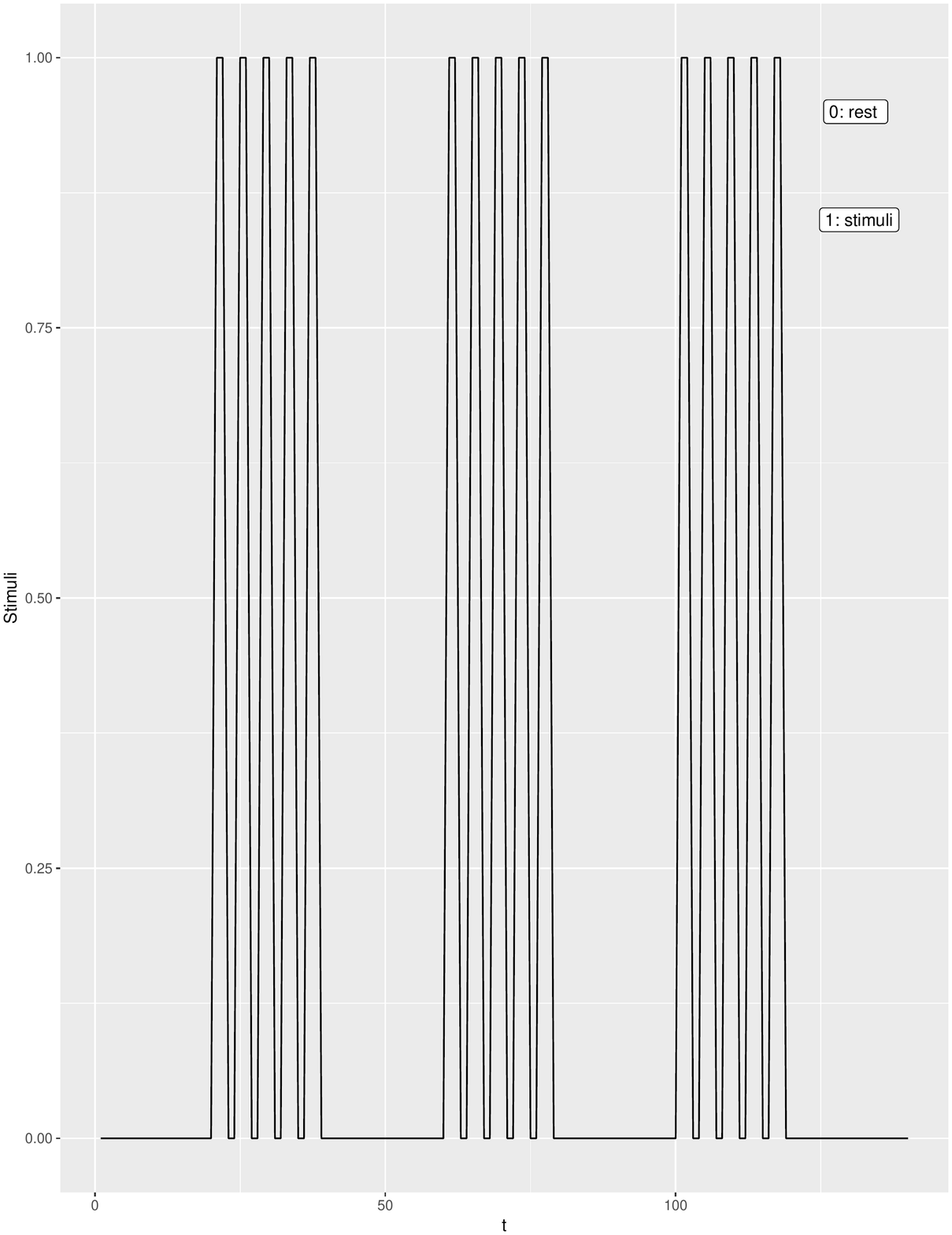}\\
(c) Event related design& (d) randomized event related design\\
\includegraphics[width=.35\textwidth]{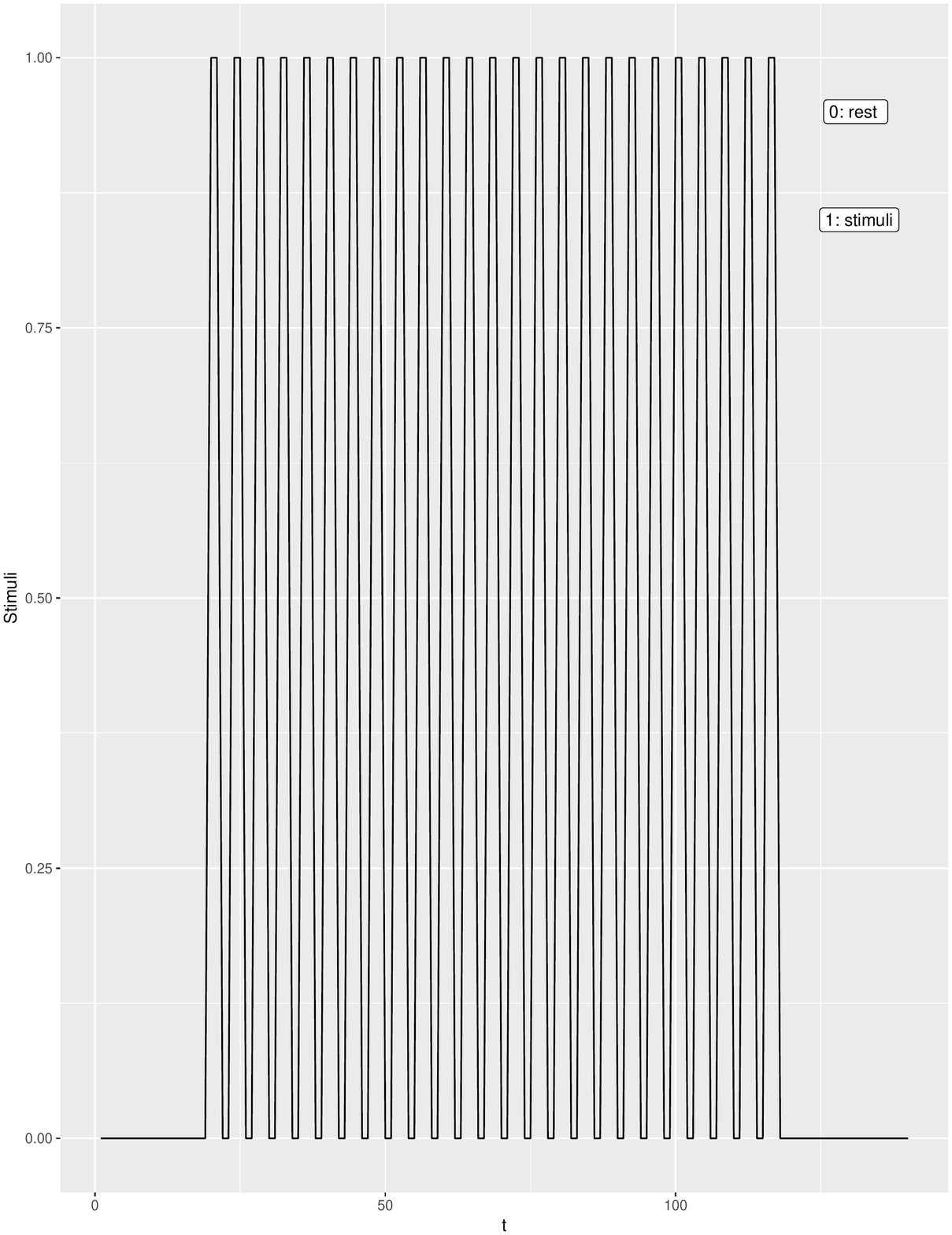}&\includegraphics[width=.35\textwidth]{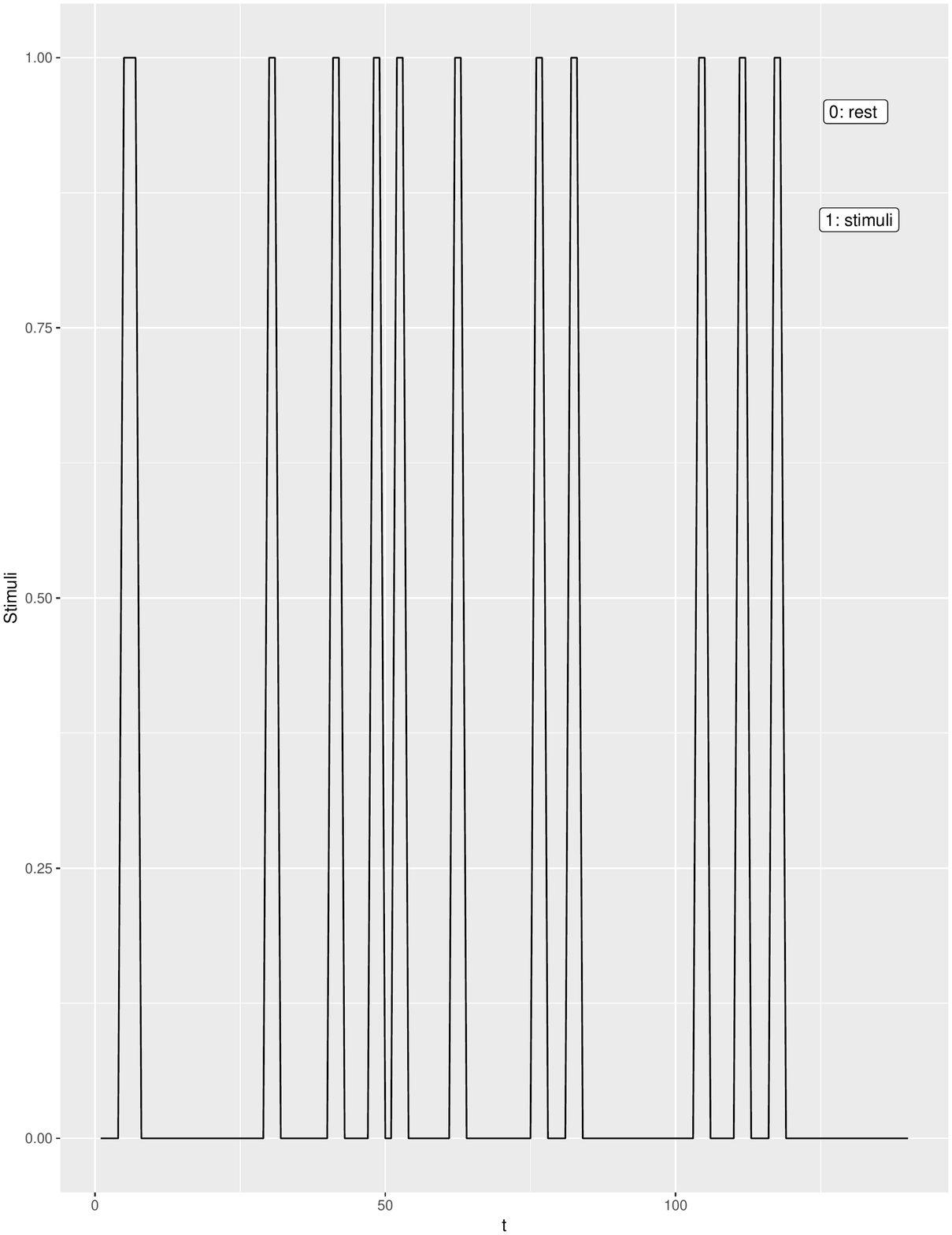}\\
\end{tabular}
\end{center}
  \caption{Some alternatives of designs broadly used in fMRI experiments.}
  \label{fig:tres} 
\end{figure}
\end{table}

\subsection{The blood-oxygenation-level dependent (BOLD) signal and the hemodynamic response function (HRF)}

\begin{figure}[H]
  \centering
  \includegraphics[width=.40\textwidth]{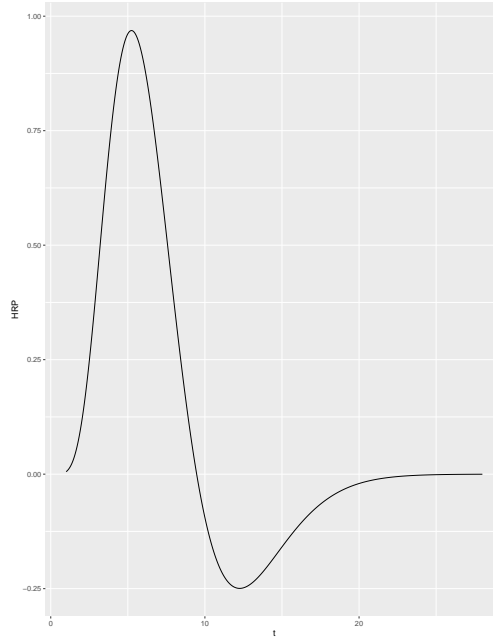} 
  \caption{Hemodynamic Response Function:$h$.}
  \label{fig:cuatro} 
\end{figure}

Above, we briefly described what an MRI is in terms of numeric information.  Now we explain how the neural activation, at least in an indirect way, can be pictured in a sequence of MRI or in an fMRI experiment, and in order to do so, we refer to \cite{poldrack2011handbook}: "The most common method of fMRI takes advantage of the fact that when neurons in the brain become active, the amount of blood flowing through that area is increased. This phenomenon has been known for more than 100 years, though the mechanisms that cause it remain only partly understood. What is particularly interesting is that the amount of blood that is sent to the area is more than is needed to replenish the oxygen that is used by the activity of the cells. Thus, the activity-related increase in blood flow caused by neuronal activity leads to a relative surplus in local blood oxygen. The signal measured in fMRI depends on this change in oxygenation and is referred to as the blood-oxygenation-level dependent, or BOLD signal." In figure~\ref{fig:cuatro} we can see the shape of the hemodynamic response function (HRF), which is the picture over time of the change of the BOLD signal. Thus, when the neuronal activity increases in a specific region of the brain, the BOLD signal also does, and the electromagnetic field generated by the MRI scanner captures those changes, allowing the phenomenon to be transformed into numeric information.

\subsection{The observed and expected BOLD signal}

One of the main objectives in an fMRI experiment is to identify a brain reaction in response to some controlled external stimuli, in other words, to look for changes in the BOLD signal related to some experimental manipulation. In order to do that, one can use any of the designs mentioned above. For example, in the figure \ref{fig:cinco} (left panel), we can see the observed BOLD signal and the block design stimuli for a particular volxel. In that case, the goal will be to find the voxel time series (i.e., the observed BOLD signal) that matches this pattern (block design). We can see, though, that the BOLD signal does not follow the block stimuli very well, due to slow physiological response. 

In this thesis, we do not deal so much with the properties of the BOLD response because it is not on the scope of this work, but we use its properties to create a predictor that will model the BOLD signal as accurately as possible.  If the reader is interested to know more about the BOLD response and its properties, refer to to \cite{poldrack2011handbook}. Thus, the stimulus time series, $f$ (such as in figure \ref{fig:cinco}, left panel) is blended with an HRF, $h$ (such as figure~\ref{fig:cuatro}), creating a shape that more closely represents the shape of the BOLD response, which is usually called ``expected BOLD response.''  See figure \ref{fig:cinco} right panel. This operation is given by the convolution 
 
\begin{equation}\label{equ2}
x(t)=(h\ast f)(t)=\int h(\tau)f(t-\tau)d\tau 
\end{equation}

\begin{table}[H]
\begin{figure}[H]
  \centering
\begin{center}
\begin{tabular}{cc}
\includegraphics[width=.40\textwidth]{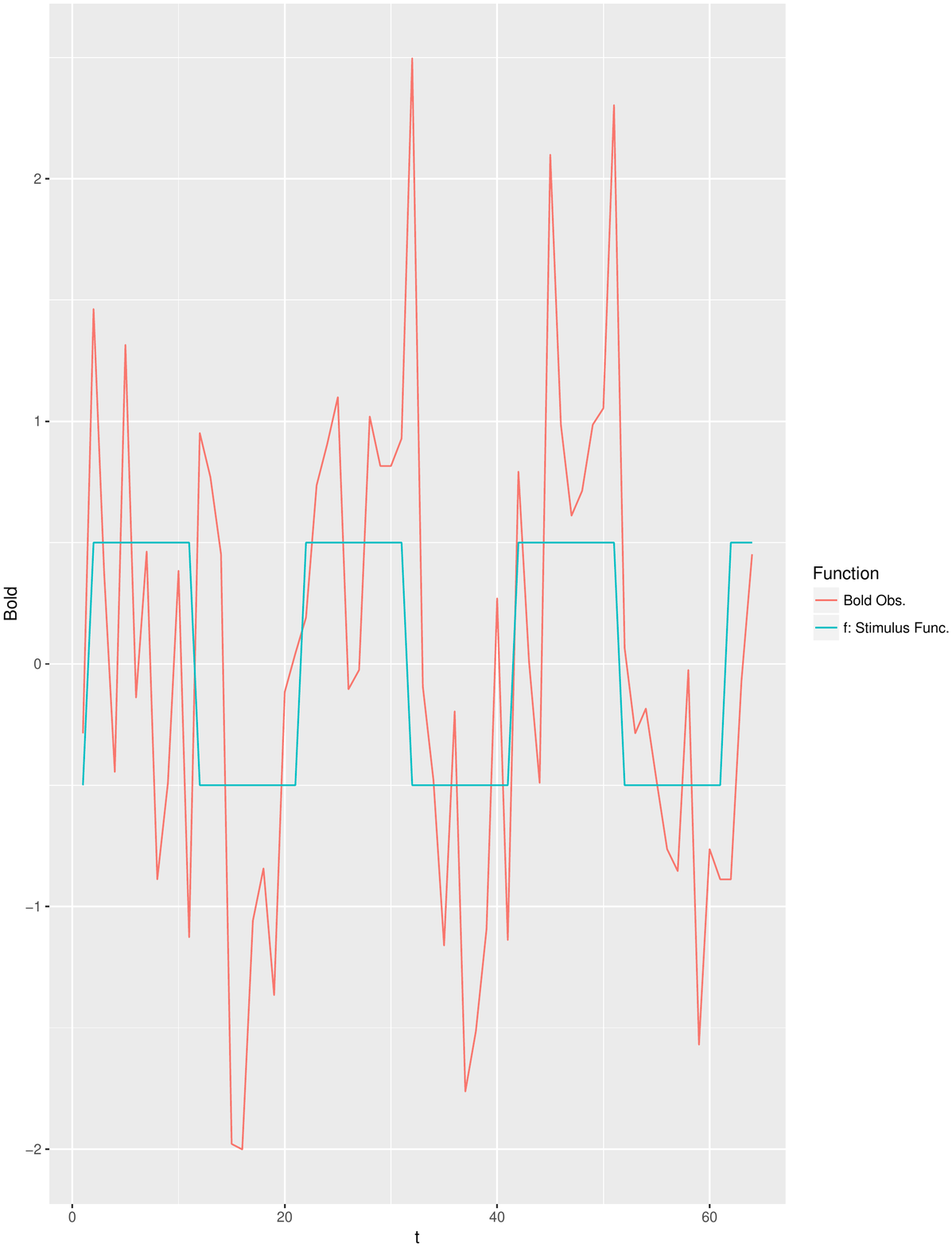}&\includegraphics[width=.40\textwidth]{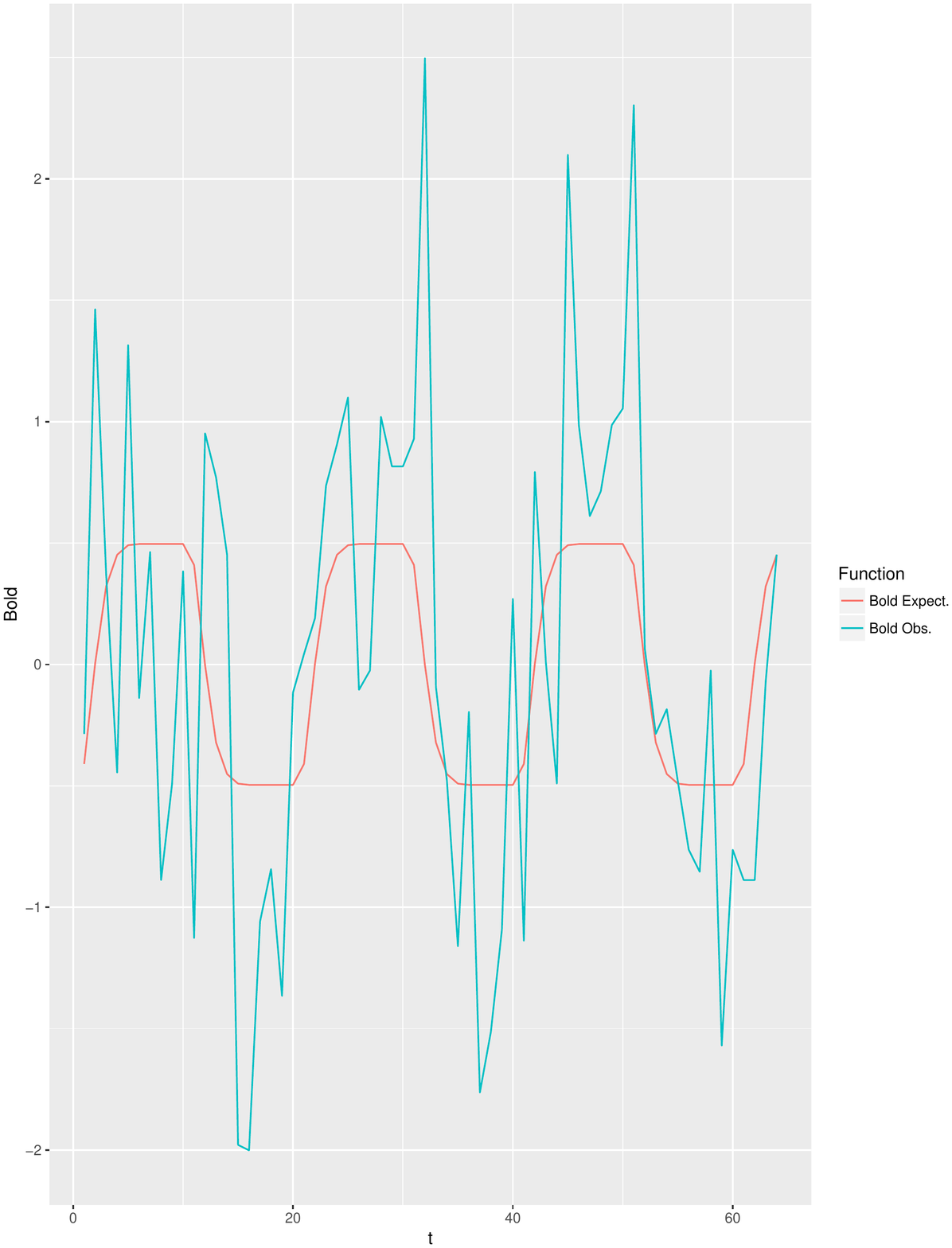}\\
\end{tabular}
\end{center}
  \caption{Left panel: Observed BOLD response (red line) and the stimulus time series (blue line). Right panel: Observed BOLD response (blue line) and the expected BOLD response (red line).}
  \label{fig:cinco} 
\end{figure}
\end{table}

Choosing an appropriate HRF is the key to capturing the best shape possible, which will ensure a good fit of the model being used (in our case, it will be a Bayesian multivariate dynamic linear model) on the BOLD observed when the signal is present (\citep{poldrack2011handbook}).

\section{Statistical modeling of the BOLD signal}

The most widely-used statistical model among users of the fMRI technique is the so-called General Linear Model (GLM), which is simply a normal linear regression model. There are several reasons that help to explain the popularity and success of this model in the fMRI field, but two very important ones are: its simplicity and robustness; and the fact that it can can be found as a standard tool in the most popular packages for fMRI data analysis. The usual specification of this model is given by

\begin{equation}
Y_v(t) = x(t)\beta_v + \epsilon_v, \quad \epsilon_v \sim N(0,\sigma^2),
\end{equation}\label{equ3}

for $v=1, \ldots,V$ and $t=1,\ldots,T$, where $V$ is the number of voxels in the fMRI array and $T$ is the number of observations in each time series. $Y_v(t)$ and $x(t)$ are the observed (obtained from the scanner) and expected ( obtained from equation~\ref{equ2}) BOLD response, respectively. Then the parameter $\beta_v$ will inform about the time series voxels ($Y_v$) that match with expected BOLD response $x(t)$. In this type of modeling, a key feature of any fMRI dataset is totally ignored: the spatio-temporal relationships. Thus, in the model \ref{equ3}, independent observations inside each voxel (temporal independence) are assumed, as is independence among voxels (spatial independence), both unrealistic suppositions. One of the main consequences of performing this type of analysis is getting incorrect inferences about the $\beta_v$ parameter and a high rate of false positives, in other words, identifying a brain activation when it really does not exist. A common practice to fix this problem is to use some sort of corrections, like the Bonferroni correction or spatial extent methods (\cite{worsley1995analysis}), among others. For instance, in figure~\ref{fig:seis} we can see an example of an activation of the visual cortex. In the left panel, the inference is performed without any type of correction and in the right panel, a Bonferroni correction is used. From this example, we might think that this kind of method could help solve the problem of detecting a false activation pattern. However, \cite{eklund2012does} and \cite{eklund2016cluster} evaluate the most common software packages for fMRI analysis using real data, and they find that in general, those correction methods do not work very well. Specifically, they use resting-state data and a total of 3 million random task group analyses to compute empirical familywise error rates. For a nominal familywise error rate of 5\%, parametric statistical methods are shown to be conservative for voxelwise inference (e.g., using the Bonferroni method) and invalid for clusterwise inference (e.g., using spatial extent methods).

\begin{table}[H]
\begin{figure}[H]
  \centering
\begin{center}
\begin{tabular}{cc}
(a)&(b)\\
\includegraphics[width=.40\textwidth]{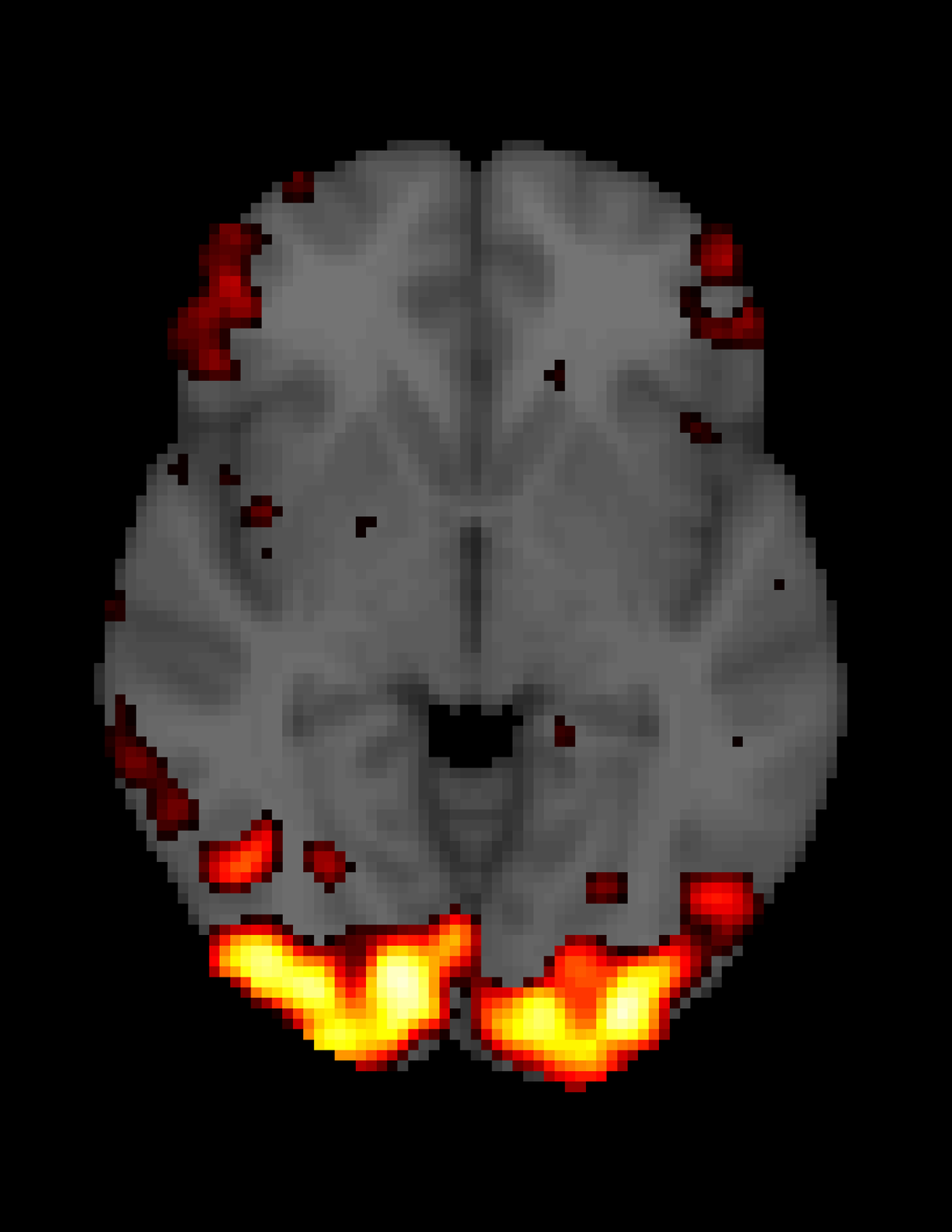}&\includegraphics[width=.40\textwidth]{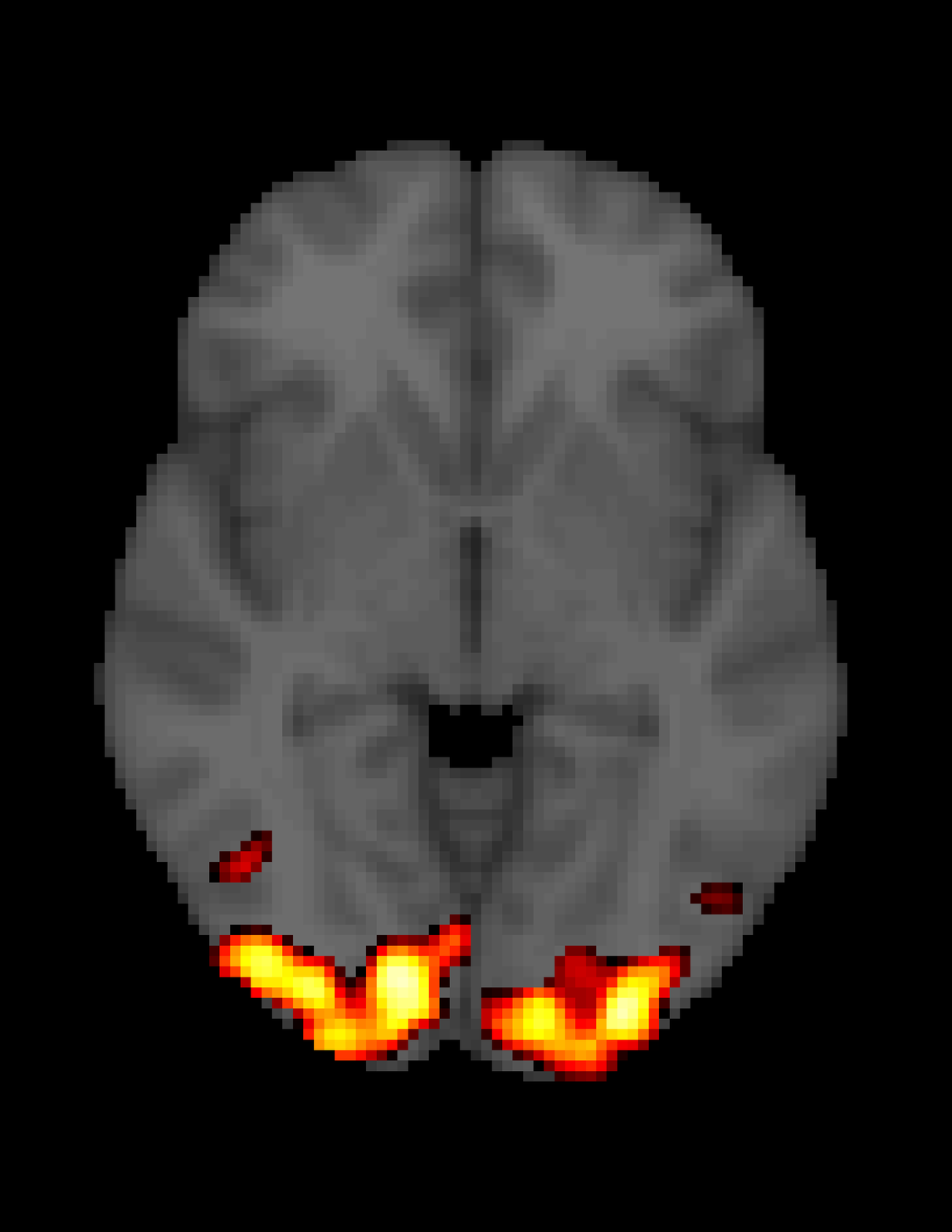}\\
\end{tabular}
\end{center}
  \caption{(a) Visual cortex activation without Bonferroni correction (a high rate of false activations or detected activations outside the visual cortex). (b) Visual cortex activation with Bonferroni correction (activation detected only in the visual cortex).}
  \label{fig:seis} 
\end{figure}
\end{table}

\subsection{Group analysis}

In fMRI studies, it is very common to perform group experiments, where the main interest could be to study either group activation or compare the activation response between two groups (e.g., patients versus controls). In order to do so, the analysis is divided in two stages. In the first stage, for each individual in each group, a statistical model (usally a GLM as in equation \ref{equ3} or a Multivariate DLM, as will be the case in this thesis) is used to model the BOLD response as described above. In the second stage, the $\beta_{vij}$ parameters obtained in the first are used to compute the group effect. Thus, for the $j-th$ group and the $v-th$ voxel the average effect is given by

\begin{equation*}
\bar{\beta}_{vj}=\frac{\sum \limits_{i=1}^{n_j} \beta_{vij}}{n_j},
\end{equation*}

where $n_j$ is the sample size of the group $j$, for $j=1,..,m$. For example, when $m=2$ one could perform any of the following tests: $\left\{ \bar{\beta}_{v1}>\bar{\beta}_{v2} \right\}$, $\left\{ \bar{\beta}_{v2}>\bar{\beta}_{v1} \right\}$, $\left\{ \bar{\beta}_{vj}>0\right\}$ and $\left\{ \bar{\beta}_{vj}=0\right\}$. The type of test to be performed rely on the interest of the researcher.

\subsection{About software packages for fMRI analysis}

The most common software packages for fMRI analysis are SPM, FSL and AFNI. Among them, only SPM has a complete option for Bayesian modeling for both first-stage and second-stage analysis. For the first-level analysis, the user can use spatial priors for regression coefficients and regularized voxel-wise $AR(p)$ models for the fMRI noise processes. The second stage uses the empirical Bayes algorithm with global shrinkage priors. However, those stages are not connected.  In other words, the second stage does not use the results obtained in the first stage. Thus, if the user wants to perform Bayesian modeling in the second stage, then he or she must have already estimated a frequentist model in the first stage and used those estimations as an input for the second, which means that with SPM it is not possible to perform fully Bayesian modeling for group fMRI analysis. In the case of FSL there is only a Bayesian option for the second stage or group comparison. In the AFNI package, there are only frequentist modeling options.

        % associado ao arquivo: 'cap-introducao.tex'
%% ------------------------------------------------------------------------- %%
\chapter{Concepts and Foundations}
\label{cap:2}

In this chapter, we review some concepts and theoretical results that are necessary for the achievement of the goals we want to reach in this thesis. Most of them are known results in the literature of Bayesian analysis. First, we review the Bayesian multivariate dynamic linear model (MDLM), which will be useful for modeling the observed BOLD response. We define the model and show the update theorem as in \cite{west1997bayesian}. Something that is worth mentioning here is that we also propose a procedure to perform inference on the dynamic parameter $\theta_t$, which is presented along with the other known results. Second, we present the functional ANOVA model (\cite{kaufman2010bayesian}) which will be useful to perform the group comparison in an fMRI experiment. In that work, \cite{kaufman2010bayesian} defines a generalization of the functional ANOVA model, performs the computation of the variance components as it is done in \cite{gelman2005analysis}, and propose graphical displays to analyze functional parameters of the functional ANOVA model. They leave the inferential procedure for the functional parameters as future work or an open problem. We propose a simple procedure based on Monte Carlo integration to compute posterior probabilities and perform the inference needed to identify differences between groups.

\section{Bayesian multivariate dynamic linear model}

The general theory of the MDLM is presented in \cite{quintana1987multivariate} and \cite{west1997bayesian}. Despite this model having been conceived for forecasting, in this thesis it is used for a different purpose: size effect estimation. In other words, we focus our interest on the estimation of, and inference on, the state parameter $\mathbf{\Theta}_t$. The framework developed in the references above is as follows. Suppose that we have a vector $\mathbf{Y}_t$, which can be modeled in terms of observation and evolution or state equations as follows.

\begin{equation} \label{chap2:equ1}
\begin{array}{lccl}
\text{Observation:}\ &\mathbf{Y}_t & =& F^{'}_t\mathbf{\Theta}_t + \mathbf{\nu}^{'}_t \\
 \text{Evolution:}\ &\mathbf{\Theta}_t& =& G_t\mathbf{\Theta}_{t-1} + \Omega_t.\\
\end{array}
\end{equation}

Where, for each $t$ we have
\begin{itemize}
\item $\mathbf{Y}_t= \left(Y_{t1}, \ldots, Y_{tq}\right)^{'}$, the $q-\text{vector}$ of observations at time $t$;
\item $\mathbf{\nu}^{'}_t=\left(\nu_{t1}, \ldots, \nu_{tq}\right)^{'}$, the $q-\text{vector}$ of observational errors at time $t$;
\item $\mathbf{\Theta}_t= \left[\mathbf{\theta}_{t1}, \ldots, \mathbf{\theta}_{tq}\right]$, the $p \times q$ matrix whose columns are the state vectors related to each of the $q-\text{observational equations}$ in \ref{chap2:equ1}.
\item $\Omega_t=\left[\omega_{t1}, \ldots,\omega_{tq}\right]$, the $p \times q$ matrix whose columns are the evolution errors of each of the $q-\text{evolution equations}$ in \ref{chap2:equ1}.
\end{itemize}

\begin{figure}[H]
  \centering
  \includegraphics[width=.40\textwidth]{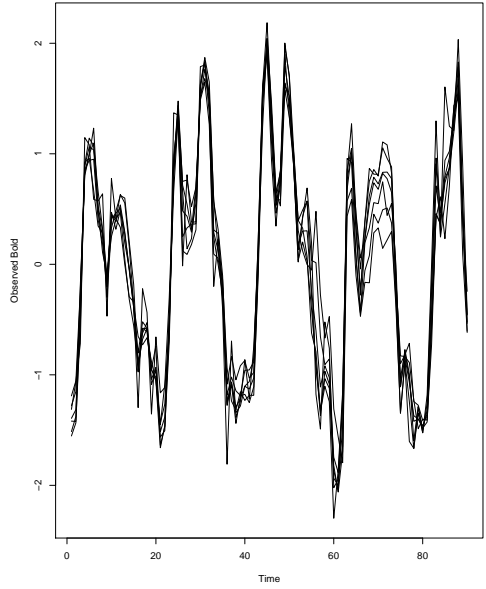} 
  \caption{Cluster of neighboring series from the visual cortex obtained under a block design applying visual stimulus.}
  \label{fig:21} 
\end{figure}

In the model given by \ref{chap2:equ1},  both $F^{'}_t$ and $G_t$ are common to each of the $q$ univariate DLM. This is a key aspect of this model for the case studied in this thesis, because, as we will show in the next chapter, $F^{'}_t$ and $G_t$ are the same for every voxel in our fMRI one-subject analysis. Another important aspect related to model \ref{chap2:equ1} highlighted by \cite{west1997bayesian} is that this model is appropiate in applications when several similar series are to be analyzed. In figure~\ref{fig:21}, we can see a cluster of temporal series related to an fMRI experiment where it is clear they are similar and evolve together. This is a common behavior of fMRI time series when neighborhoods of voxels are analyzed.

Now to proceed, it is necessary to identify the distributions of the observational error vector $\nu_t$ and the evolution error matrix $\Omega_t$. The former is multivariate normal,

\begin{equation*}
\nu_t \sim N\left[\mathbf{0}, V_t \mathbf{\Sigma}\right],
\end{equation*}

independently over time, where $\mathbf{\Sigma}$ defines the cross-sectional covariance structure for the multivariate model. The latter is a matrix-variate normal distribution (\cite{dawid1981some}), described as follows.\\

The random matrix $\Omega_t$ has a matrix normal distribution with mean matrix $\mathbf{0}$, left variance matrix $\mathbf{W}_t$ and right variance matrix $\mathbf{\Sigma}$. The density function is given by

\begin{equation*}
p(\Omega_t)=k(\mathbf{W}_t, \mathbf{\Sigma})\exp{ \left(-\frac{1}{2}trace\left[\Omega_t^{'}\mathbf{W}_t^{-1}\Omega_t\mathbf{\Sigma}^{-1}\right]\right)},
\end{equation*}

where

\begin{equation*}
k(\mathbf{W}_t, \mathbf{\Sigma})=(2*\pi)^{q*n/2}|\mathbf{W}_t|^{-q/2}|\mathbf{\Sigma|^{-n/2}}.
\end{equation*}

The distribution for $\Omega_t$ is given by

\begin{equation*}
\Omega_t \sim N\left[\mathbf{0}, \mathbf{W}_t, \mathbf{\Sigma}\right].
\end{equation*}

Suppose also that the initial prior for $\mathbf{\Theta}_0$ and $\mathbf{\Sigma}$ is matrix normal/inverse Wishart (\citep{quintana1987multivariate}[Chapter 3]), namely

\begin{equation}\label{chap2:equ2}
(\mathbf{\Theta}_0, \mathbf{\Sigma}|D_0) \sim NW_{n_0}^{-1}\left[\mathbf{m}_0,\mathbf{C}_0, \mathbf{S}_0 \right],
\end{equation}

for some known defining parameters $\mathbf{m}_0$, $\mathbf{C}_0$, $\mathbf{S}_0$ and $n_0$. $D_t$ are the data observed at time $t$.  Then, for all times $t>1$, the following results apply.

\subsection{Update Theorem and Posterior Inference}

\begin{theorem}[Actualization theorem]\label{theorem1}
One-step forecast and posterior distributions in the model \ref{chap2:equ1} are given, for each $t$, as follows.

\begin{itemize}
\item[(a)] Posteriors at $t-1$:\\ For some $\mathbf{m}_{t-1}$, $\mathbf{C}_{t-1}$, $\mathbf{S}_{t-1}$ and $n_{t-1}$,

\begin{equation*}
(\mathbf{\Theta}_{t-1}, \mathbf{\Sigma}|D_{t-1}) \sim NW_{n_{t-1}}^{-1}\left[\mathbf{m}_{t-1},\mathbf{C}_{t-1}, \mathbf{S}_{t-1} \right],
\end{equation*}

\item[(b)] Priors at $t$:

\begin{equation*}
(\mathbf{\Theta}_{t}, \mathbf{\Sigma}|D_{t-1}) \sim NW_{n_{t-1}}^{-1}\left[\mathbf{a}_{t},\mathbf{R}_{t}, \mathbf{S}_{t-1} \right],
\end{equation*}

where $\mathbf{a}_{t}= \mathbf{G}_t\mathbf{m}_{t-1}$ and $R_{t}=\mathbf{G}_t\mathbf{C}_{t-1}\mathbf{G}_t^{'} + \mathbf{W}_t$.

\item[(c)] One-step forecast:
 \begin{equation*}
(\mathbf{Y}_t| \mathbf{\Sigma}, D_{t-1})\sim N\left[\mathbf{f}_t, Q_t\mathbf{\Sigma}\right],
\end{equation*}
with marginal 

 \begin{equation*}
(\mathbf{Y}_t| D_{t-1})\sim T_{n_{t-1}}\left[\mathbf{f}_t, Q_t\mathbf{S}_{t-1}\right],
\end{equation*}
where $\mathbf{f}_t^{'}=\mathbf{F}_t^{'}\mathbf{a}_t$ and $Q_t=V_t + \mathbf{F}_t^{'}\mathbf{R}_t\mathbf{F}_t$.

\item[(d)] Posterior at $t$: 

\begin{equation*}
(\mathbf{\Theta}_{t}, \mathbf{\Sigma}|D_{t}) \sim NW_{n_{t}}^{-1}\left[\mathbf{m}_{t},\mathbf{C}_{t}, \mathbf{S}_{t} \right],
\end{equation*}

with $\mathbf{m}_{t}=\mathbf{a}_{t}+ \mathbf{A}_{t}\mathbf{e}_{t}^{'}$ and $\mathbf{C}_{t}=\mathbf{R}_{t}-\mathbf{A}_{t}\mathbf{A}_{t}^{'}Q_t$, $n_t=n_{t-1}+1$ and $\mathbf{S}_{t}=n_{t}^{-1}\left[n_{t-1}\mathbf{S}_{t-1} + \mathbf{e}_{t}\mathbf{e}_{t}^{'}/Q_t\right]$, where $\mathbf{A}_{t}=\mathbf{R}_{t}\mathbf{F}_{t}/Q_t$ and $\mathbf{e}_{t}=\mathbf{Y}_{t} - \mathbf{f}_{t}$.
\end{itemize} 

\end{theorem}

Full details of the proof appear in \cite{quintana1987multivariate}.\\

As a consequence of the update theorem, we have that the posterior marginal distribution of $\mathbf{\Theta}_{t}$ is given by

\begin{equation}\label{chap2:equ3}
(\mathbf{\Theta}_{t}|D_{t}) \sim T_{n_{t}}\left[\mathbf{m}_{t},\mathbf{C}_{t}, \mathbf{S}_{t} \right],
\end{equation}

where $T_n\left[\mathbf{m}, \mathbf{C}, \mathbf{S} \right]$ is called the matrix $T$ distribution (\cite{dawid1981some}). \\

\subsubsection*{Posterior Inference}

As we can see from the above results, there is a posterior distribution $p(\mathbf{\Theta}_{t}|D_{t})$ for each time $t$, for $t= 1, \ldots, T$. We mentioned at the beginning of this chapter that our main interest with the MDLM is to perform inference on $\mathbf{\Theta}$ rather than  forecast $\mathbf{Y}_t$. Thus, any necessary inference to be performed could be based on posterior probabilities. For example, \cite{west1997bayesian} define a measure of evidence against of the hypothesis $\mathbf{\theta}_{1t}=0$ as 

\begin{equation*}
\alpha = Pr\left[F_{q,n_t} \geq q^{-1}\mathbf{m}_{t1}^{'}\mathbf{C}_{t1}^{-1}\mathbf{m}_{t1},\right]
\end{equation*}

where $\mathbf{\theta}_{1t}$, $\mathbf{m}_{t1}$ and $\mathbf{C}_{t1}$ are subsets of $q$ elements from $\mathbf{\Theta}_{t}$, $\mathbf{m}_{t}$ and $\mathbf{C}_{t}$ respectively, and $F_{q,n_t}$ denotes a random quantity having the standard $F$ distribution with $q$ degrees of freedom in the numerator and $n_t$ in the denominator. Thus, a small value of $\alpha$ indicates rejection of the hypothesized value $\mathbf{\theta}_{1t}=0$ as unlikely. To our knowledge, this last reference is perhaps the only one in the DLM literature that defines a formal testing procedure for the state parameters $\mathbf{\theta}_{t}$. However, a question arises from this test procedure: must this test be performed for all $t\in\{1, \ldots, T\},$ or is it enough to test for a subset $t\in\{m, \ldots, T\}$, for $m>1$? We propose two ways to test $H_0:\mathbf{\theta}_{1t}\in\mathbf{\theta_{0t}}$, where $\mathbf{\theta_{0t}}$ is a subset of the parameter space associated to $\mathbf{\Theta}_t$. The first is a simple test procedure based on the last posterior distribution $p(\mathbf{\Theta}_{T}|D_{T})$, taking advantage of the sequential update procedure from the update theorem. Thus, the last posterior distribution $p(\mathbf{\Theta}_{T}|D_{T})$ can be seen as the most recent belief update of $\mathbf{\Theta}$, and the measure of evidence against $H_0$ is a simple posterior probability. The second test procedure is based on the posterior distributions $p(\mathbf{\Theta}_{m}|D_{m}), p(\mathbf{\Theta}_{m+1}|D_{m+1}), \ldots, p(\mathbf{\Theta}_{T}|D_{T})$, for $m \in \{1, \ldots, T\}$. The choice of $m$ may depend on the particularities of the application. We discuss that choice in more detail in the next chapter. We define the next algorithm to draw curves $\theta=\left(\theta_m, \theta_{m+1}, \ldots, \theta_T \right)$ (or  on-line estimated trajectory of $\theta_t$).

\begin{algorithm}
\caption{Our algorithm}\label{euclid}
\begin{algorithmic}[1]
\Procedure{MyProcedure}{}
\State Draw $\theta^{(k)}_t$ from $p(\theta_t|D_t)$ for $t=1,\ldots,T$
\State Draw $\nu^{(k)}$ from $p(\nu_t|D_t)$ for $t=1,\ldots,T$
\State Compute $y^{(k)}_t=F_t^{'}\theta_t^{(k)} + \nu^{*}_t$ for $t=1,\ldots,T$
\State Compute $p(\theta^{(k)}_t|D_t)$ for $t=1,\ldots,T$ and take $\tilde{\theta}^{(k)}_t=E(\theta_t^{(k)}|D_t)$
\State Let $\tilde{\boldsymbol{\theta}}^{(k)}=(\tilde{\theta}_m^{(k)}, \ldots, \tilde{\theta}_T^{(k)})$
\EndProcedure
\end{algorithmic}
\end{algorithm}

 Then, our measure of evidence is computed as
\begin{equation*}
p(\boldsymbol{\theta}\in \boldsymbol{\theta}_{0})= E(1_{(\boldsymbol{\theta}\in \boldsymbol{\theta}_{0})})\approx \frac{\sum\limits_{k=1}^{N} 1_{(\tilde{\boldsymbol{\theta}}^{(k)}\in \mathbf{\theta_{0}})}}{N},
\end{equation*}

where $\boldsymbol{\theta}_{0}$ is a subset of the parameter space associated with $\mathbf{\Theta}_t$. It can be noticed that this last procedure is only suitable for a composite hypothesis.  In the next chapter, we explain why composite hypothesis are more appropiate in order to detect voxel activation. However, we could also define sharp hyphoteses of the form $H_0:\mathbf{\theta}_{1t}=\mathbf{\theta_{0t}}$ using the Fully Bayesian Significance Test (FBST) (\cite{pereira2008can}). In the R package we intend to build with the results obtained from this thesis, the user will be able to choose the best test procedure --composite or sharp-- according to his or her needs.

\section{Gaussian Process ANOVA Model}

In this section, we present the Bayesian functional ANOVA model introduced by \cite{kaufman2010bayesian}, which is used here to compare batches of curves associated with on-line estimated trajectories of $\theta_t$. In the next chapter, we explain in more detail why and how this comparison is made, but for now let's suppose that we have the curve $\theta_{ig}=\left(\theta_{m,ig}, \theta_{m+1,ig}, \ldots, \theta_{T,ig} \right)$, for $i=1,\ldots,n_g$, $m \in \{1, \ldots, T\}$ and $g \in \{ A, B\}$. Then, we are interested in comparing the batch of curves from group $A$ to the batch of curves from group $B$. In figure~\ref{cap2:fig2}, we can see an example of two batches of curves of on-line estimated trajectories of $\theta_{ig}$. Each curve is obtained fitting the model \ref{chap2:equ1} to the same voxel (or neighborhood of voxels) for each subject belonging to each group. It is worth mentioning that this comparison is performed for each voxel from the fMRI data array.

\begin{table}[H]
\begin{figure}[H]
  \centering
\begin{center}
\begin{tabular}{cc}
Group A&Group B\\
\includegraphics[width=.40\textwidth]{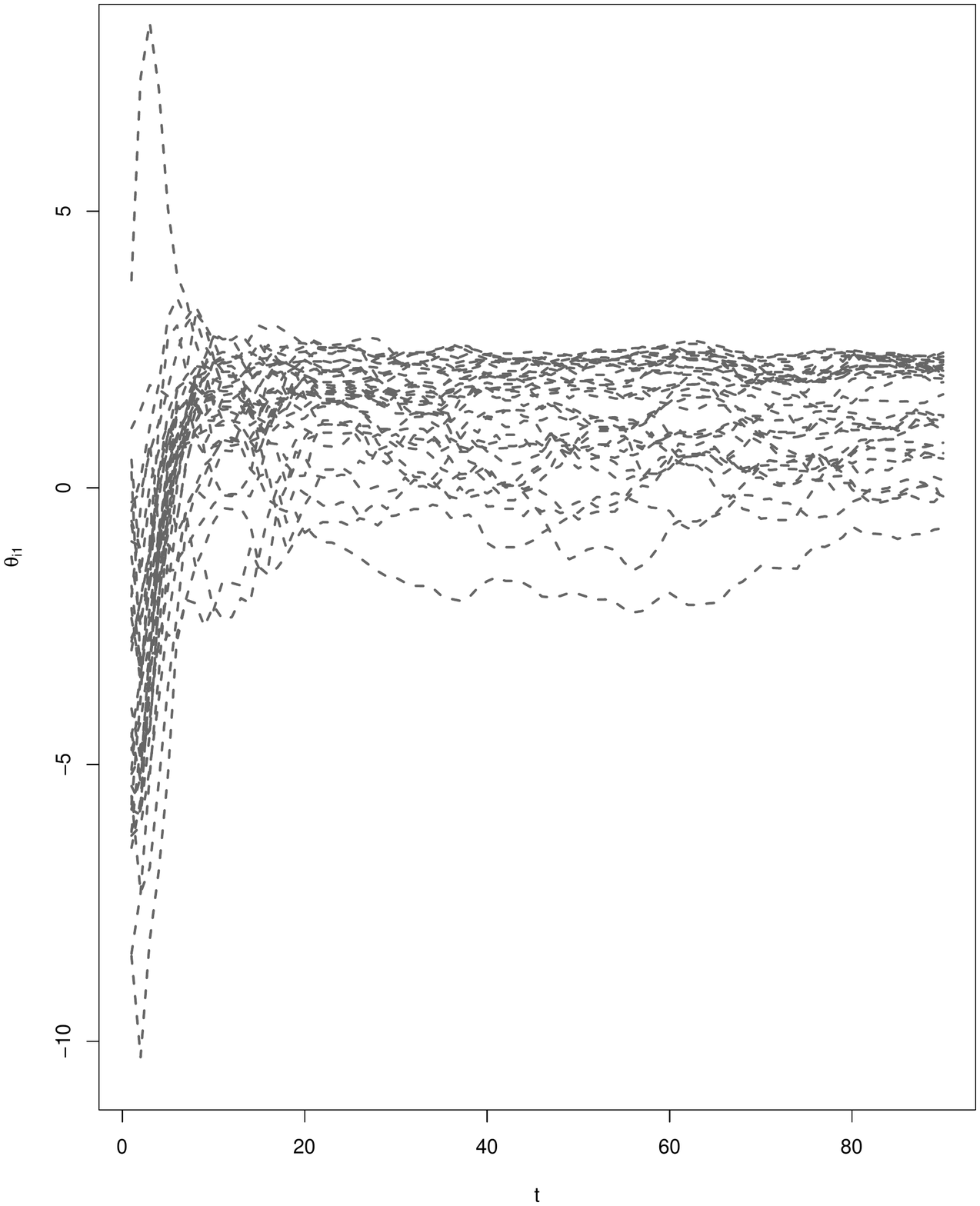}&\includegraphics[width=.40\textwidth]{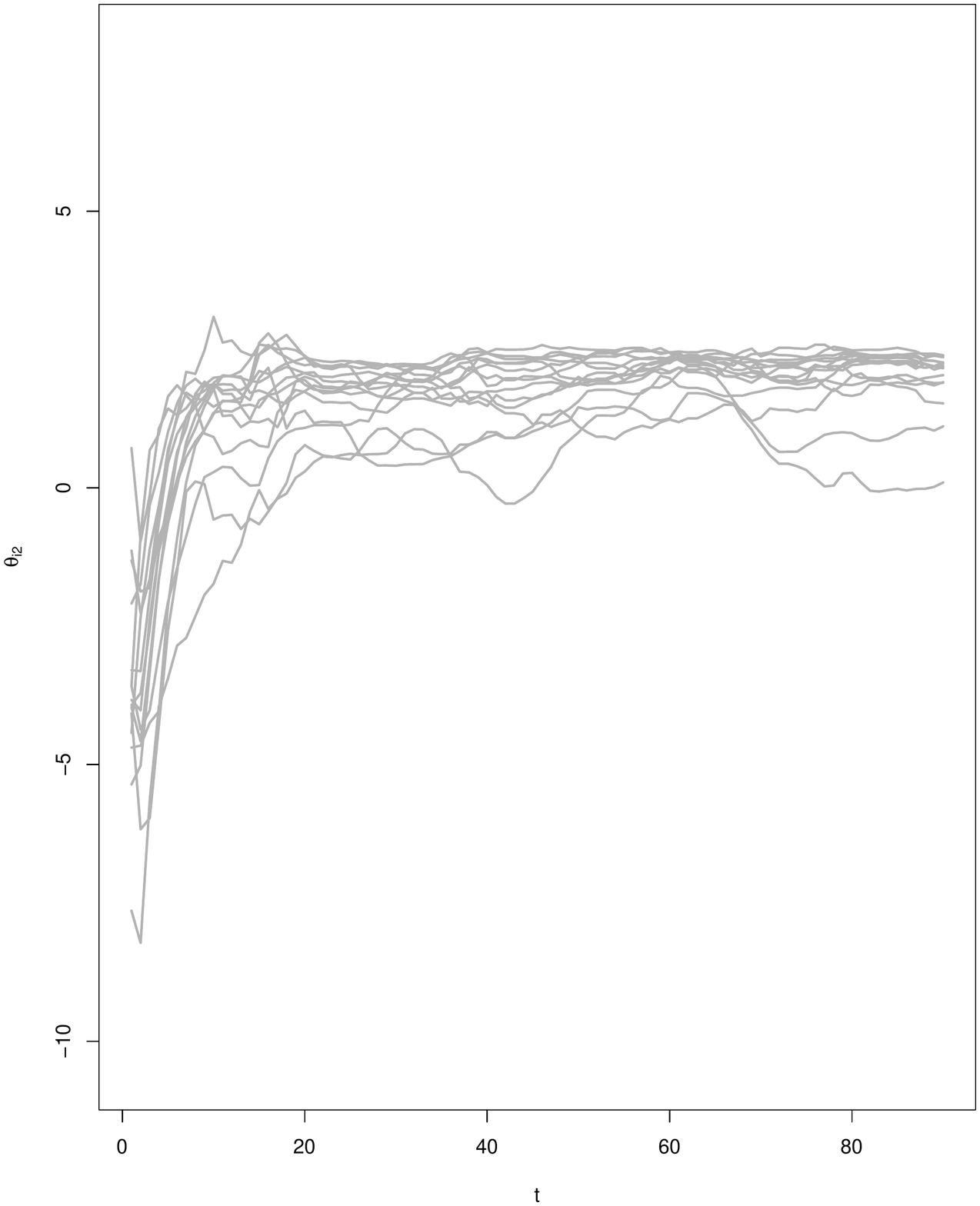}\\
\multicolumn{2}{c}{Mean group curves.}\\
\multicolumn{2}{c}{\includegraphics[width=.40\textwidth]{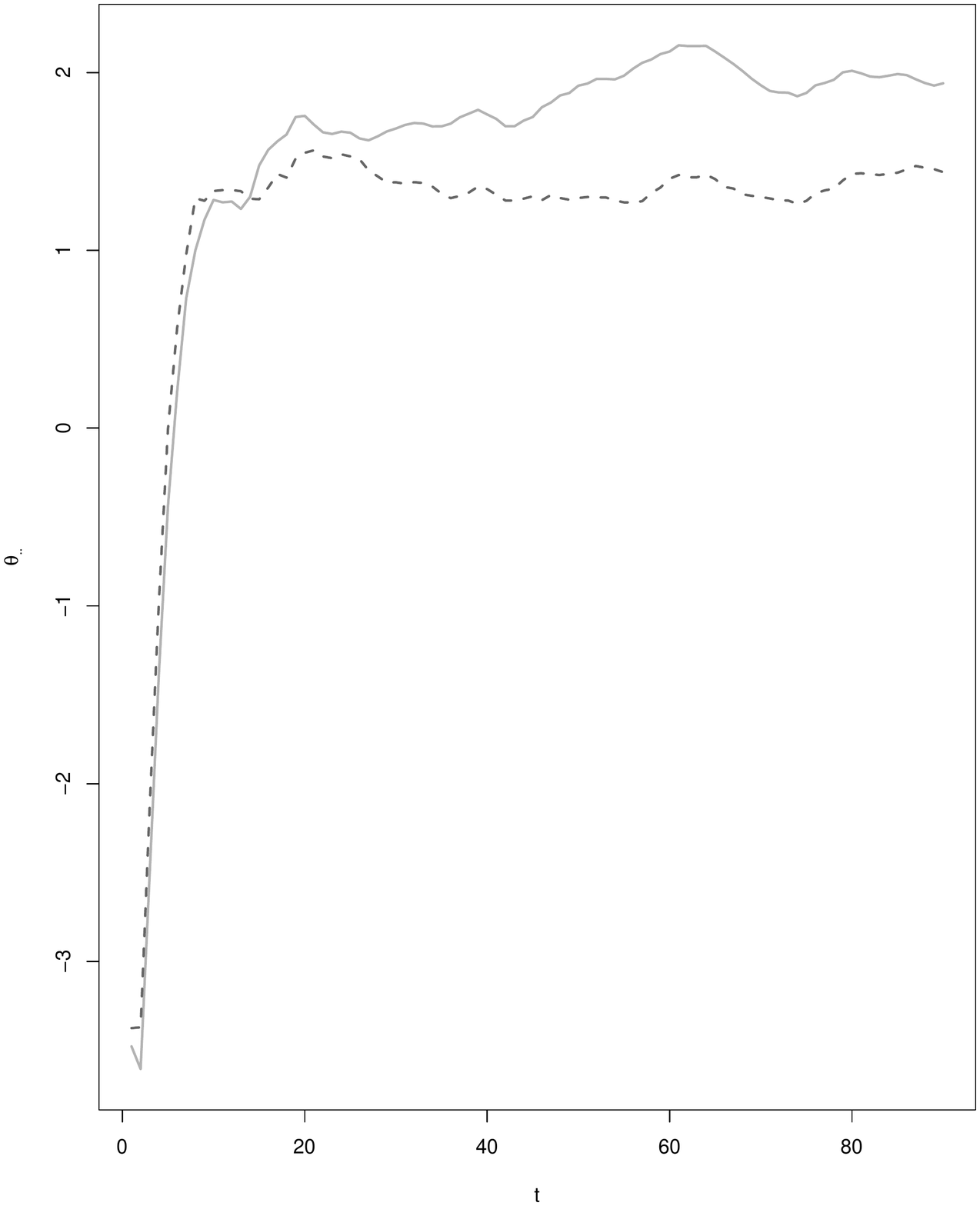}}\\
\end{tabular}
\end{center}
  \caption{Top left to right panels: the on-line estimated trajectories of the parameter $\theta_t$ in a fixed voxel for two different groups of subjects. Bottom panel: the average online estimated trajectory for each group.}
  \label{cap2:fig2} 
\end{figure}
\end{table}

Before describing the Gaussian process ANOVA model, we have to begin with the definition of an important concept.\\
\textbf{Definition}: A Gaussian process is a stochastic process such that any finite subcollection of random variables has a multivariate normal distribution.  In particular, a Gaussian process is a stochastic process parametrized by its mean function $\mu(\cdot)$ and covariance function $K(\cdot, \cdot)$, which we denote $GP(\mu, K)$.\\
The fundamentals of Gaussian processes applying to functional regression models can be found in \cite{shi2011gaussian} and \cite{rasmussen2006gaussian}. Thus, following the ideas of \cite{kaufman2010bayesian}, we model $\theta_{ig}=\left(\theta_{t_1,ig}, \theta_{t_2,ig}, \ldots, \theta_{t_p,ig} \right)^{'}$ as a finite set of observations from an underlying smooth realization of a stochastic process defined for $t\geq 0$, where $\theta_{ig}$ represent the $i^{th}$ response at group $g$. Let $\mu_{i}(t)= \mu(t)+\alpha_i(t)$. Then the first stage of the model is

\begin{equation*}
\theta_{ig}|\{\mu_i \}, \sigma_{\epsilon}^2, \tau_{\epsilon} \widesim{indep} GP(\mu_i, \sigma_{\epsilon}^2  R_{\tau_{\epsilon}})
\end{equation*}

for $i=1,\ldots,n_g$ and $g \in \{ A, B\}$. We now specify Gaussian process prior distributions for $\mu$ and $\{\alpha_i\}$, taking each batch of functions to be independent of the other batches and independent of the residuals a priori, and assigning each batch its own set of higher-level parameters. Thus, for the second stage of the model we have

\begin{equation}
\mu|\phi,\sigma^2_{\mu}, \tau_{\mu} \sim GP(\phi 1_p, \sigma^2_{\mu}R_{\tau_{\mu}}),
\end{equation}

where $1_p$ is a vector of length $p$ with all its entries equal to one and $\phi$, $\sigma^2_{\mu}$, $\tau_{\mu}$ are unknown hyperparameters. The prior distribution for $\{\alpha_i\}$ satisfy the constraints $\sum_{i\in\{ A, B\}} \alpha_i(t)=0$ for all $t$. Specifically, a prior distribution for $\{\alpha_i \}$ is defined such that each $\alpha_i$ is marginally a mean zero Gaussian process, and

\begin{equation}
Cov(\alpha_i(t), \alpha_{i^{'}}(t^{'}) )=\begin{cases} 
      \left(1-\frac{1}{2} \right)\sigma_{\alpha}^{2}R_{\tau_{\alpha}}(t,t^{'}) & i=i^{'} \\
      -\frac{1}{2}\sigma_{\alpha}^{2}R_{\tau_{\alpha}}(t,t^{'}) & i\neq i^{'}     
   \end{cases}
\end{equation}

Then, if we express the joint distribution of $\mathbf{\alpha}_A$ and $\mathbf{\alpha}_B$ as $p(\mathbf{\alpha}_A, \mathbf{\alpha}_B|\sigma_{\alpha}^{2}, \tau_{\alpha})=p(\mathbf{\alpha}_B|\mathbf{\alpha}_A, \sigma_{\alpha}^{2}, \tau_{\alpha})p(\mathbf{\alpha}_A|\sigma_{\alpha}^{2}, \tau_{\alpha})$ we get that  

\begin{equation}
\alpha_A|\sigma^2_{\alpha}, \tau_{\mu} \sim GP(\mathbf{0}, \sigma^2_{\alpha}R_{\tau_{\alpha}}),
\end{equation}

and the distribution of $\mathbf{\alpha}_B|\mathbf{\alpha}_A, \sigma_{\alpha}^{2}, \tau_{\alpha}$ is degenerate, reflecting the sum-to-zero constraint.

\subsection{Posterior Sampling and Inference}

Now we define the prior distributions for the hyperparameters $\sigma^{2}_{\cdot}$, $\tau^{2}_{\cdot}$ and $\phi$ in order to compute the posterior distribution of the unknown quantities, specifically the posterior distribution of $\{\alpha_i\}$ which will inform us about the difference between the two groups $A$ and $B$. As in \citep{kaufman2010bayesian}, we use uniform prior distributions. In the next chapter, we will explain further about the details of such prior distributions.  For the moment, let $p(\sigma^{2}_{\cdot})=I(\sigma^{2}_{\cdot})_{[a_1, b_1]}$, $p(\tau^{2}_{\cdot})=I(\tau_{\cdot})_{[a_2, b_2]}$  and $p(\phi)=I(\phi)_{[a_3, b_3]}$. In this case, we cannot get a closed form for the posterior distribution, so we use an MCMC algorithm to generate posterior samples from it. Specifically, we use the Gibbs sampler for the distributions with a closed form for their full conditional distributions, and the general Metropolis-Hastings algorithm for the distributions not available in closed form. 

\subsubsection*{Full Conditional Distributions}

The full conditional posterior distributions for $\phi$,  $\mathbf{\mu}$, $\{\mathbf{\alpha_i}\}$, $\sigma_{\mu}^2$ and $\sigma_{\alpha}^{2}$ are given by

\begin{equation*}
\begin{array}{rcl}
\phi|\mu,\sigma_{\mu}^{2}, \tau_{\mu}^2 &\sim & N(M_1, V_1^{-1}),\\
\mu|\phi, \alpha_A, \sigma_{\mu}^{2}, \tau_{\mu},\sigma_{\alpha}^{2}, \tau_{\alpha}, Data &\sim & N_p(M_2, V_2^{-1}),\\
\alpha_A|\mu, \sigma_{\mu}^{2}, \tau_{\mu},\sigma_{\alpha}^{2}, \tau_{\alpha}, Data&\sim & N_p(M_3, V_3^{-1}),\\
\sigma_{\epsilon}^2|\mu, \alpha_A, \tau_{\epsilon}, Data&\sim & TruncatedIG_{(a_1,b_1)}((n_A+n_B)p/2-1; (D_{0A}+D_{0B})/2)\\
\sigma_{\mu}^2|\mu, \phi, \tau_{\mu}&\sim & TruncatedIG_{(a_1,b_1)}(p/2-1; D_1/2),\\
\sigma_{\alpha}^{2}|\alpha_A,\tau_{\alpha}&\sim & TruncatedIG_{(a_1,b_1)}(p/2+1; D_2),\\
\end{array}
\end{equation*}
where 

\begin{equation*}
\begin{array}{l}
M_1=V_1^{-1}(1_pR_{\tau_{\mu}}^{-1}\mu); V_1=1_p^{'}R_{\tau_{mu}}^{-1}1_p\sigma_{\mu}^{-2},\\
M_2=V_2^{-1}\left[R_{\tau_{\epsilon}}^{-1} \left(\sum\limits_{i=1}^{n_A}(\theta_{iA}-\alpha_A) + \sum\limits_{i=1}^{n_B}(\theta_{iB}-\alpha_B)  \right)\frac{1}{\sigma_{\epsilon}^{2}}+ R_{\tau_{\mu}}^{-1}1_p\frac{\phi}{\sigma_{\mu}^{2}}\right];  V_2=\frac{(n_A+n_B)}{\sigma_{\epsilon}^{2}}R_{\tau_{\epsilon}}^{-1} +R_{\tau_{\mu}}^{-1}\frac{1}{\sigma_{\mu}^{2}}, \\
M3=V_3^{-1}\left(R_{\tau_{\epsilon}}^{-1}\sum\limits_{i=1}^{n_A}(\theta_{iA}-\mu)\frac{1}{\sigma_{\epsilon}^2} \right), V_3=\frac{n_A}{\sigma_{\epsilon}^2}  R_{\tau_{\epsilon}}^{-1} + \frac{2}{\sigma_{\alpha}^2}R_{\tau_{\alpha}}^{-1},\\
D_{0g}=\sum\limits_{i=1}^{n_g}(\theta_{ig}-\mu-\alpha_g)^{'}R_{\tau_{\epsilon}}^{-1}(\theta_{ig}-\mu-\alpha_g), \ \ \textbf{for} \ \ g\in\{A, B\}\\
D_1=\left(\mu - 1_p\phi\right)^{'}R_{\tau_{\mu}}^{-1}\left(\mu - 1_p\phi\right),\\
D_2=\alpha_A^{'}R_{\tau_{\alpha}}^{-1}\alpha_A,
\end{array}
\end{equation*}

and $R_{\tau_{\cdot}}$ is a member of a particular class of correlation functions indexed by $\tau_{\cdot}$. The $Data$ is this case corresponds to the array composed of all the curves $\{\theta_{ig}\}$. For $\tau_{\epsilon}$, $\tau_{\mu}$ and $\tau_{\alpha}$ we have

\begin{equation*}
p(\tau_{\epsilon}|\mu,\alpha_A, \sigma_{\epsilon}^2, Data)\propto \exp\left\{-\frac{1}{2\sigma_{\epsilon}^2}(D_{0A}+D_{0B})\right\}|R_{\tau_{\epsilon}}|^{-(n_A+n_B)/2}p(\tau_{\epsilon}),
\end{equation*}

\begin{equation*}
p(\tau_{\mu}|\mu, \phi, \sigma_{\mu}^2)\propto \exp\left\{-\frac{1}{2\sigma_{\mu}^2}D_{1}\right\}|R_{\tau_{\mu}}|^{-1/2}p(\tau_{\mu}),
\end{equation*}
 
\begin{equation*}
p(\tau_{\alpha}|\alpha_A, \sigma_{\alpha}^2)\propto \exp\left\{-\frac{1}{\sigma_{\alpha}^2}D_{2}\right\}|R_{\tau_{\alpha}}|^{-1/2}p(\tau_{\alpha}).
\end{equation*}

\subsubsection*{Inference}

\cite{kaufman2010bayesian} create graphical displays to analyze the posterior distribution of $\{\alpha_i\}$.  In particular, they create plots of intervals of high posterior probability for that functional parameter. This model can also incorporate formal testing, but \cite{kaufman2010bayesian} leave this for future work. We propose measures based on posterior probabilities to test the hypothesis $g(\alpha_g) \in C^{0}$, where $C^{0}$ is the critical region associated with the testing procedure. For example, if one wants to test $\alpha_A-\alpha_B>0$, then 

\begin{equation*}
p(\alpha_A-\alpha_B>0)= E(1_{\{\alpha_A-\alpha_B>0\}})\approx \frac{\sum\limits_{k=1}^{N} 1_{\{\alpha_A^{(k)}-\alpha_B^{(k)}>0\}}}{N},
\end{equation*}

for a sample of size $N$ from the posterior distribution.

         % associado ao arquivo: 'cap-conceitos.tex'
\chapter{fMRI group data analysis}
\label{cap:3}
In this chapter, we present our modeling procedure for group fMRI data analysis. First, we present the individual case, which is a necessary step in performing the group analysis. In this first stage of the analysis, we employ the MDLM presented in chapter 2. For each voxel (taking the "appropriate" neighborhood) from each subject we fit the model \ref{chap2:equ1} and compute the posterior distribution \ref{chap2:equ3} using the theorem \ref{theorem1}. That posterior distribution can be used to perform inference at the individual level.  In other words, it can be used to detect brain activation for a particular subject. Here we propose three different ways to perform that inference. This same posterior distribution is also used as an input for the stage of group analysis. In this case, for every voxel in the fMRI array, we combine the posterior distribution \ref{chap2:equ3} through the subjects using a linear transformation. It is worth mentioning that in this thesis, we only focus on cases with one and two groups, but the ideas can be easily extended to more general cases. As in the individual case, we also propose three different ways to perform the inference related to brain activation for each group and to the differences in brain activation between the two groups.

\section{Voxel-wise Individual Analysis}
\label{chap3:sec1}

As we mentioned in chapter~\ref{cap:introducao}, the usual fMRI dataset is a four-dimensional array that contains the observed BOLD response associated with an fMRI experiment. There are different types of fMRI experiments, but in this work, we focus specifically on experiments where a stimulus is presented, following one of the designs showed in figure~\ref{fig:tres}. For example, in figure~\ref{cap3:fig1}, we can see some plots related to an experiment where a visual stimulus was presented. The graph on the left panel shows the expected BOLD response (obtained using the convolution \ref{equ2}) associated with a block design.  In the center and right panels are the observed BOLD response from two different voxels within and outside the visual cortex, respectively. Thus, the aim is to identify the fMRI time series that matches the observed BOLD response. For that purpose, we model the observed BOLD response as a linear function of the expected BOLD response using the model  
\ref{chap2:equ1}. Let $y_{\scaleto{[i,j,k],t,1\mathstrut}{6pt}}^{*}$ and $x_t$ be the observed and expected BOLD response, respectively at position $\{i,j,k\}$ and time $t$, for $i=1, \ldots, d_1$, $j=1,\ldots,d_2$, $k=1,\ldots,d_3$ and $t=1,\ldots, T$. Something that can be noticed from this last definition is that $x_t$ is supposed to be the same for all the locations in the brain image, in other words, it is supposing that the BOLD response is the same in all brain regions. This may be an unrealistic assumption, but it is one that works well in practice. We could model $x_t$ as a function of the location, trying to obtain a more accurate detection of brain reaction, but we leave this for future work. Let the vector

\begin{equation}\label{chap3:equ0}
 \mathbf{Y}_{\scaleto{[i,j,k]t\mathstrut}{6pt}}=\left( \begin{array}{c}
y_{\scaleto{[i,j,k],t,1\mathstrut}{6pt}}^{*}\\ 
y_{\scaleto{[i+1,j,k],t,2\mathstrut}{6pt}}\\ 
y_{\scaleto{[i-1,j,k],t,3\mathstrut}{6pt}}\\
y_{\scaleto{[i,j+1,k],t,4\mathstrut}{6pt}}\\
y_{\scaleto{[i,j-1,k],t,5\mathstrut}{6pt}}\\ 
y_{\scaleto{[i,j,k+1],t,6\mathstrut}{6pt}}\\ 
y_{\scaleto{[i,j,k-1],t,7\mathstrut}{6pt}}
\end{array}\right)_{\scaleto{1\times q\mathstrut}{5pt}}
\end{equation}

represent the cluster or neighborhood of size $q=7$ of the voxel at position $\{i,j,k\}$. Thus, we model $\mathbf{Y}{\scaleto{[i,j,k]t\mathstrut}{6pt}}$ using the MDLM \ref{chap2:equ1}, where $F_t^{'}=(x_t, z_t)$ and $z_t$ represent the additional covariates one can include in the model. What we intend with this type of modeling is to capture the covariance structure within the cluster of voxels through the matrix $\mathbf{\Sigma}$. The criterion to define that cluster form is based on the Euclidean distance, where the distance between the voxel $V^{*}$ and the neighboring voxel $V_{\sim}$ is given by $d(V^{*}, V_{\sim})\leq r$. In figure \ref{cap3:fig0} (right panel), we can see a graphical illustration of a neighborhood of voxels, where $r=1$.

\begin{table}[H]
\begin{figure}[H]
  \centering
\begin{center}
\begin{tabular}{ccc}
\includegraphics[width=.25\textwidth]{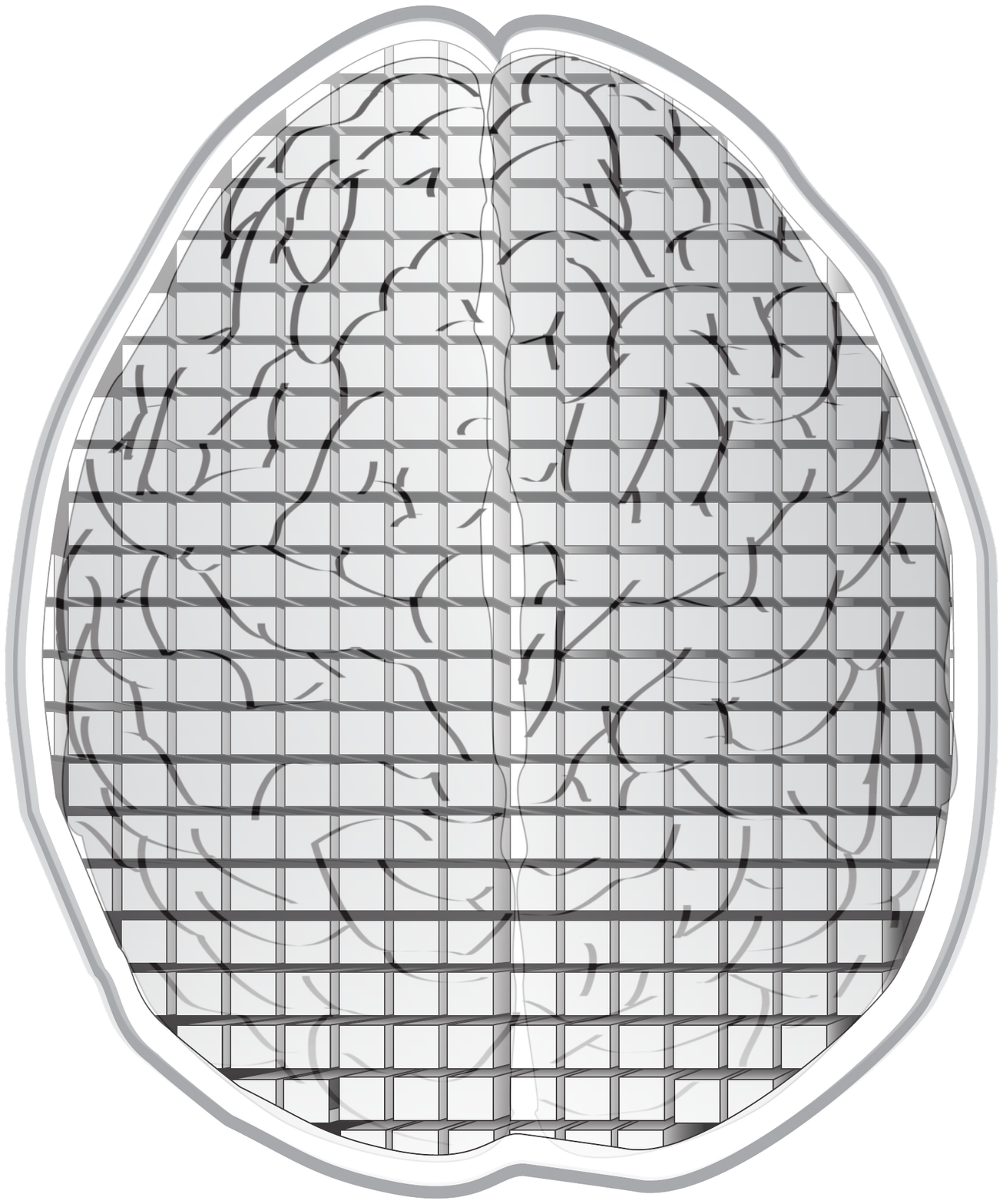}&\includegraphics[width=.25\textwidth]{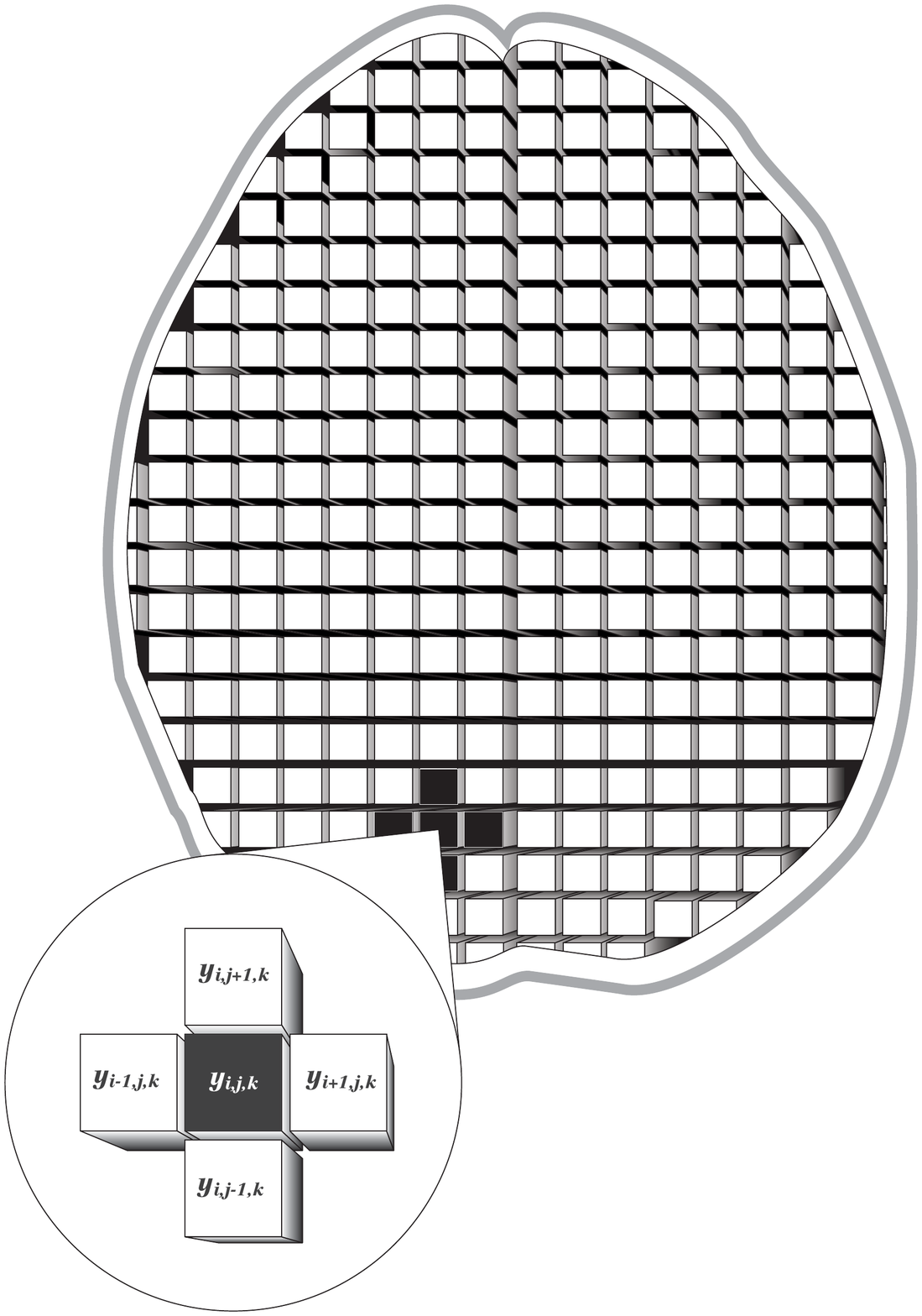}&\includegraphics[width=.25\textwidth]{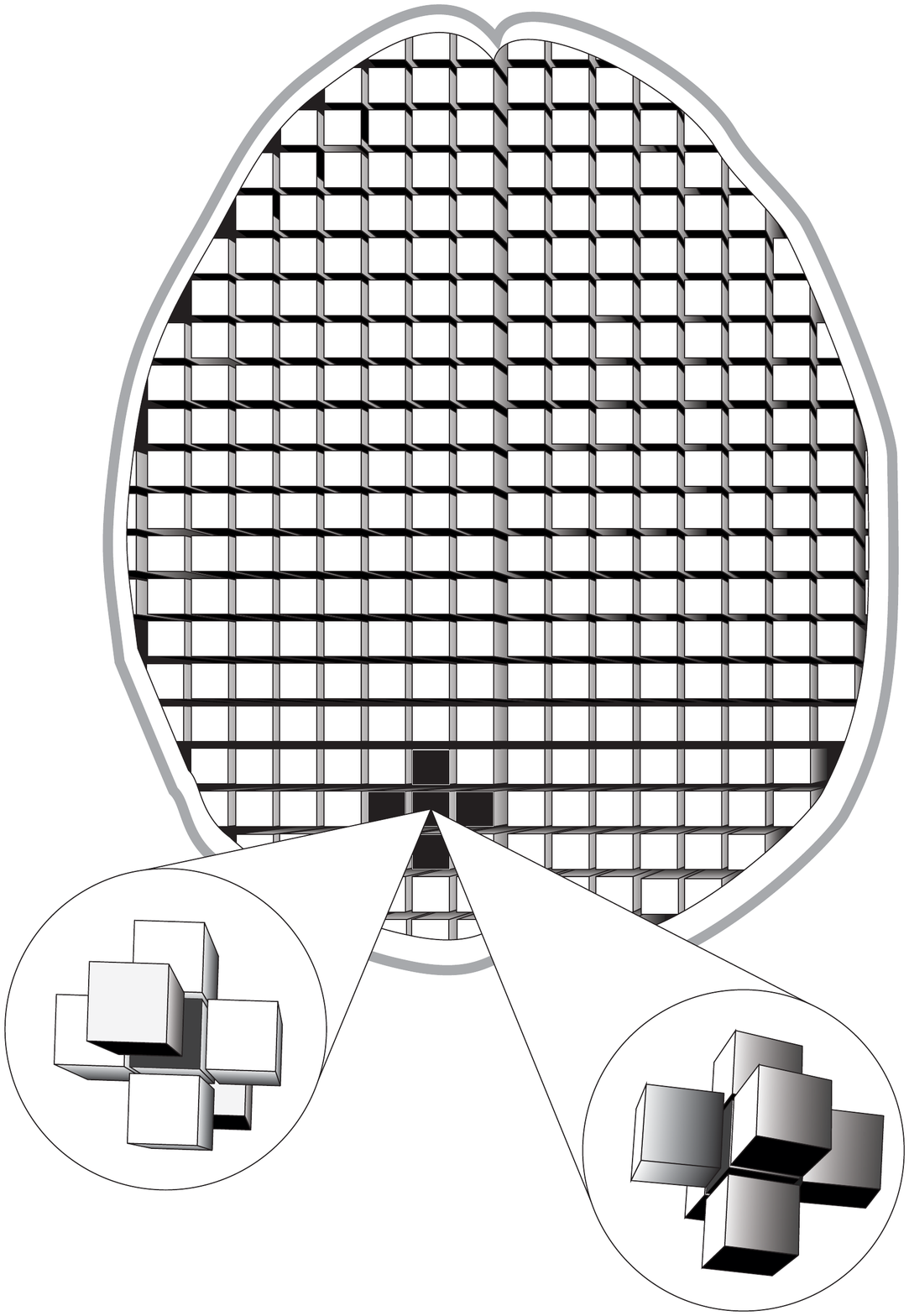}\\
\end{tabular}
\end{center}
  \caption{Grafical illustration of a neighborhood of voxels, for $r=1$.}
  \label{cap3:fig0} 
\end{figure}
\end{table} 

\begin{table}[H]
\begin{figure}[H]
  \centering
\begin{center}
\begin{tabular}{ccc}
\includegraphics[width=.25\textwidth]{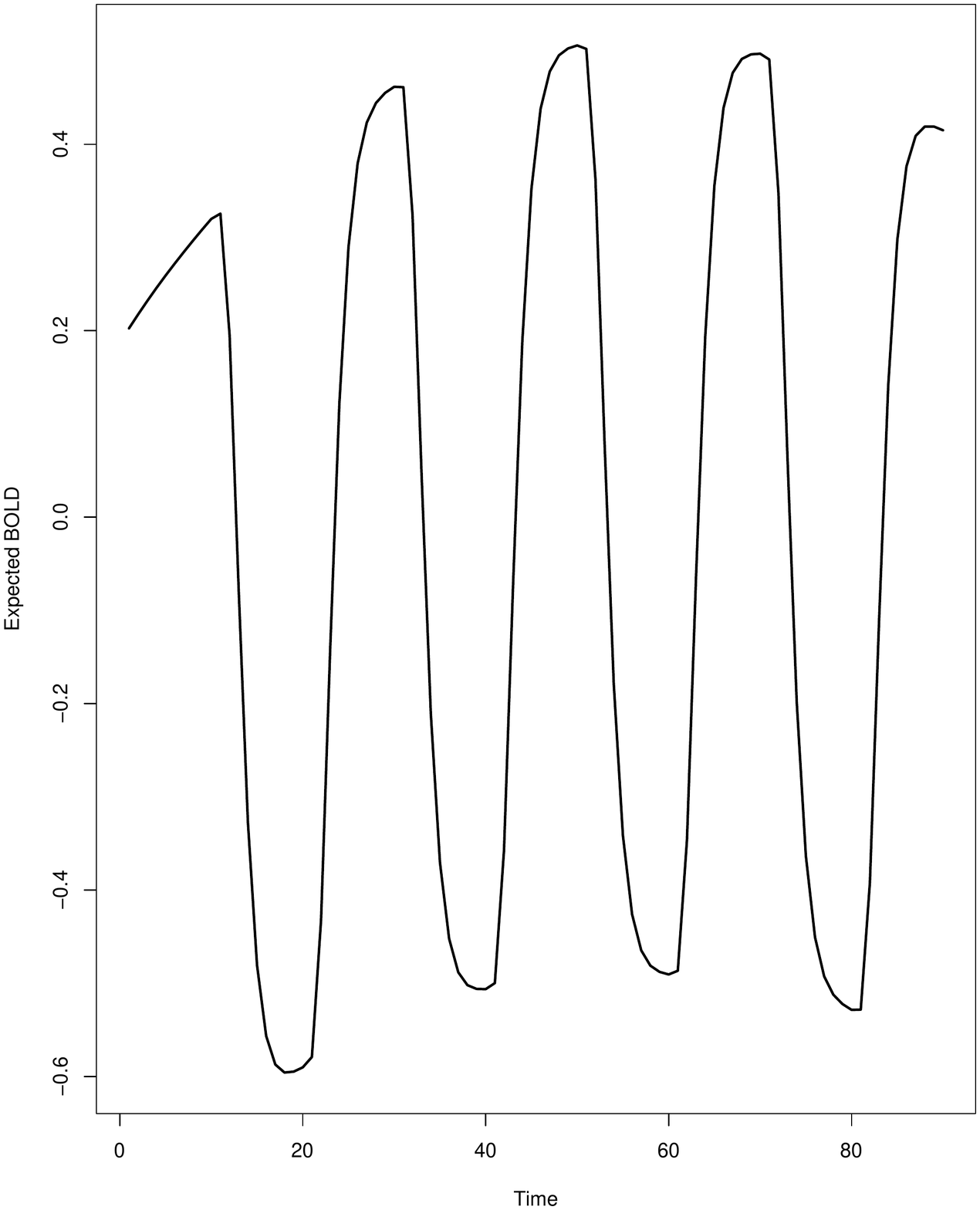}&\includegraphics[width=.25\textwidth]{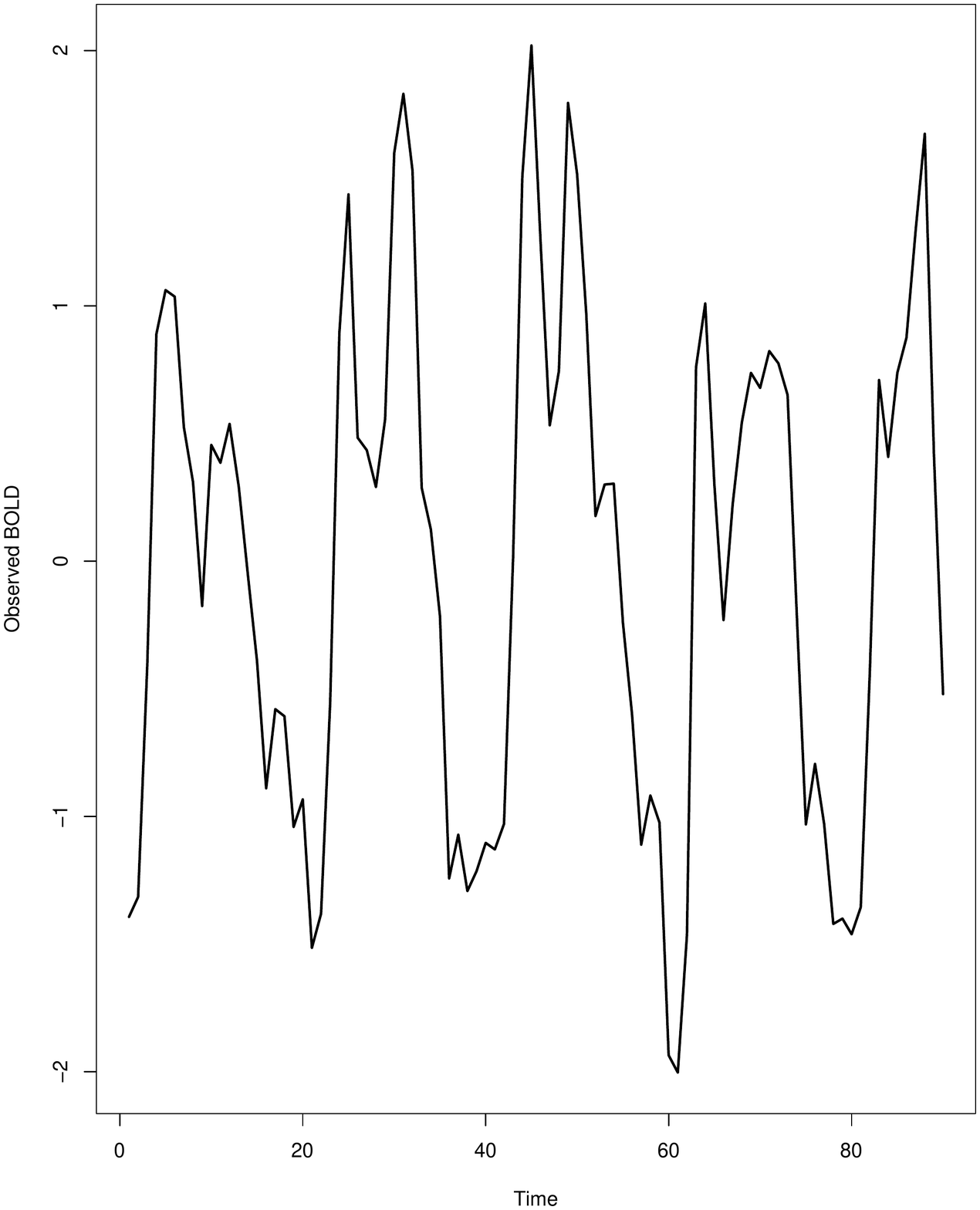}&\includegraphics[width=.25\textwidth]{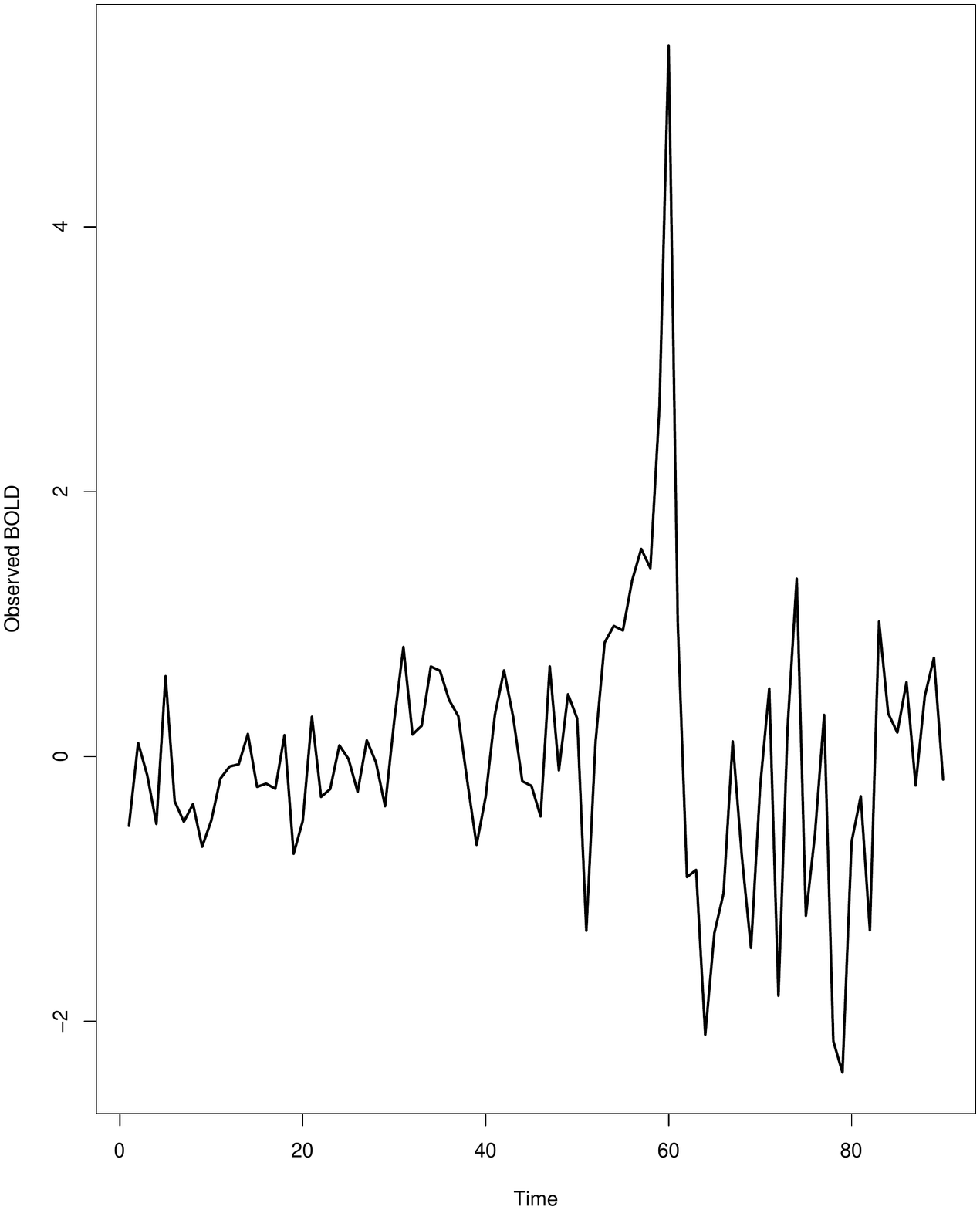}\\
\end{tabular}
\end{center}
  \caption{Left panel, expected Bold response. On the center and right panels are the fMRI time series for two different voxels from within and outside the visual cortex respectively.}
  \label{cap3:fig1} 
\end{figure}
\end{table} 

\newpage

\begin{figure}[H]
  \centering
\begin{center}
\includegraphics[width=.50\textwidth]{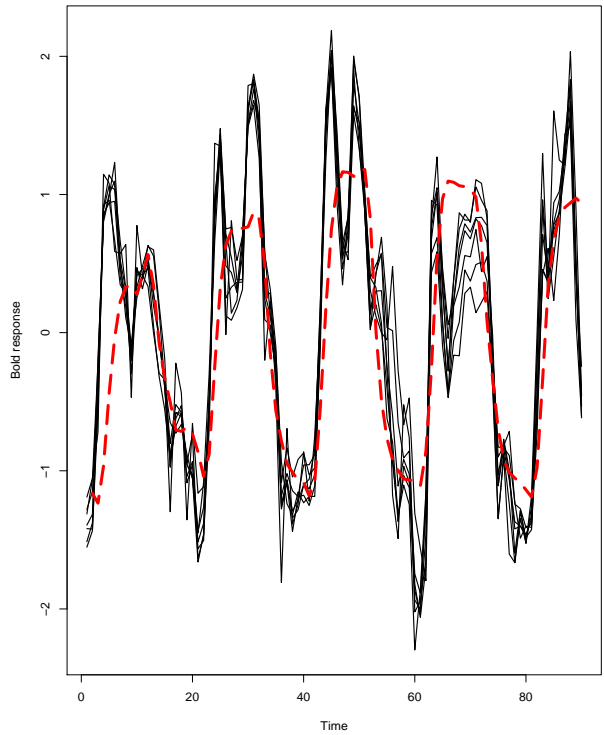}
\end{center}
  \caption{Black lines represent the observed BOLD response for a cluster centered on the voxel $V^{*}$. The red longdash line represents the fitted BOLD response for the voxel $V^{*}$.}\label{cap3:fig2}  
\end{figure}

In this way, for every cluster of voxels taking $r=1$ (i.e., clusters composed of at least seven time series), the model \ref{chap2:equ1} is fitted. This procedure is performed for every location in the fMRI array, building for every voxel its own cluster of neighbors. For every parameter $\theta_{\scaleto{i,j,k,t,l,v\mathstrut}{6pt}}^{*}$, the appropiate inference is performed, taking into account its cluster information through $\mathbf{\Sigma}$. In figure~\ref{cap3:fig2}, we show an example of a cluster of voxels from the visual cortex, where the MDLM \ref{chap2:equ1} was fitted to those time series.

Let the matrix

\begin{equation}\label{chap3:equ1}
 \mathbf{\Theta}_{\scaleto{[i,j,k]t\mathstrut}{6pt}}=\left( \begin{array}{ccccccc}
\theta_{\scaleto{i,j,k,t,1,1\mathstrut}{6pt}}^{*}&\theta_{\scaleto{i,j,k,t,1,2\mathstrut}{6pt}}&\theta_{\scaleto{i,j,k,t,1,3\mathstrut}{6pt}}&\theta_{\scaleto{i,j,k,t,1,4\mathstrut}{6pt}}&\theta_{\scaleto{i,j,k,t,1,5\mathstrut}{6pt}}&\theta_{\scaleto{i,j,k,t,1,6\mathstrut}{6pt}}&\theta_{\scaleto{i,j,k,t,1,7\mathstrut}{6pt}}\\ 
\vdots & \vdots & \vdots & \vdots & \vdots & \vdots & \vdots\\
\theta_{\scaleto{i,j,k,t,p,1\mathstrut}{6pt}}^{*}&\theta_{\scaleto{i,j,k,t,p,2\mathstrut}{6pt}}&\theta_{\scaleto{i,j,k,t,p,3\mathstrut}{6pt}}&\theta_{\scaleto{i,j,k,t,p,4\mathstrut}{6pt}}&\theta_{\scaleto{i,j,k,t,p,5\mathstrut}{6pt}}&\theta_{\scaleto{i,j,k,t,p,6\mathstrut}{6pt}}&\theta_{\scaleto{i,j,k,t,p,7\mathstrut}{6pt}}
\end{array}\right)
\end{equation}

represent the states or parameters in model \ref{chap2:equ1}, whose columns are the $p\times 1$ parameter vectors related to each of the $q$ voxels inside the cluster. Our aim is to perform inference on the vector $\boldsymbol{\theta}_{\scaleto{i,j,k,t\mathstrut}{6pt}}^{*}= \left(\theta_{\scaleto{i,j,k,t,1,1\mathstrut}{6pt}}^{*}, \ldots, \theta_{\scaleto{i,j,k,t,p,1\mathstrut}{6pt}}^{*}\right)^{'}$ taking advantage of the information brought by the remaining parameters in the matrix \ref{cap3:fig1}. In order to do so, we instead define the row vectors   

\begin{equation*}
\begin{array}{cc}
\boldsymbol{\theta}_{\scaleto{i,j,k,t,1\mathstrut}{6pt}}&=\left(\theta_{\scaleto{i,j,k,t,1,1\mathstrut}{6pt}}^{*},\theta_{\scaleto{i,j,k,t,1,2\mathstrut}{6pt}},\theta_{\scaleto{i,j,k,t,1,3\mathstrut}{6pt}},\theta_{\scaleto{i,j,k,t,1,4\mathstrut}{6pt}},\theta_{\scaleto{i,j,k,t,1,5\mathstrut}{6pt}},\theta_{\scaleto{i,j,k,t,1,6\mathstrut}{6pt}},\theta_{\scaleto{i,j,k,t,1,7\mathstrut}{6pt}}\right)\\
\vdots & \\
\boldsymbol{\theta}_{\scaleto{i,j,k,t,p\mathstrut}{6pt}}&=\left( \theta_{\scaleto{i,j,k,t,p,1\mathstrut}{6pt}}^{*},\theta_{\scaleto{i,j,k,t,p,2\mathstrut}{6pt}},\theta_{\scaleto{i,j,k,t,p,3\mathstrut}{6pt}},\theta_{\scaleto{i,j,k,t,p,4\mathstrut}{6pt}},\theta_{\scaleto{i,j,k,t,p,5\mathstrut}{6pt}},\theta_{\scaleto{i,j,k,t,p,6\mathstrut}{6pt}},\theta_{\scaleto{i,j,k,t,p,7\mathstrut}{6pt}}
 \right)
\end{array}
\end{equation*}

and perform the inference on each one of these. Now we only need to identify the posterior distributions for each of those vector parameters, but before doing so, we have to make some necessary considerations of the posterior distribution $p(\mathbf{\Theta}_{t}|D_{t})$.

\subsubsection{Inference at the individual level}

From \citep{quintana1987multivariate}, we know that the posterior distribution of $(\mathbf{\Theta}_{t}|D_{t})$ is the matrix $T$ distribution with $p\times q$ mean matrix $\mathbf{m}_t$, $p\times p$ left variance matrix $\mathbf{C}_t$, and $q\times q$ right variance matrix $\mathbf{S}_t$. Thus, 

\begin{equation}\label{chap3:equ2}
(\mathbf{\Theta}_{t}|D_{t})\sim T_{n_t}[\mathbf{m}_t, \mathbf{C}_t, \mathbf{S}_t],
\end{equation}

where $n_{t}=n_{t-1}+1$. Thus, a reasonable approximation for the posterior distribution \ref{chap3:equ2} when $n_t\geq 30$ is given by

\begin{equation}\label{chap3:equ3}
(\mathbf{\Theta}_{t}|D_{t})\widesim{approx} N[\mathbf{m}_t, \mathbf{C}_t, \mathbf{S}_t].
\end{equation}

We have two main considerations to justify working only with the posterior distributions for $n_t\geq 30$. The first is that as we use vague prior distributions at $t=0$, the sequential update process from theorem~\ref{theorem1} takes some period of time before reaching a posterior distribution dominated by the data. Then, the first posterior distributions (i.e., for $t<30$) could be considered irrelevant for the analysis. The second consideration is simply that dealing with normal distributions simplifies the mathetical work with linear transfomations, which will be quite common in this thesis for the inferential procedure on the matrix parameter $\mathbf{\Theta}_{t}$. \\

\subparagraph{About the hypothesis testing} 
In the fMRI literature, it is common to see sharp hypothesis-testing procedures of the form $H_0: \boldsymbol{\theta}_{\scaleto{i,j,k,t,1\mathstrut}{6pt}}=\mathbf{0}$. But, from our point of view, that does not make sense, because one is interested in identifying the observed BOLD response that matches with the expected BOLD response, and the test does not correspond with that premise. In figure~\ref{cap3:fig3}, left panel, we can see an example of a plot of an observed BOLD response that actually matches with the expected BOLD response, and in the right panel, we can see a scatter plot for those two variables. In the figure \ref{cap3:fig4} we can see another case of the same plots for an artificial example, where the observed BOLD response does not match with the expected BOLD response. It is obvious that that sharp testing procedures could fail in the last example because there is a downward slope, but those curves do not match at all. Then one could falsely infer the existence of an actual neural activation that in reality was not there. In that sense, we proppose testing $H_0: \boldsymbol{\theta}_{\scaleto{i,j,k,t,1\mathstrut}{6pt}}\geq\mathbf{0}$, based on the posterior probability $p(\boldsymbol{\theta}_{\scaleto{i,j,k,t,1\mathstrut}{6pt}}|D_{\scaleto{i,j,k,t\mathstrut}{6pt}})$. For example, if $p(\boldsymbol{\theta}_{\scaleto{i,j,k,t,1\mathstrut}{6pt}}>\mathbf{0}|D_{\scaleto{i,j,k,t\mathstrut}{6pt}})>\alpha$ (e.g., $\alpha=0.95$), then one can conclude that the observed and expected BOLD response match, in other words, one can conclude that there is a neural activation.

\begin{table}[H]
\begin{figure}[H]
  \centering
\begin{center}
\begin{tabular}{cc}\\
\multicolumn{2}{c}{Match case: activated voxel}\\
\includegraphics[width=.40\textwidth]{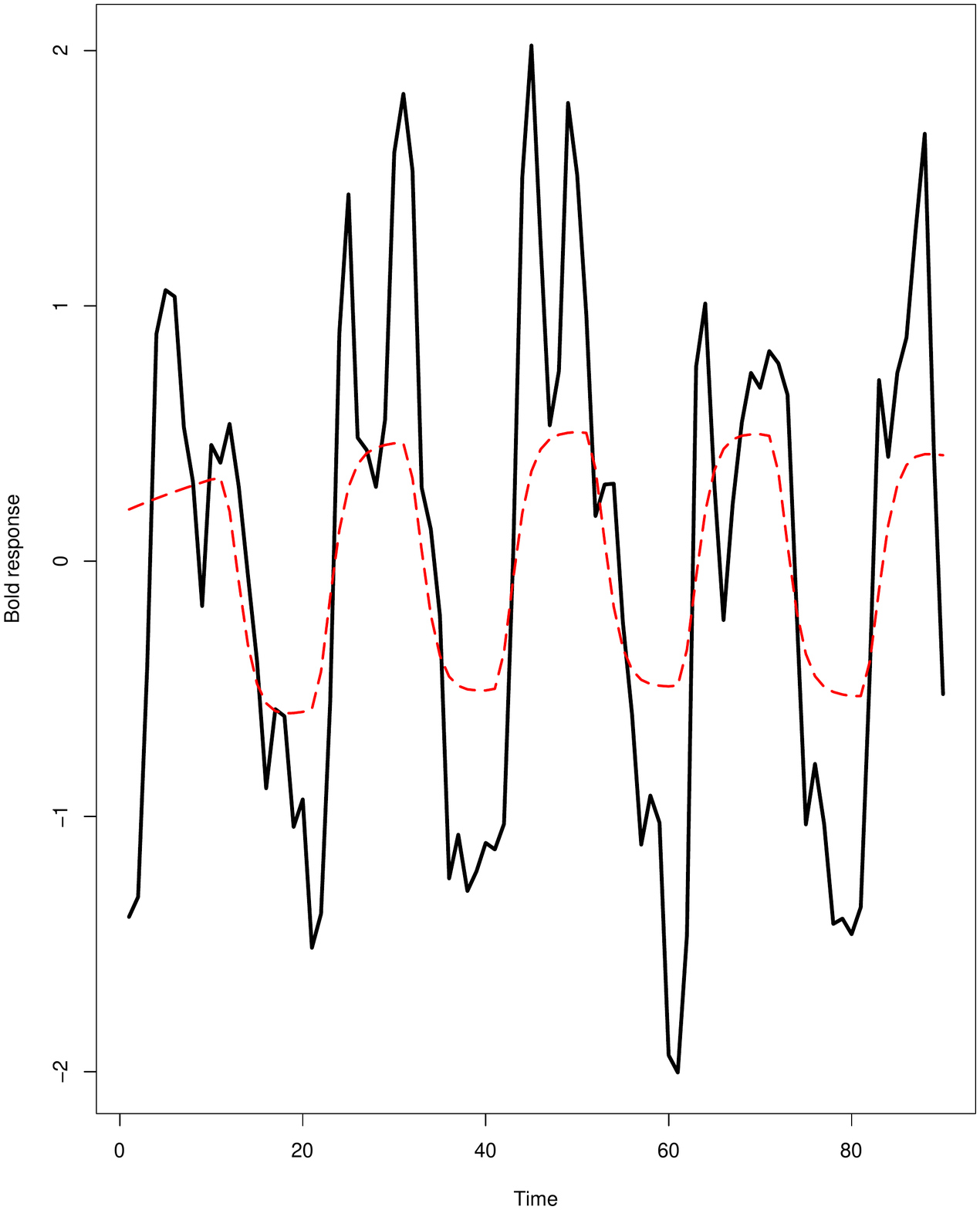}&\includegraphics[width=.40\textwidth]{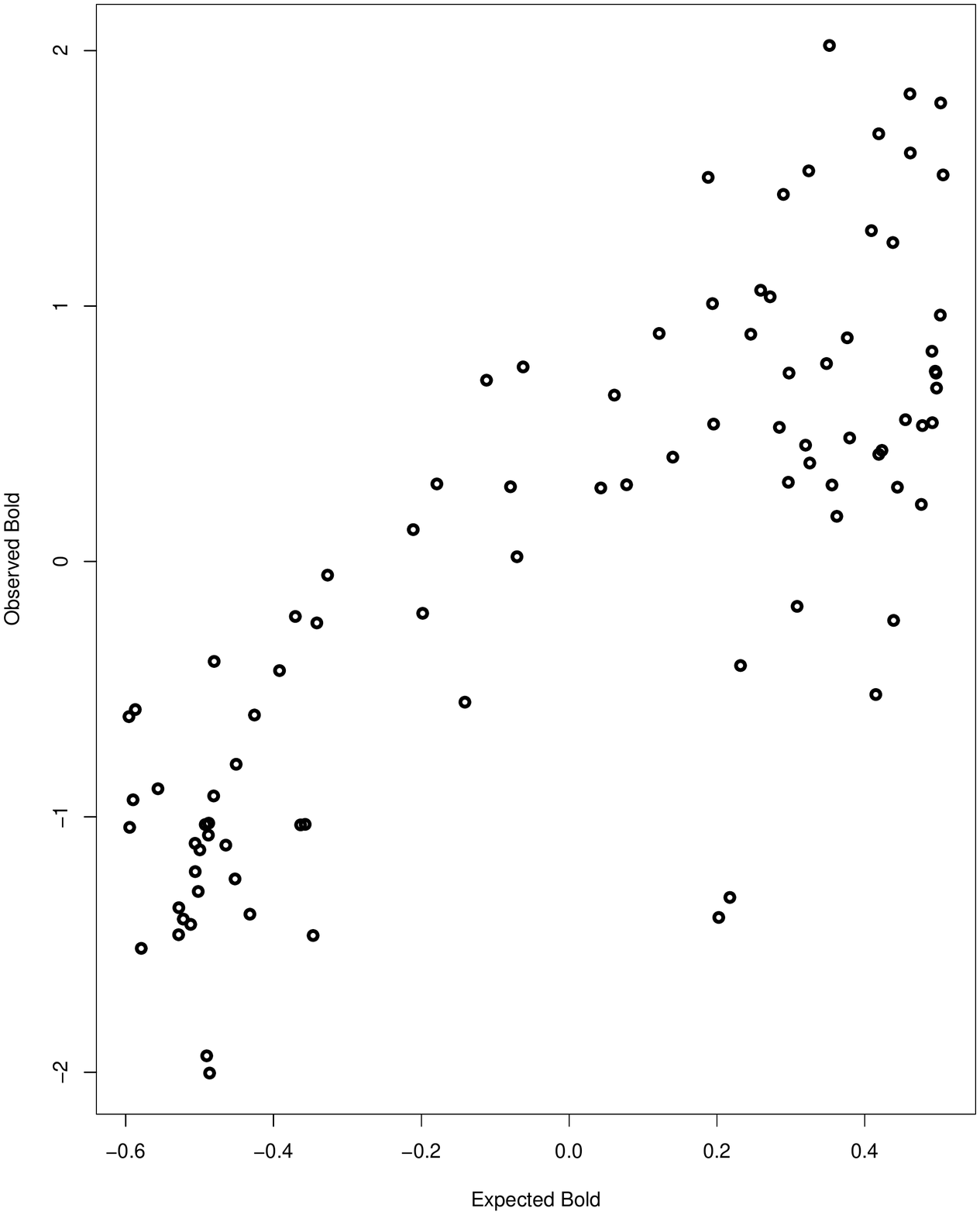}\\
\end{tabular}
\end{center}
  \caption{Left panel: observed BOLD response (black line) and expected BOLD response (red longdash line). Right panel: Scatter plot, expected BOLD response vs. observed BOLD response}
  \label{cap3:fig3} 
\end{figure}
\end{table}

\begin{table}[H]
\begin{figure}[H]
  \centering
\begin{center}
\begin{tabular}{cc}\\
\multicolumn{2}{c}{No match case: non-activated voxel}\\
\includegraphics[width=.40\textwidth]{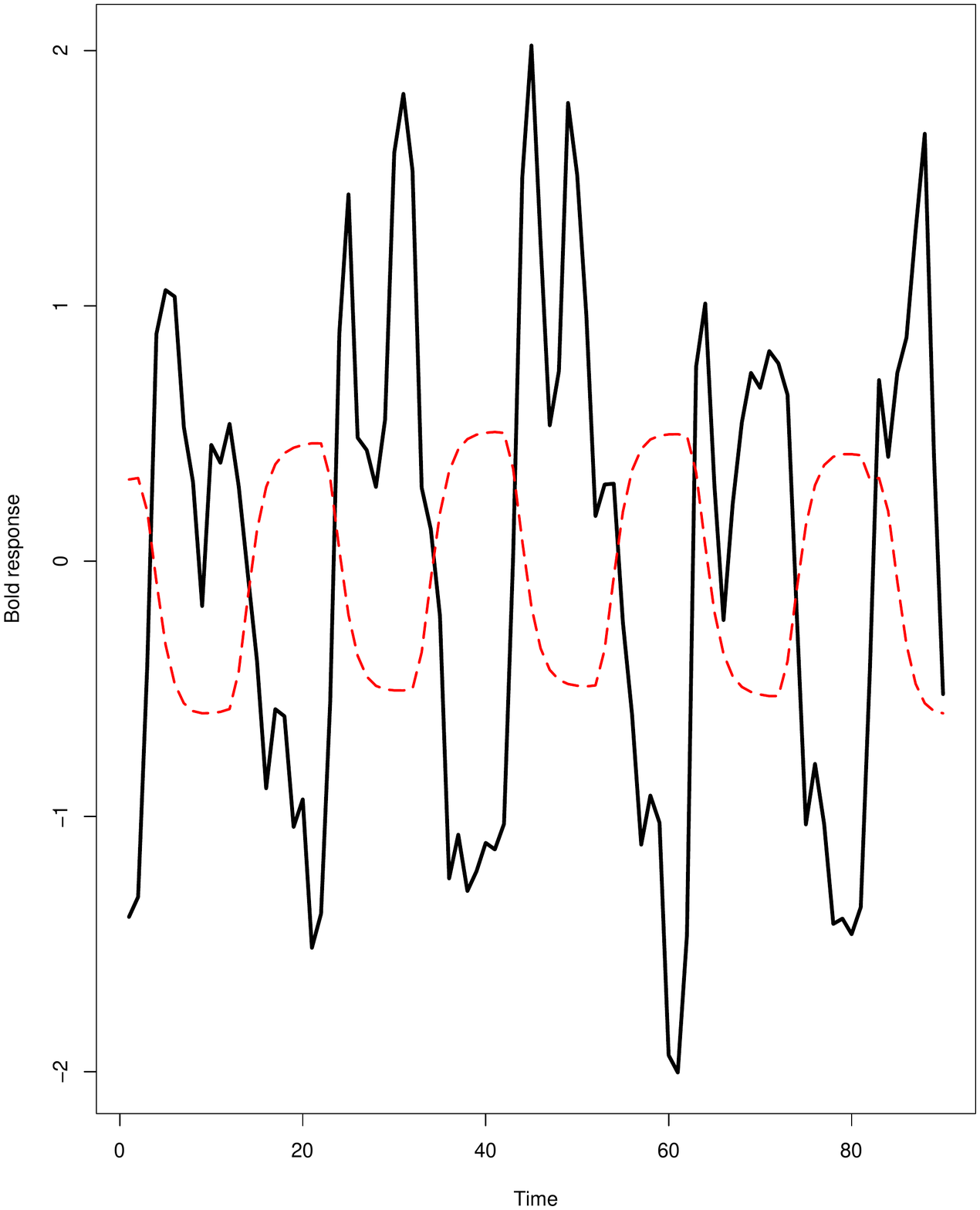}&\includegraphics[width=.40\textwidth]{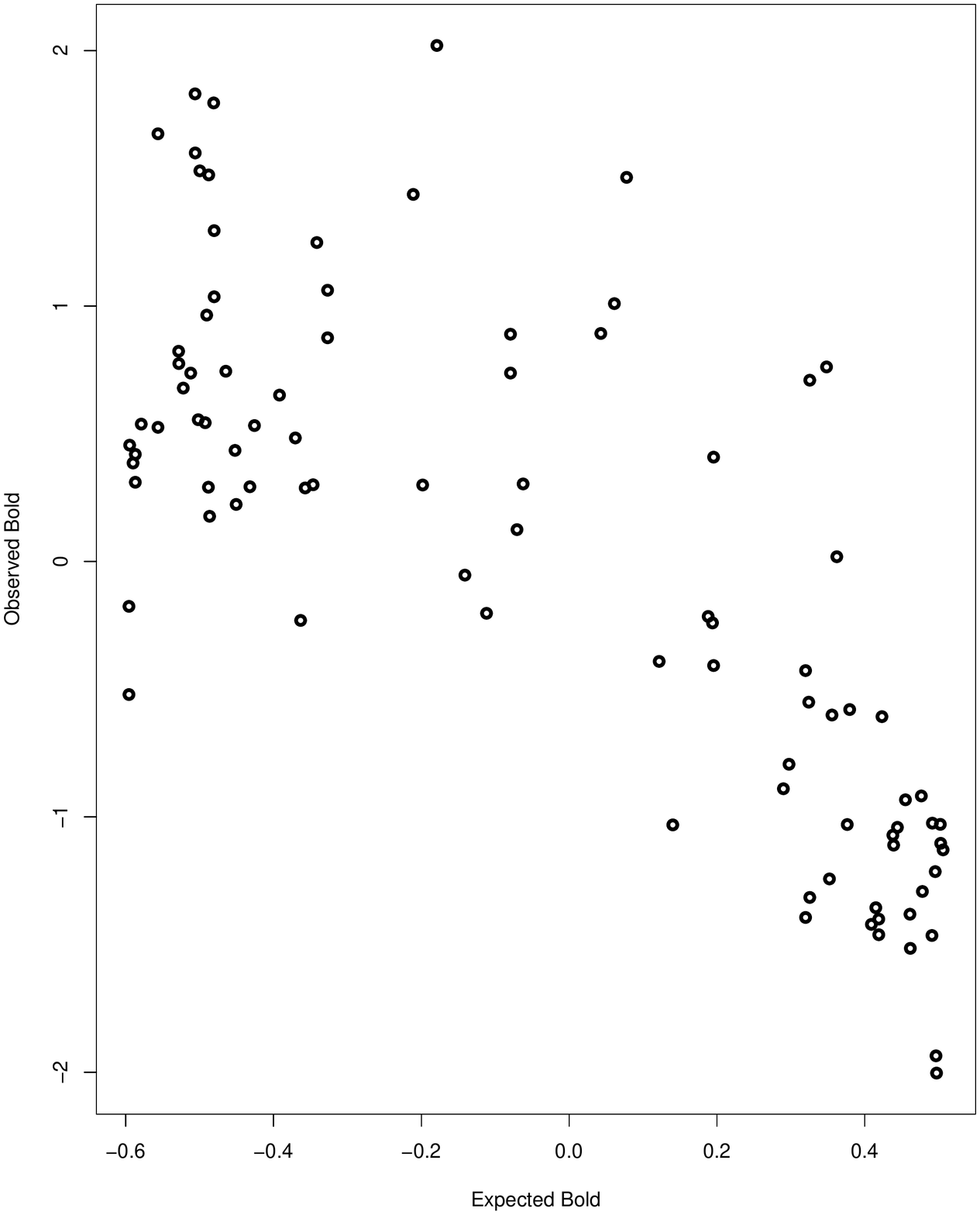}\\
\end{tabular}
\end{center}
  \caption{Left panel: observed BOLD response (black line) and expected BOLD response (red longdash line). Right panel: Scatter plot, expected BOLD response vs. observed BOLD response}
  \label{cap3:fig4} 
\end{figure}
\end{table}

\subparagraph{Inference using the latest posterior distribution}
In this case, we assume that $p(\boldsymbol{\theta}_{\scaleto{i,j,k,T,l\mathstrut}{6pt}}|D_{\scaleto{i,j,k,T\mathstrut}{6pt}})$, where $D_{\scaleto{i,j,k,T\mathstrut}{6pt}}=\{\mathbf{Y}_{\scaleto{[i,j,k]1\mathstrut}{6pt}}, \mathbf{Y}_{\scaleto{[i,j,k]2\mathstrut}{6pt}}, \ldots, \mathbf{Y}_{\scaleto{[i,j,k]T\mathstrut}{6pt}} \}$, contains all the relevant information about the fMRI experiment for a particular voxel.  In other words, we assume that the latest updated posterior distribution at $t=T$ brings all the historical information of the observed BOLD response that is necessary in order to detect neural activation. Thus, we perform three different optional tests: the marginal test $H_0:\theta_{\scaleto{i,j,k,T,l,1\mathstrut}{6pt}}^{*}>0$, the joint test $H_0:\boldsymbol{\theta}_{\scaleto{i,j,k,T,l\mathstrut}{6pt}}>0$, and the linear transformation or average test $H:\bar{\theta}_{\scaleto{i,j,k,T,l\mathstrut}{6pt}}>0$, where $\bar{\theta}_{\scaleto{i,j,k,T,l\mathstrut}{6pt}}=\frac{1}{q}\sum \limits_{v=1}^{q}\theta_{\scaleto{i,j,k,T,l,v\mathstrut}{6pt}}$,  for $l=1,\ldots,p$, where $p$ is the number of parameters in each voxel model. From \ref{chap3:equ3} and the properties of the matrix normal distribution, we obtain the following distributions:

\begin{equation}\label{chap3:pos1}
\theta_{\scaleto{i,j,k,T,l,1\mathstrut}{6pt}}^{*}|D_{\scaleto{i,j,k,T\mathstrut}{6pt}}\sim N(m_{\scaleto{i,j,k,T,l,1\mathstrut}{6pt}}^{*}, C_{\scaleto{T,l,l\mathstrut}{6pt}}*S_{\scaleto{T,1,1\mathstrut}{6pt}}),
\end{equation}

\begin{equation}\label{chap3:pos2}
\boldsymbol{\theta}_{\scaleto{i,j,k,T,l\mathstrut}{6pt}}|D_{\scaleto{i,j,k,T\mathstrut}{6pt}}\sim N(\boldsymbol{m}_{\scaleto{i,j,k,T,l\mathstrut}{6pt}}, C_{\scaleto{T,l,l\mathstrut}{6pt}}*\boldsymbol{S}_{\scaleto{T\mathstrut}{6pt}}),
\end{equation}

\begin{equation}\label{chap3:pos3}
\bar{\theta}_{\scaleto{i,j,k,T,l\mathstrut}{6pt}}|D_{\scaleto{i,j,k,T\mathstrut}{6pt}}\sim N(\bar{m}_{\scaleto{i,j,k,T,l\mathstrut}{6pt}}, \bar{S}_{\scaleto{T,l\mathstrut}{6pt}}),
\end{equation}

where $C_{\scaleto{T,l,l\mathstrut}{6pt}}$ and $S_{\scaleto{T,v,v\mathstrut}{6pt}}$ are the elements on the main diagonal of the matrices $\boldsymbol{C}_T$ and $\boldsymbol{S}_{\scaleto{T\mathstrut}{6pt}}$ respectively, and $\bar{m}_{\scaleto{i,j,k,T,l\mathstrut}{6pt}}=\frac{1}{q}\sum \limits_{v=1}^{q}m_{\scaleto{i,j,k,T,l,v\mathstrut}{6pt}}$, $\bar{S}_{\scaleto{T,l\mathstrut}{6pt}}=\frac{1}{q^2}\left[\sum \limits_{v=1}^{q}C_{T,l,l}S_{T,v,v} + \sum \limits_{v\neq v^{'}} C_{T,l,l}S_{T,v,v^{'}} \right]$. Thus, for instance, a measure of evidence againsts the hypothesis $H_0:\bar{\theta}_{\scaleto{i,j,k,T,l\mathstrut}{6pt}}>0$ is defined by $\alpha=Pr\left[\bar{\theta}_{\scaleto{i,j,k,T,l\mathstrut}{6pt}}>0\right]$. Then, a small value of $\alpha$ indicates rejection of a possible voxel activation.

\begin{figure}[H]
  \centering
\begin{center}
\includegraphics[width=.50\textwidth]{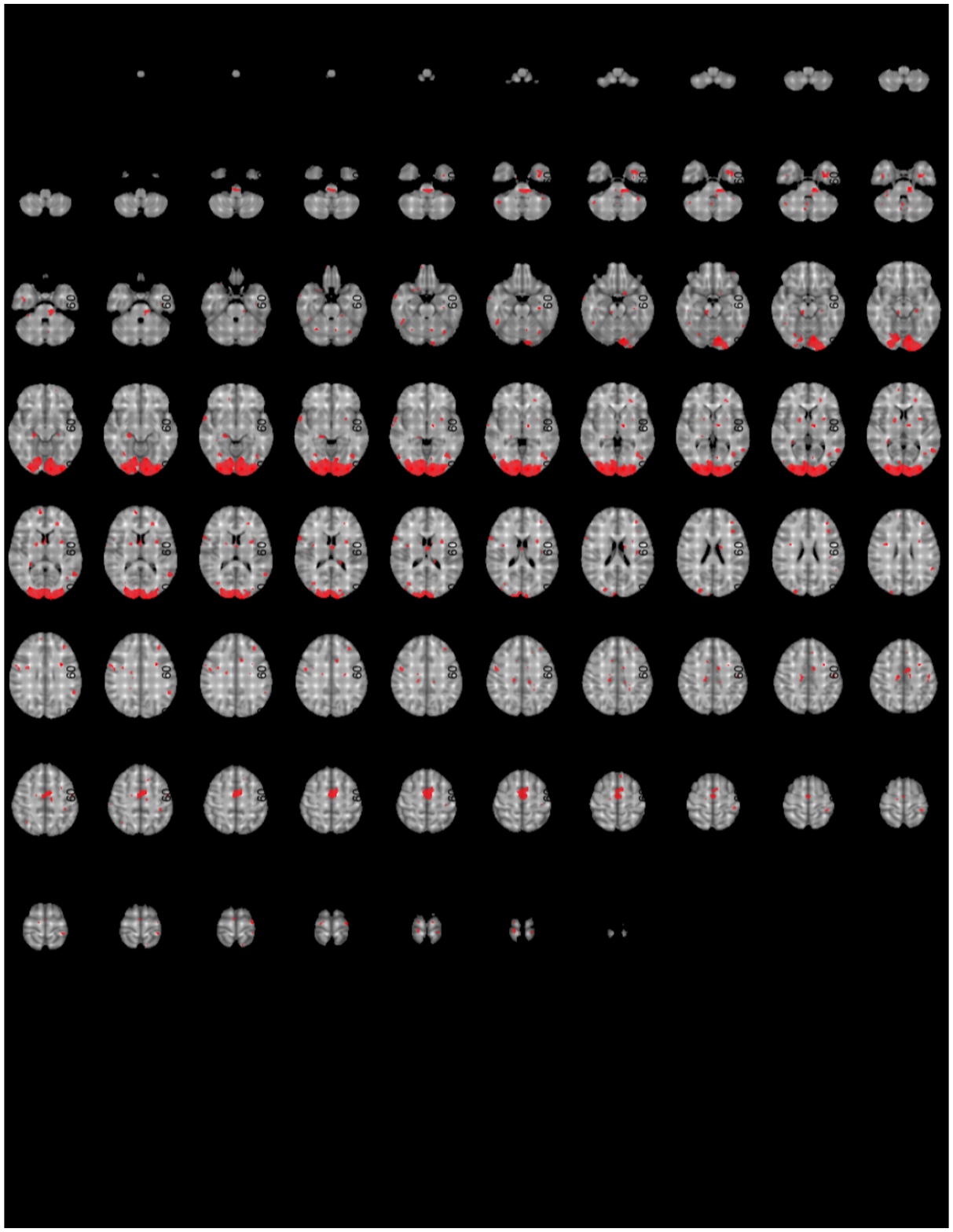}
\end{center}
  \caption{Posterior Probability Map obtained after performing the marginal test on every voxel.}\label{cap3:fig5} 
\end{figure}

\begin{figure}[H]
  \centering
\begin{center}
\includegraphics[width=.50\textwidth]{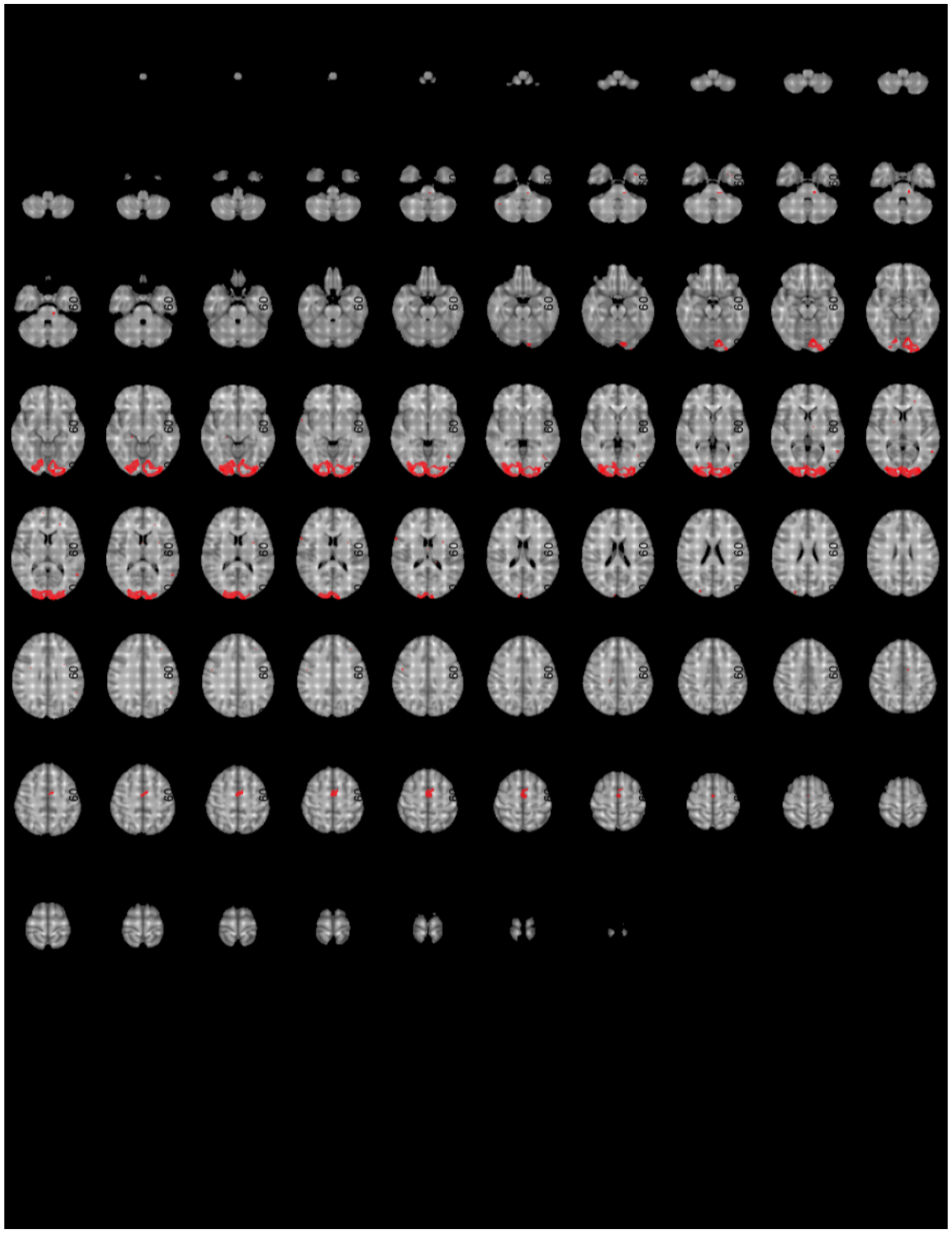}
\end{center}
  \caption{Posterior Probability Map obtained after performing the joint test on every voxel.}\label{cap3:fig6} 
\end{figure}

\begin{figure}[H]
  \centering
\begin{center}
\includegraphics[width=.50\textwidth]{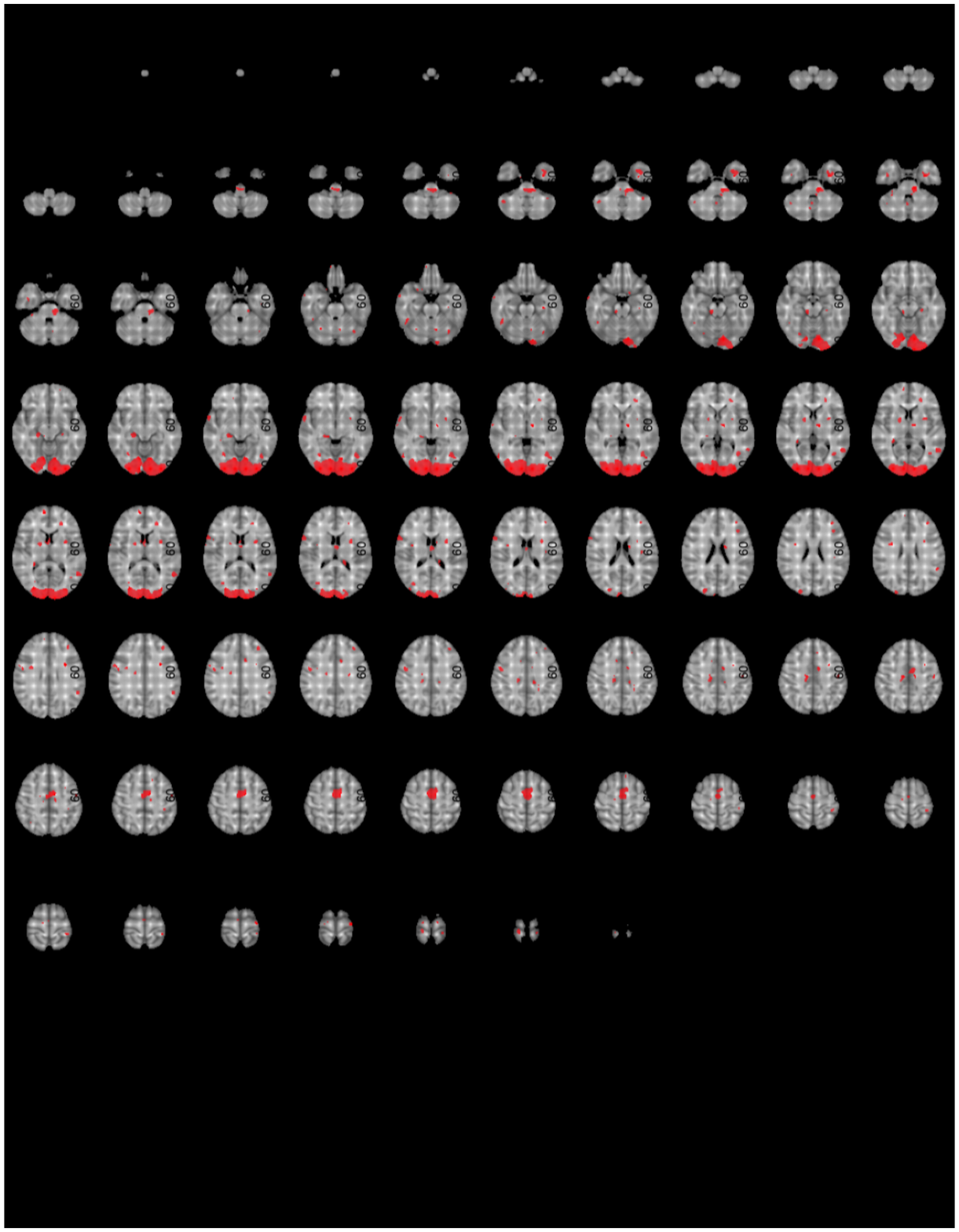}
\end{center}
  \caption{Posterior Probability Map obtained after performing the average test on every voxel.}\label{cap3:fig7} 
\end{figure}

\subparagraph{Example} To illustrate the ideas developed so far, we present a example of an fMRI experiment where a block design was used. In this experiment, a visual stimulus consisting of a full chess board flashing at a frequency of 8Hz was presented, and the response was compared with the response to a dark condition (resting state). Visual stimulation was presented over six on-off cycles in 30-second blocks of the flashing chessboard alternating with 30-second blocks of dark condition. The data were acquired from 35 healthy adults and 15 patients with a particular pathology. In this example, we only analyze the data of one of the subjects belonging to the control group. In figure~\ref{cap3:fig1}, we can see the expected BOLD response and the observed BOLD response for two voxels from inside and outside the visual cortex, respectively. 
In  figures~\ref{cap3:fig5}, \ref{cap3:fig6}, and \ref{cap3:fig7}, we can see the posterior probability maps (PPM) for the marginal, joint and average tests, respectively.
Every voxel painted red indicates brain activation.  Otherwise, there is no such brain reaction. In all three cases, an activation in the visual cortex is successfully detected, which was something expected, given the type of stimulus presented in the experiment performed. The results of the marginal and average tests are very similar, but the joint test seems to be more conservative, with a lower rate of false positives.

\subparagraph{Inference using all the posterior distributions for $t\geq 30$}
    
Suppose now we want to use all the posterior distributions for $t\geq 30$ in order to perform the inference related to a voxel activation, in other words, to perform an inference over the parameters related to the first column of the matrix $\mathbf{\Theta}$. Using the algorithm \ref{euclid}, we can draw on-line estimated trajectories of these parameters and compute a measure of evidence, as we showed in chapter~\ref{cap:2}. The only additional consideration is to change the posterior distribution in the algorithm \ref{euclid} to any of the three options \ref{chap3:pos1}, \ref{chap3:pos2}, or \ref{chap3:pos3}. In figure~\ref{cap3:fig8}, we can see in the left panel the observed BOLD response for a particular activated voxel's cluster, and in the right panel we can see the simulated BOLD response for that same voxel, obtained using step four of the algorithm \ref{euclid}.   
In figure~\ref{cap3:fig9}, left panel, we can see the on-line estimated trajectory of the first component of the vector $\boldsymbol{\theta}_{\scaleto{i,j,k,t,l\mathstrut}{6pt}}$ obtained from \ref{chap3:pos2}, for $t=1,\ldots,T$, and in the right panel, we can see the simulated on-line trajectories obtained with step six of the algorithm \ref{euclid}. In figures~\ref{cap3:fig10} and \ref{cap3:fig11}, we can see the same analysis for a non-activated voxel. This type of analysis was performed for several clusters inside and outside the visual cortex, and we are confident in the conclusion that with this algorithm applied to the model \ref{chap2:equ1}, one can identify whether a particular voxel is activated or non-activated.

\begin{table}[H]
\begin{figure}[H]
  \centering
\begin{center}
\begin{tabular}{cc}\\
\multicolumn{2}{c}{Match case: activated voxel}\\
\includegraphics[width=.35\textwidth]{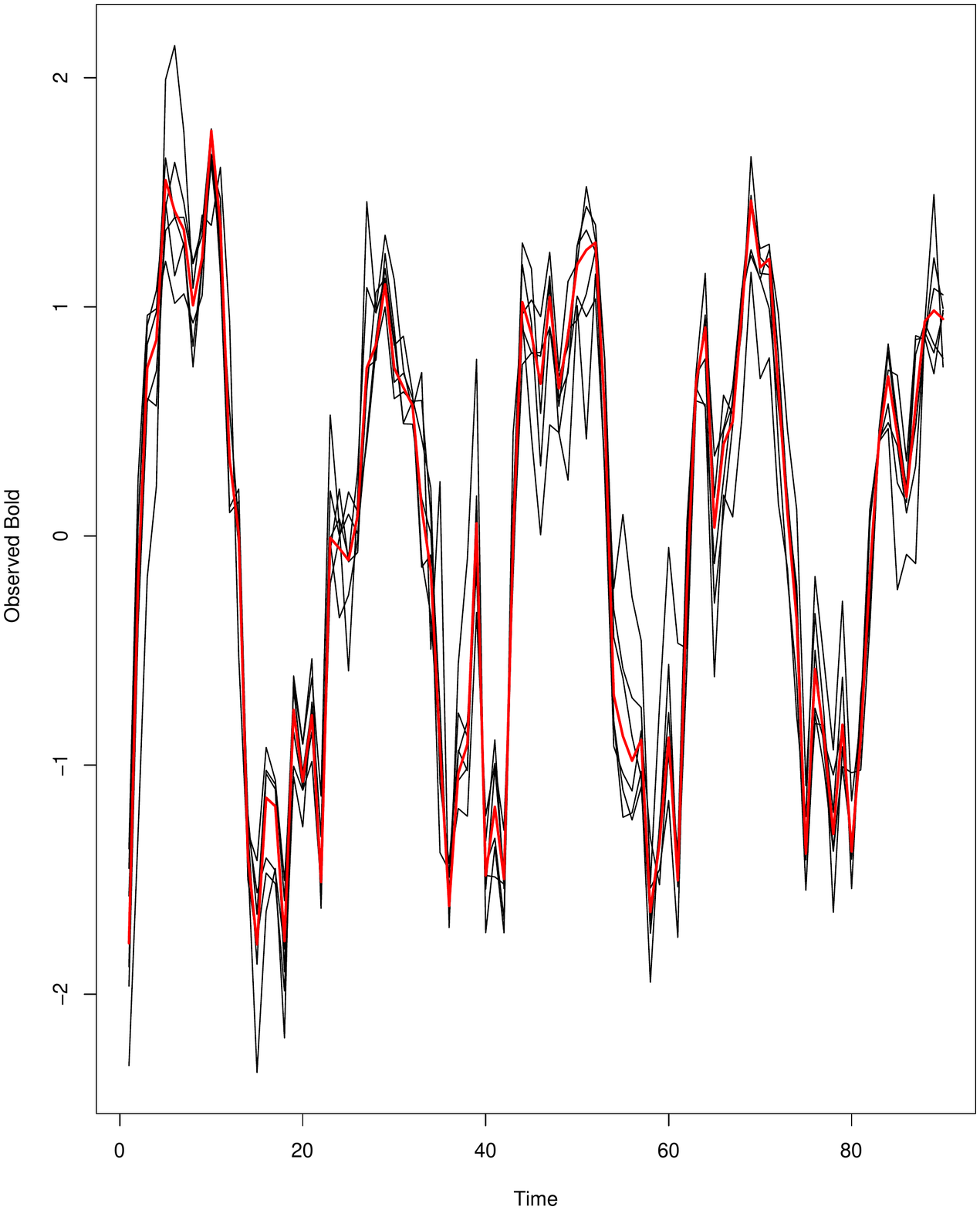}&\includegraphics[width=.40\textwidth]{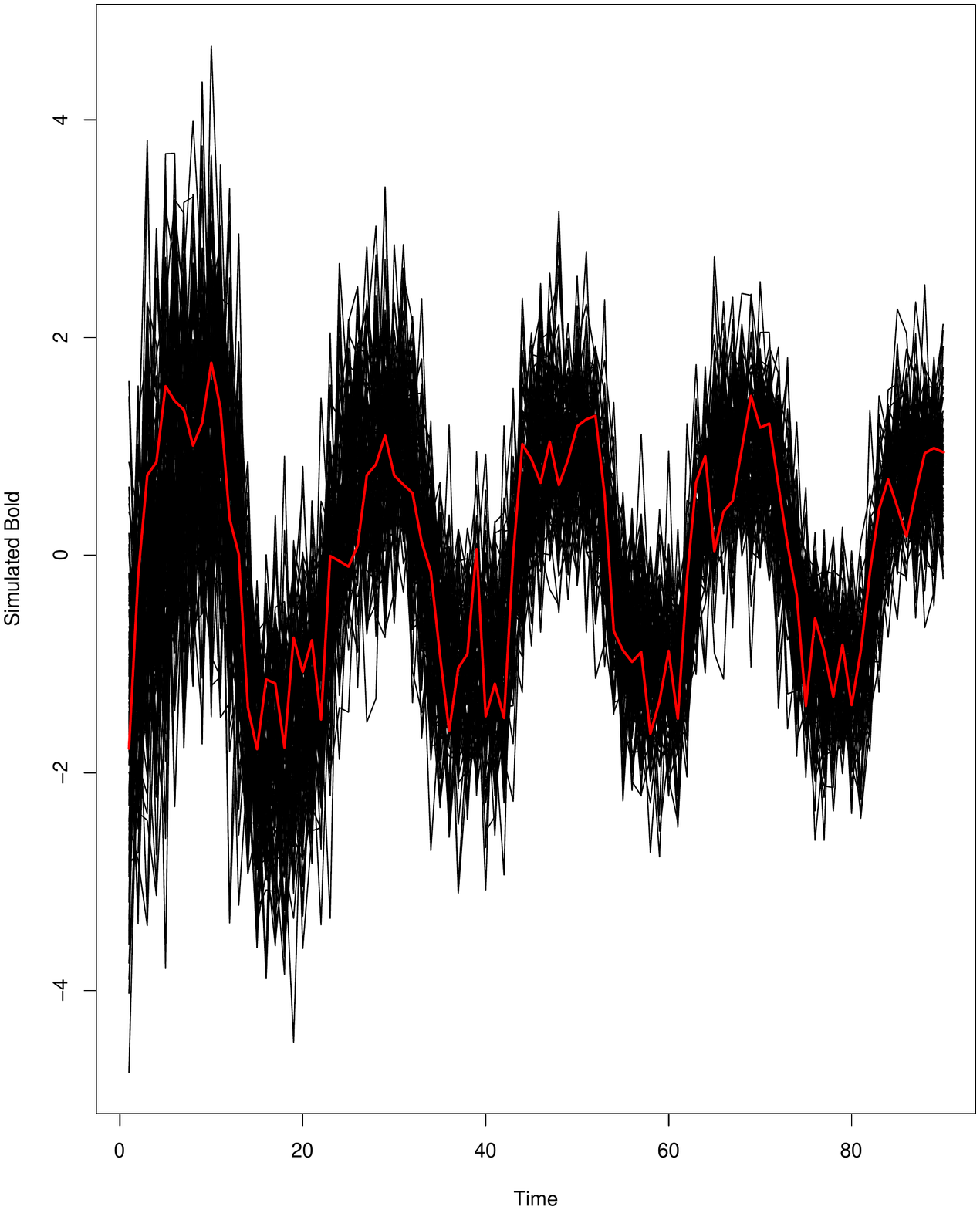}\\
\end{tabular}
\end{center}
  \caption{Left panel: Observed BOLD response for a cluster from the visual cortex. Right panel: Simulated BOLD response obtained with our proposed algorithm. The red curve in both figures is the observed BOLD response related to the first component in \ref{chap3:equ0}.}
  \label{cap3:fig8} 
\end{figure}
\end{table}

\begin{table}[H]
\begin{figure}[H]
  \centering
\begin{center}
\begin{tabular}{cc}\\
\multicolumn{2}{c}{Match case: activated voxel}\\
\includegraphics[width=.35\textwidth]{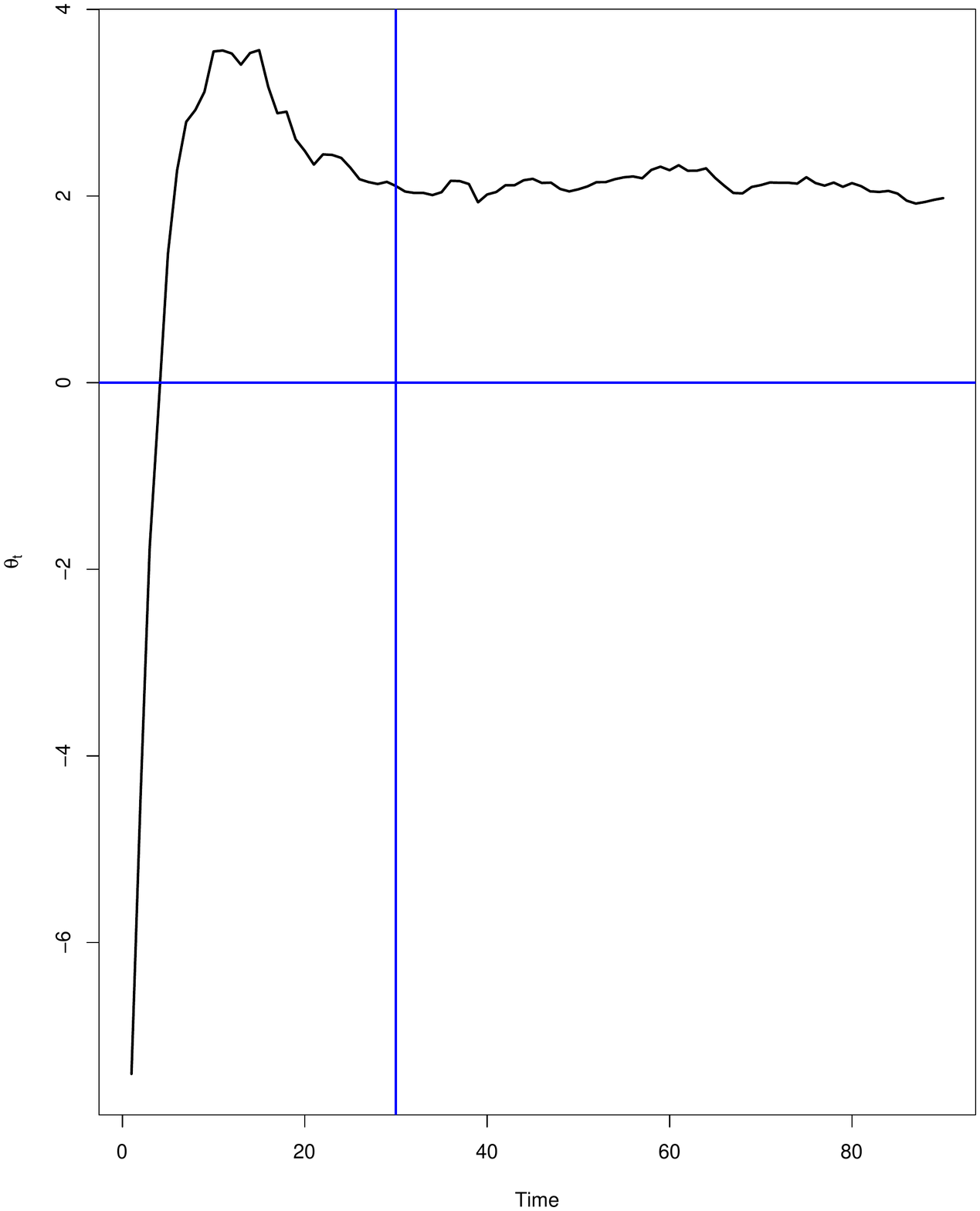}&\includegraphics[width=.35\textwidth]{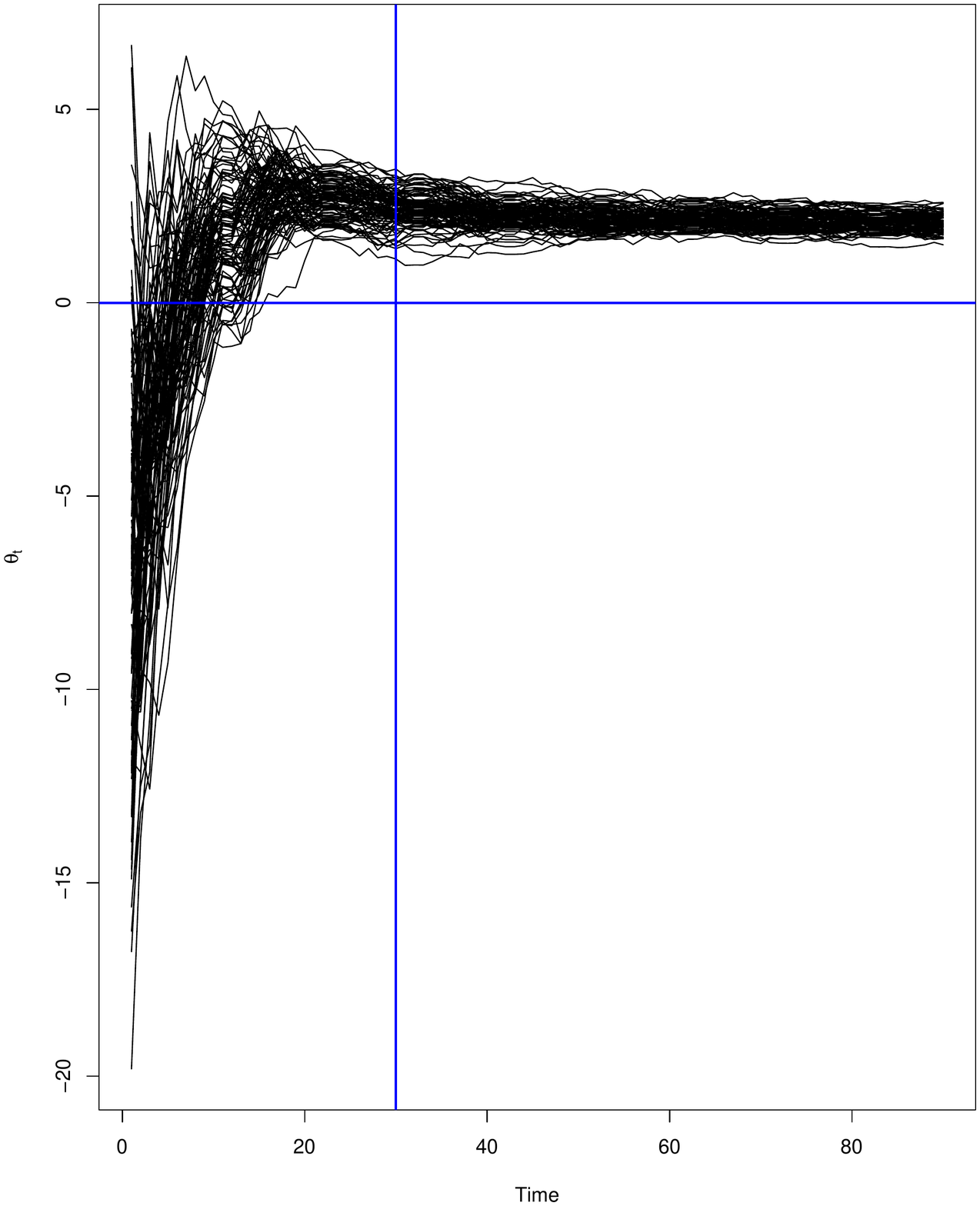}\\
\end{tabular}
\end{center}
  \caption{Left panel: the on-line estimated trajectory of the parameter $\theta_t$ for a voxel from the visual cortex. Right panel: the on-line simulated trajectories of the parameter $\theta_t$ obtained with our proposed algorithm.}
  \label{cap3:fig9} 
\end{figure}
\end{table}

\begin{table}[H]
\begin{figure}[H]
  \centering
\begin{center}
\begin{tabular}{cc}\\
\multicolumn{2}{c}{No match case: non-activated voxel}\\
\includegraphics[width=.35\textwidth]{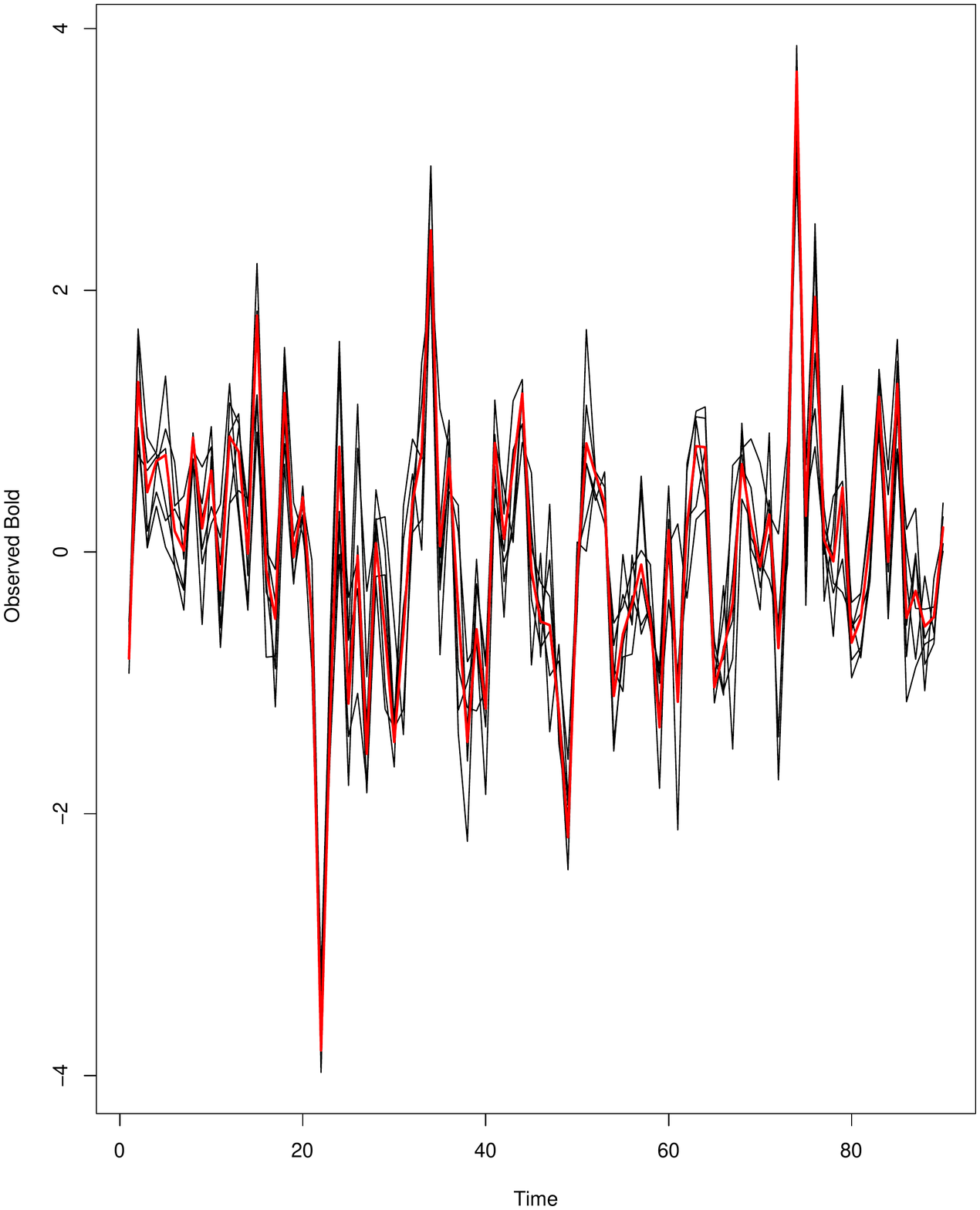}&\includegraphics[width=.35\textwidth]{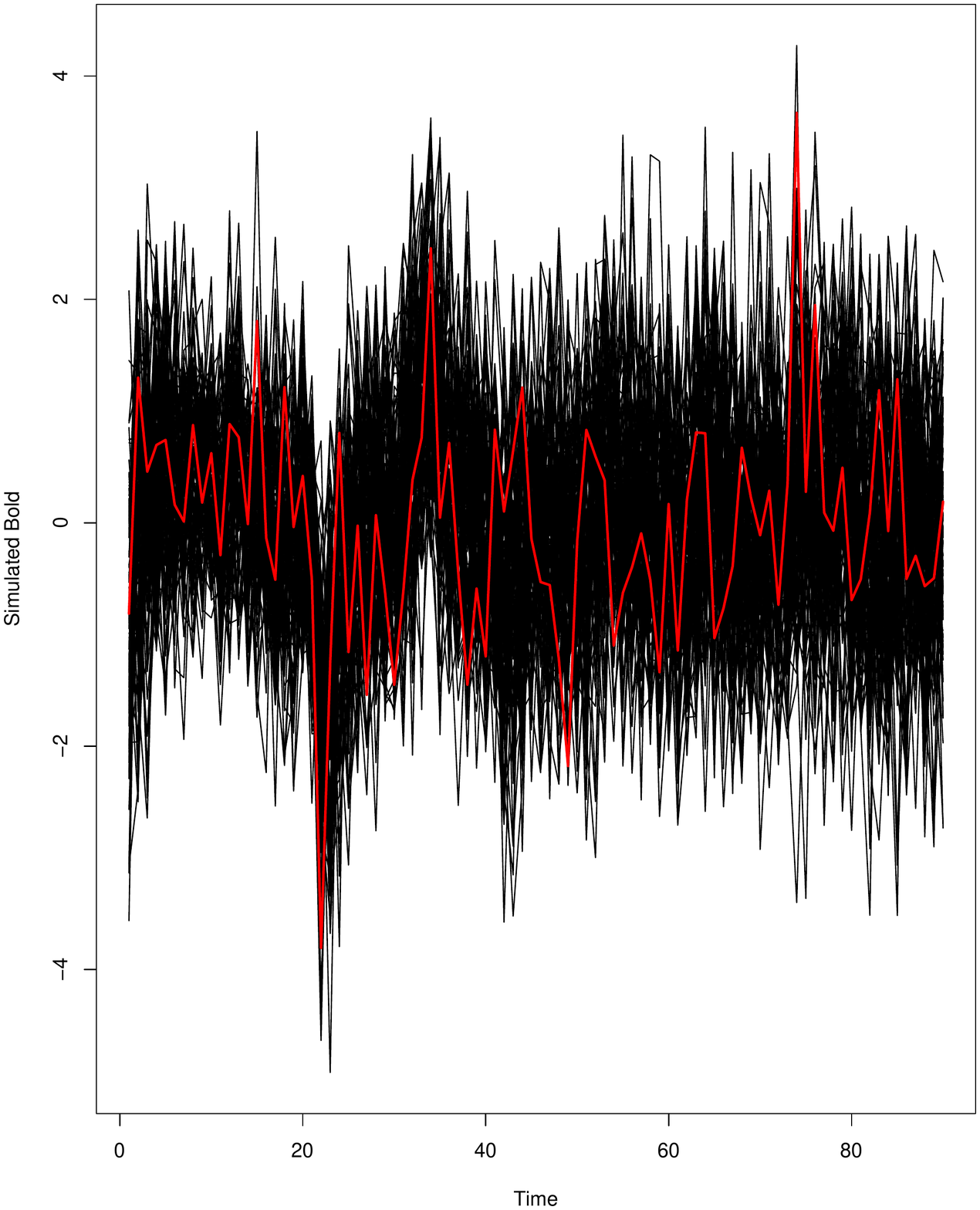}\\
\end{tabular}
\end{center}
  \caption{Left panel: Observed BOLD response for a cluster from outside the visual cortex. Right panel: Simulated BOLD response obtained with our proposed algorithm. The red curve in both figures is the observed BOLD response related to the first component in \ref{chap3:equ0}.}
  \label{cap3:fig10} 
\end{figure}
\end{table}

\begin{table}[H]
\begin{figure}[H]
  \centering
\begin{center}
\begin{tabular}{cc}\\
\multicolumn{2}{c}{No match case: non-activated voxel}\\
\includegraphics[width=.35\textwidth]{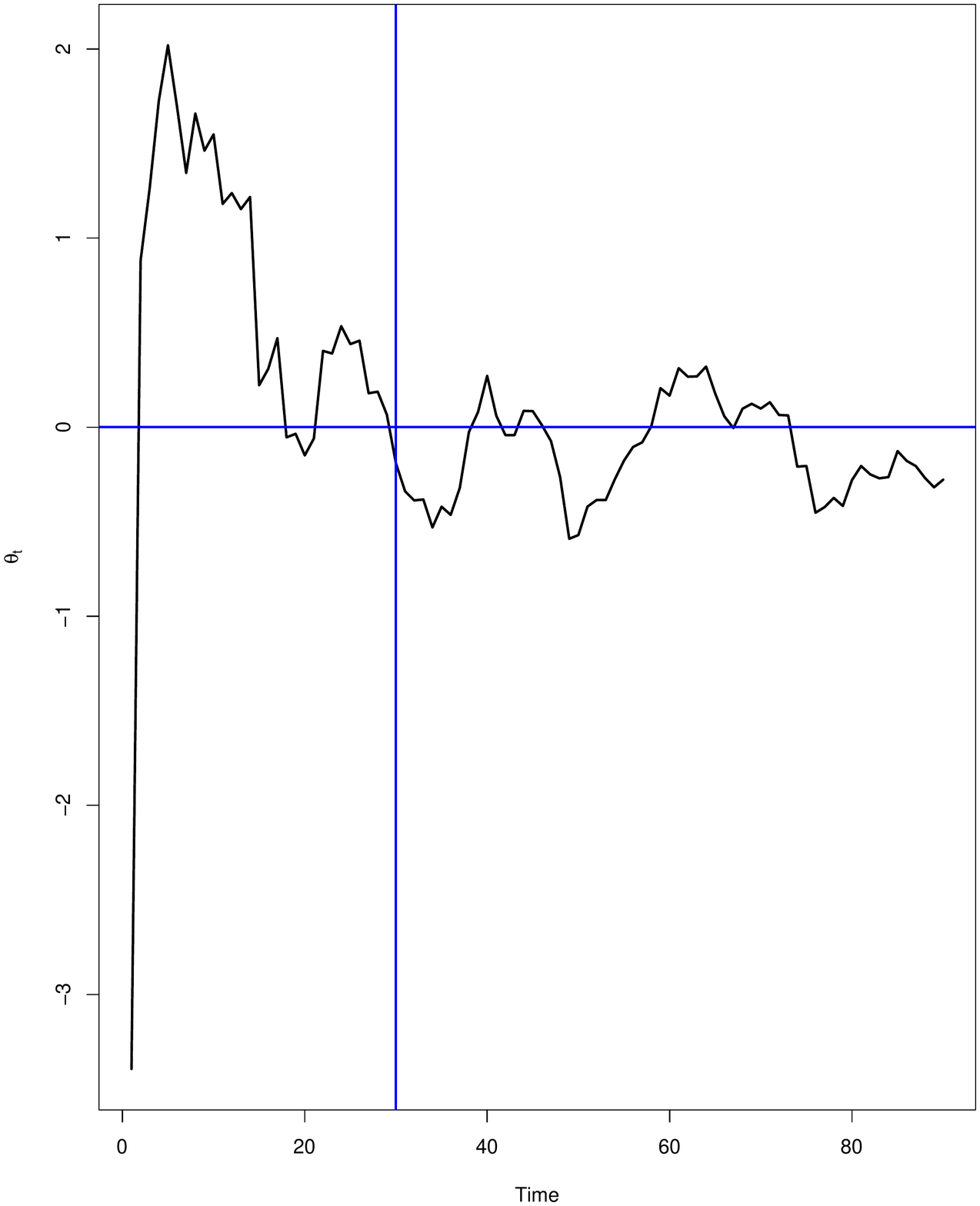}&\includegraphics[width=.35\textwidth]{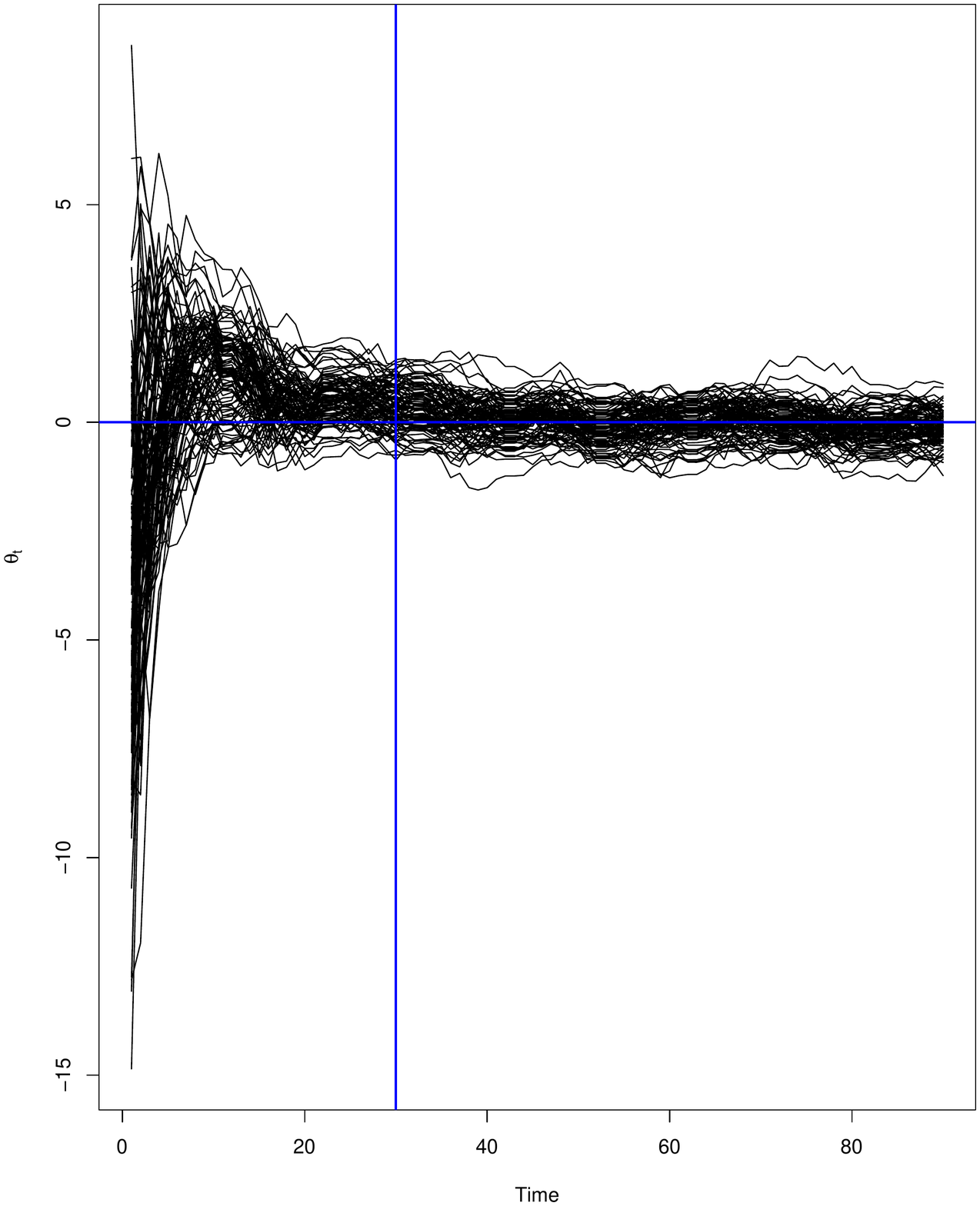}\\
\end{tabular}
\end{center}
  \caption{Left panel: the on-line estimated trajectory of the parameter $\theta_t$ for a voxel from outside the visual cortex. Right panel: the on-line simulated trajectories of the parameter $\theta_t$obtained with our proposed algorithm.}
  \label{cap3:fig11} 
\end{figure}
\end{table}

\section{Voxel-wise group analysis}
\label{chap3:sec2}
We now describe the fRMI group analysis for two possible cases. The first is single-group analysis, which is useful when the interest is to detect an average group activation. The second one is two-group comparison, which is helpful when the aim is to compare voxel activation between two groups, let's say, patients vs. controls. Here we take any of the posterior distributions \ref{chap3:pos1}, \ref{chap3:pos2}, or \ref{chap3:pos3}, depending on the case, as an input for this stage. For instance, let's go back to the last example where 35 controls were part of an fMRI experiment where a visual simulus was presented. Also, suppose that we are interested in using the posterior distribution \ref{chap3:pos2} as an input for this stage, then the average group effect is given by $\boldsymbol{\bar{\theta}}_{\scaleto{i,j,k,T,l\mathstrut}{6pt}}^{A}=\frac{1}{n_A}\sum\limits_{s=1}^{n_A}\boldsymbol{\theta}_{\scaleto{i,j,k,T,l,s\mathstrut}{6pt}}^{A}$, where $n_A$ is the number of subjects in the control group. For the comparison case we have, for example $\boldsymbol{\bar{\theta}}_{\scaleto{i,j,k,T,l\mathstrut}{6pt}}^{AB}=\frac{1}{n_A}\sum\limits_{s=1}^{n_A}\boldsymbol{\theta}_{\scaleto{i,j,k,T,l,s\mathstrut}{6pt}}^{A} - \frac{1}{n_B}\sum\limits_{s=1}^{n_B}\boldsymbol{\theta}_{\scaleto{i,j,k,T,l,s\mathstrut}{6pt}}^{B}$, where $n_B$ is the number of subjects in the patient group. As in the individual case, we describe two different ways to perform the analysis, taking only the latest distribution at $t=T$ or taking all the posterior distributions for $t\geq 30$.

\subparagraph{Inference using the latest posterior distribution:} Applying the same ideas as in the individual case, using the posterior distributions \ref{chap3:pos1}, \ref{chap3:pos2}, \ref{chap3:pos3}, and taking advantage of the properties of the normal distribution, we obtain the following distributions for the group stage:

\begin{equation}\label{chap3:pos4}
\bar{\theta}_{\scaleto{i,j,k,T,l,1\mathstrut}{6pt}}^{*g} \sim N(\bar{m}_{\scaleto{i,j,k,T,l,1\mathstrut}{6pt}}^{*g}, \bar{S}_{\scaleto{T,1,1\mathstrut}{6pt}}^{g}),
\end{equation}

\begin{equation}\label{chap3:pos5}
\boldsymbol{\bar{\theta}}_{\scaleto{i,j,k,T,l\mathstrut}{6pt}}^{g}\sim N(\boldsymbol{\bar{m}}_{\scaleto{i,j,k,T,l\mathstrut}{6pt}}^{g}, \boldsymbol{\bar{S}}_{\scaleto{T\mathstrut}{6pt}}^{g}),
\end{equation}

\begin{equation}\label{chap3:pos6}
\bar{\bar{\theta}}_{\scaleto{i,j,k,T,l\mathstrut}{6pt}}^{g}\sim N(\bar{\bar{m}}_{\scaleto{i,j,k,T,l\mathstrut}{6pt}}^{g}, \bar{\bar{S}}_{\scaleto{T,l\mathstrut}{6pt}}^{g}),
\end{equation}

where,

\begin{equation*}
\begin{array}{ll}
\bar{m}_{\scaleto{i,j,k,T,l,1\mathstrut}{6pt}}^{*g}= \frac{1}{n_g}\sum \limits_{s=1}^{n_g}m_{\scaleto{i,j,k,T,l,1,s\mathstrut}{6pt}}^{*}& \bar{S}_{\scaleto{T,1,1\mathstrut}{6pt}}^{g}=\frac{1}{n_{g}^2}\sum \limits_{s=1}^{n_g}C_{\scaleto{T,l,l,s\mathstrut}{6pt}}*S_{\scaleto{T,1,1,s\mathstrut}{6pt}},\\
\boldsymbol{\bar{m}}_{\scaleto{i,j,k,T,l\mathstrut}{6pt}}^{g}=\frac{1}{n_{g}}\sum \limits_{s=1}^{n_g}\boldsymbol{m}_{\scaleto{i,j,k,T,l,s\mathstrut}{6pt}}& \boldsymbol{\bar{S}}_{\scaleto{T\mathstrut}{6pt}}^{g}=\frac{1}{n_{g}^2}\sum \limits_{s=1}^{n_g}C_{\scaleto{T,l,l,s\mathstrut}{6pt}}*\boldsymbol{S}_{\scaleto{T,s\mathstrut}{6pt}},\\
\bar{\bar{m}}_{\scaleto{i,j,k,T,l\mathstrut}{6pt}}^{g}=\frac{1}{n_{g}}\sum \limits_{s=1}^{n_g}\bar{m}_{\scaleto{i,j,k,T,l,s\mathstrut}{6pt}}& \bar{\bar{S}}_{\scaleto{T,l\mathstrut}{6pt}}^{g}=\frac{1}{n_{g}^2}\sum \limits_{s=1}^{n_g}\bar{S}_{\scaleto{T,l,s\mathstrut}{6pt}},
\end{array}
\end{equation*}

for $g \in \{A, B\}$. As in the individual case, if one wants to test, for example, $H_0: \bar{\bar{\theta}}_{\scaleto{i,j,k,T,l\mathstrut}{6pt}}^{g}>0$, then a measure of evidence against $H_0$ is defined by $\alpha=Pr\left[\bar{\theta}_{\scaleto{i,j,k,T,l\mathstrut}{6pt}}>0\right]$. Then a small value of $\alpha$ indicates rejection of a possible group activation for a specific voxel. Extending these ideas to the comparison case, let's say cases vs. controls, we have the following distributions:

\begin{equation}\label{chap3:pos7}
\bar{\theta}_{\scaleto{i,j,k,T,l,1\mathstrut}{6pt}}^{*B} - \bar{\theta}_{\scaleto{i,j,k,T,l,1\mathstrut}{6pt}}^{*A} \sim N(\bar{m}_{\scaleto{i,j,k,T,l,1\mathstrut}{6pt}}^{*B}-\bar{m}_{\scaleto{i,j,k,T,l,1\mathstrut}{6pt}}^{*A}, \bar{S}_{\scaleto{T,1,1\mathstrut}{6pt}}^{B}+\bar{S}_{\scaleto{T,1,1\mathstrut}{6pt}}^{A}),
\end{equation}

\begin{equation}\label{chap3:pos8}
\boldsymbol{\bar{\theta}}_{\scaleto{i,j,k,T,l\mathstrut}{6pt}}^{B}-\boldsymbol{\bar{\theta}}_{\scaleto{i,j,k,T,l\mathstrut}{6pt}}^{A}\sim N(\boldsymbol{\bar{m}}_{\scaleto{i,j,k,T,l\mathstrut}{6pt}}^{B}-\boldsymbol{\bar{m}}_{\scaleto{i,j,k,T,l\mathstrut}{6pt}}^{A}, \boldsymbol{\bar{S}}_{\scaleto{T\mathstrut}{6pt}}^{B}+\boldsymbol{\bar{S}}_{\scaleto{T\mathstrut}{6pt}}^{A}),
\end{equation}

\begin{equation}\label{chap3:pos9}
\bar{\bar{\theta}}_{\scaleto{i,j,k,T,l\mathstrut}{6pt}}^{B}-\bar{\bar{\theta}}_{\scaleto{i,j,k,T,l\mathstrut}{6pt}}^{A}\sim N(\bar{\bar{m}}_{\scaleto{i,j,k,T,l\mathstrut}{6pt}}^{B}-\bar{\bar{m}}_{\scaleto{i,j,k,T,l\mathstrut}{6pt}}^{A}, \bar{\bar{S}}_{\scaleto{T,l\mathstrut}{6pt}}^{B}+\bar{\bar{S}}_{\scaleto{T,l\mathstrut}{6pt}}^{A}).
\end{equation}

Consider again the same example presented previously, where 35 controls and 15 patients participated in an fMRI experiment where a visual stimulus was presented according to a block design. In figures~\ref{cap3:fig10}, \ref{cap3:fig11}, and \ref{cap3:fig12}, we can see the PPM for the brain activation in the patient group using the posterior distributions \ref{chap3:pos4}, \ref{chap3:pos5}, and \ref{chap3:pos5}, respectively. In all three cases, an activation on the visual cortex was detected. As in the individual case, the results for the marginal and average tests are very similar and the joint test is highly conservative. 

\begin{figure}[H]
  \centering
\begin{center}
\includegraphics[width=.50\textwidth]{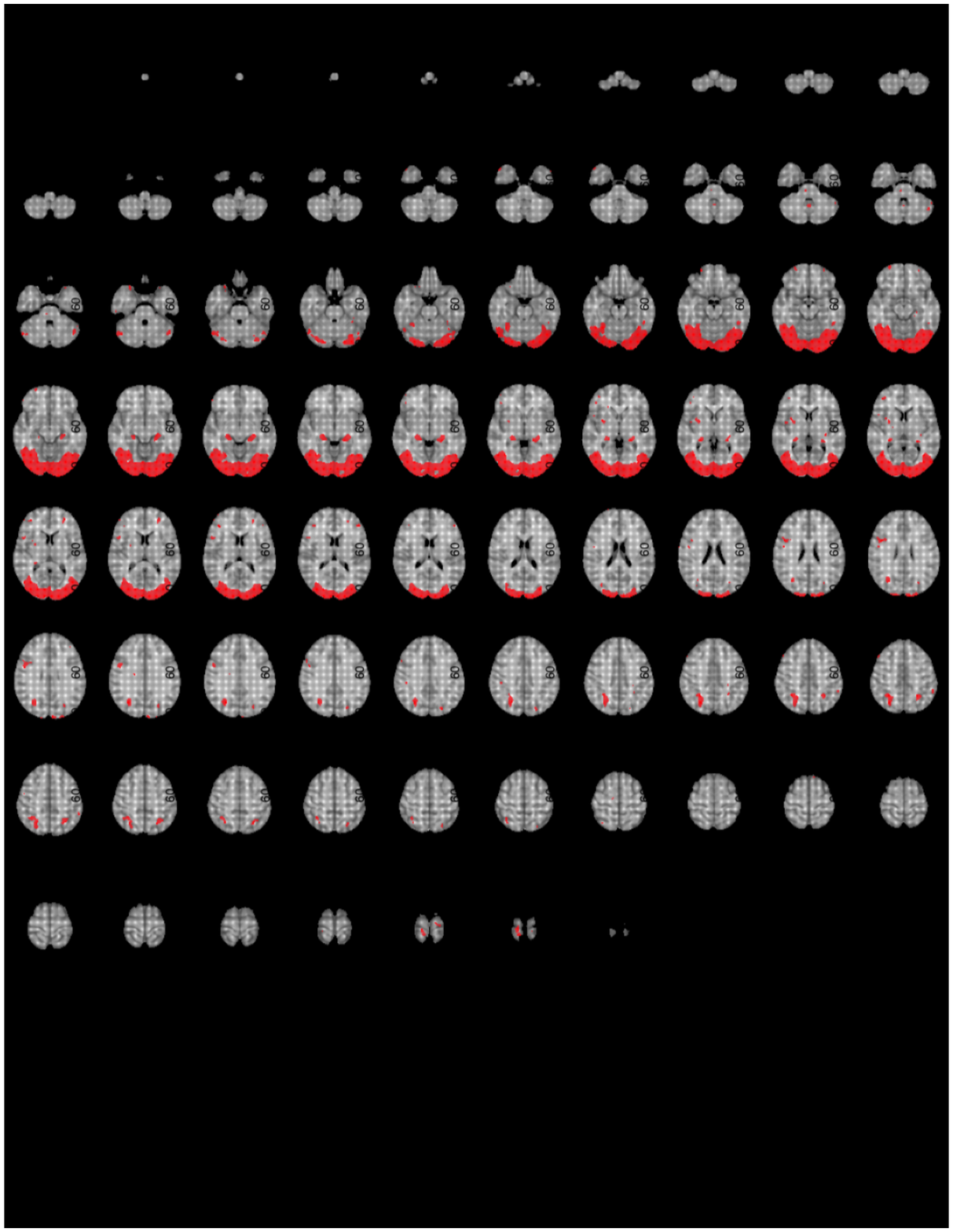}
\end{center}
  \caption{Posterior probability map for the patient group obtained after performing the marginal test on every voxel.}\label{cap3:fig10} 
\end{figure}

\begin{figure}[H]
  \centering
\begin{center}
\includegraphics[width=.50\textwidth]{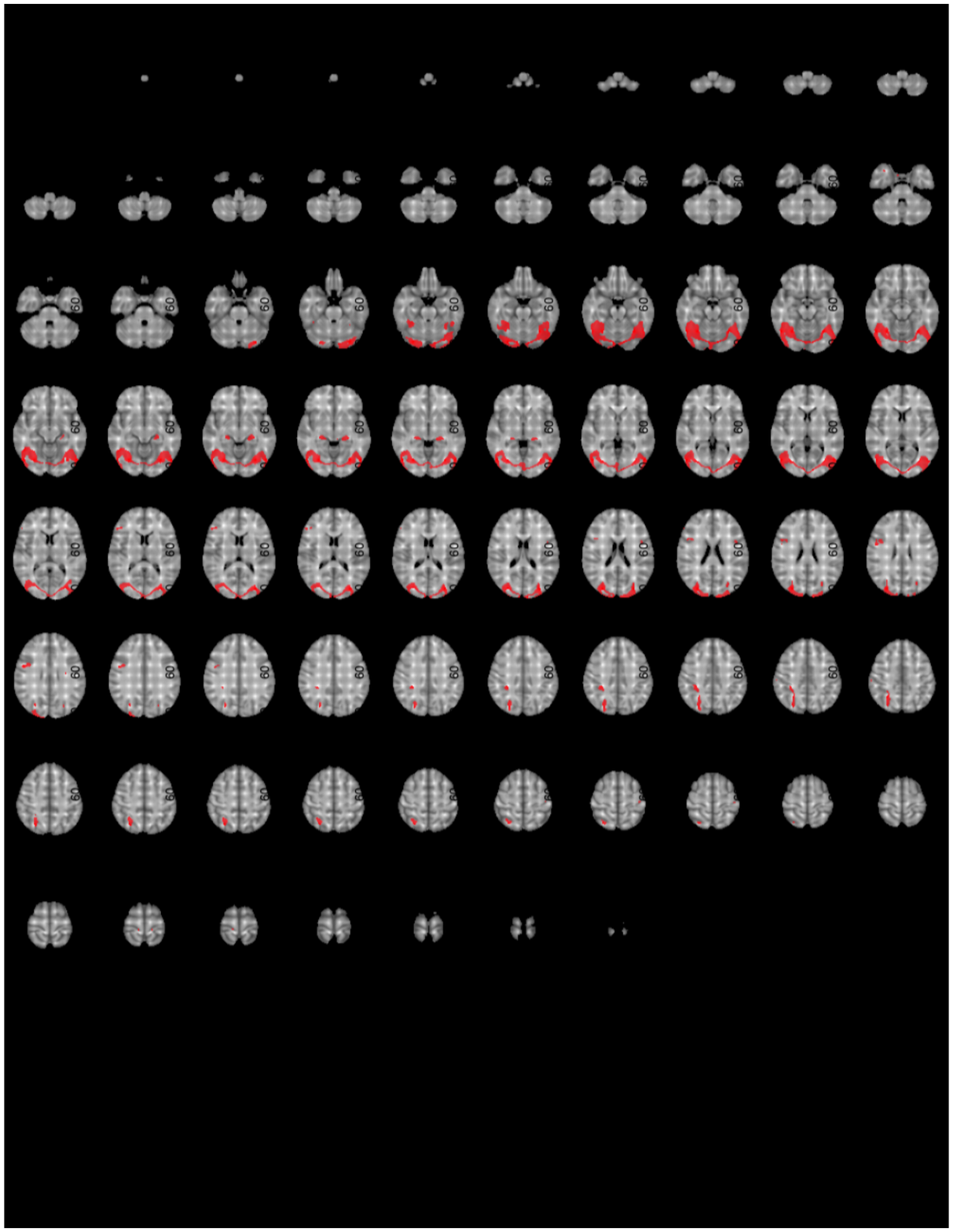}
\end{center}
  \caption{Posterior probability map for the patient group obtained after performing the joint test on every voxel.}\label{cap3:fig11} 
\end{figure}

\begin{figure}[H]
  \centering
\begin{center}
\includegraphics[width=.50\textwidth]{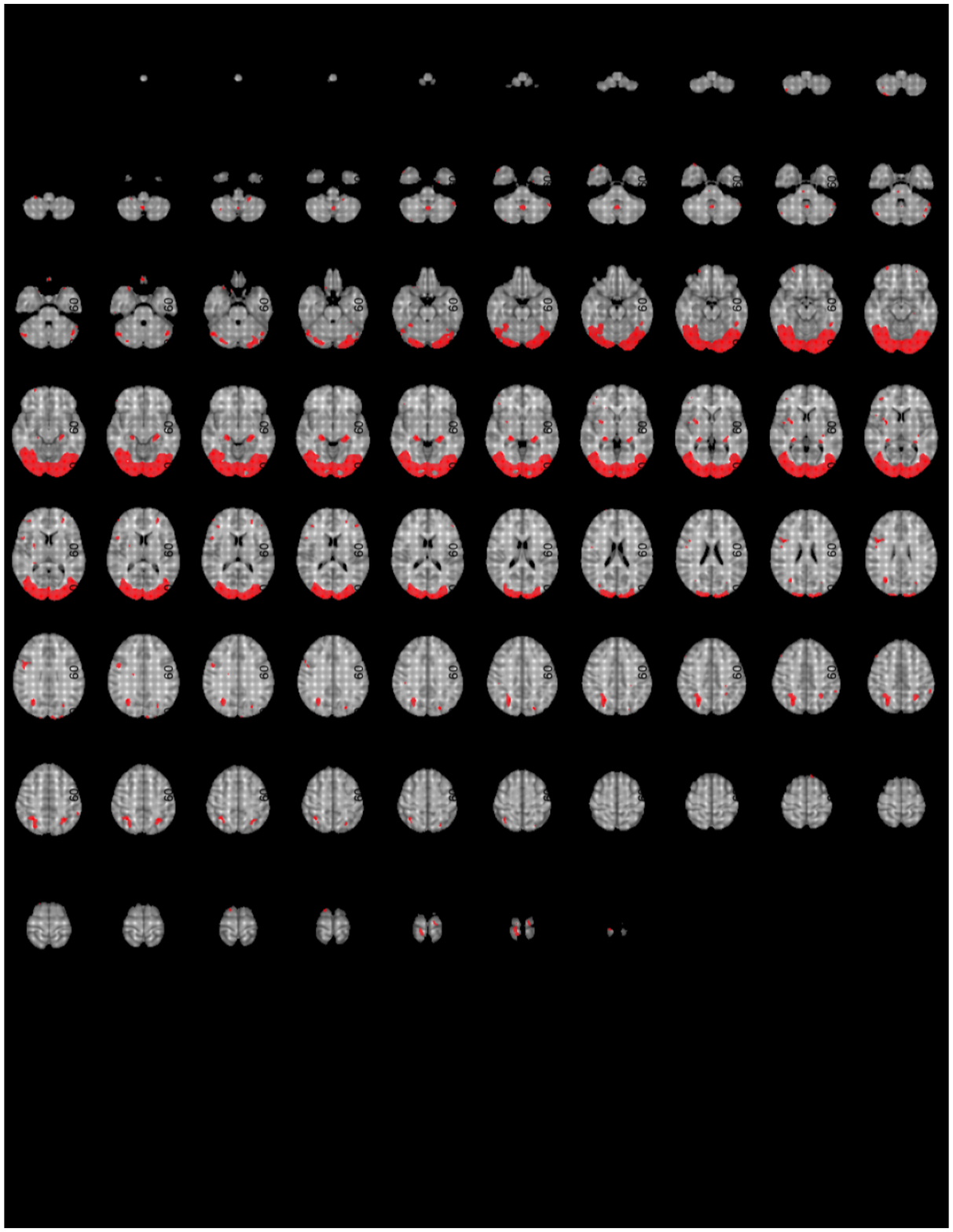}
\end{center}
  \caption{Posterior probability map for the patient group obtained after performing the average test on every voxel.}\label{cap3:fig12} 
\end{figure}

\subparagraph{Inference using all the posterior distributions for $t\geq 30$}
As in the individual case, we now use the algorithm \ref{euclid} to compute a measure of evidence to detect voxel activation using all the posterior distributions for $t\geq 30$. 
For the case of group activation, we just have to choose one of the distributions among \ref{chap3:pos4}, \ref{chap3:pos5}, and \ref{chap3:pos6} as the posterior distribution and run the algorithm. In the case of the group comparison, we just have to apply the algorithm to each group and add one more line to it:

\begin{itemize}
\item[7:] Compute $\tilde{\boldsymbol{\theta}}^{(k)AB}= \tilde{\boldsymbol{\theta}}^{(k)A}-\tilde{\boldsymbol{\theta}}^{(k)B}$
\end{itemize}

Then, compute the measure of evidence for a possible difference as

\begin{equation*}
p(\boldsymbol{\theta}>\mathbf{0})= E(1_{(\boldsymbol{\theta}>\mathbf{0})}) \approx \frac{\sum\limits_{k=1}^{N} 1_{(\tilde{\boldsymbol{\theta}}^{(k)AB}>\mathbf{0})}}{N}.
\end{equation*}

In figures \ref{cap3:fig13} and \ref{cap3:fig14}, we can see an example of the algorithm \ref{euclid} applied to two particular voxels (for the patient group) inside and outside the visual cortex respectively. It can be seen that in both cases this procedure allows for correct identification of the brain activity related to a visual stimulus.

\begin{table}[H]
\begin{figure}[H]
  \centering
\begin{center}
\begin{tabular}{cc}\\
\multicolumn{2}{c}{Match case: activated voxel}\\
\includegraphics[width=.35\textwidth]{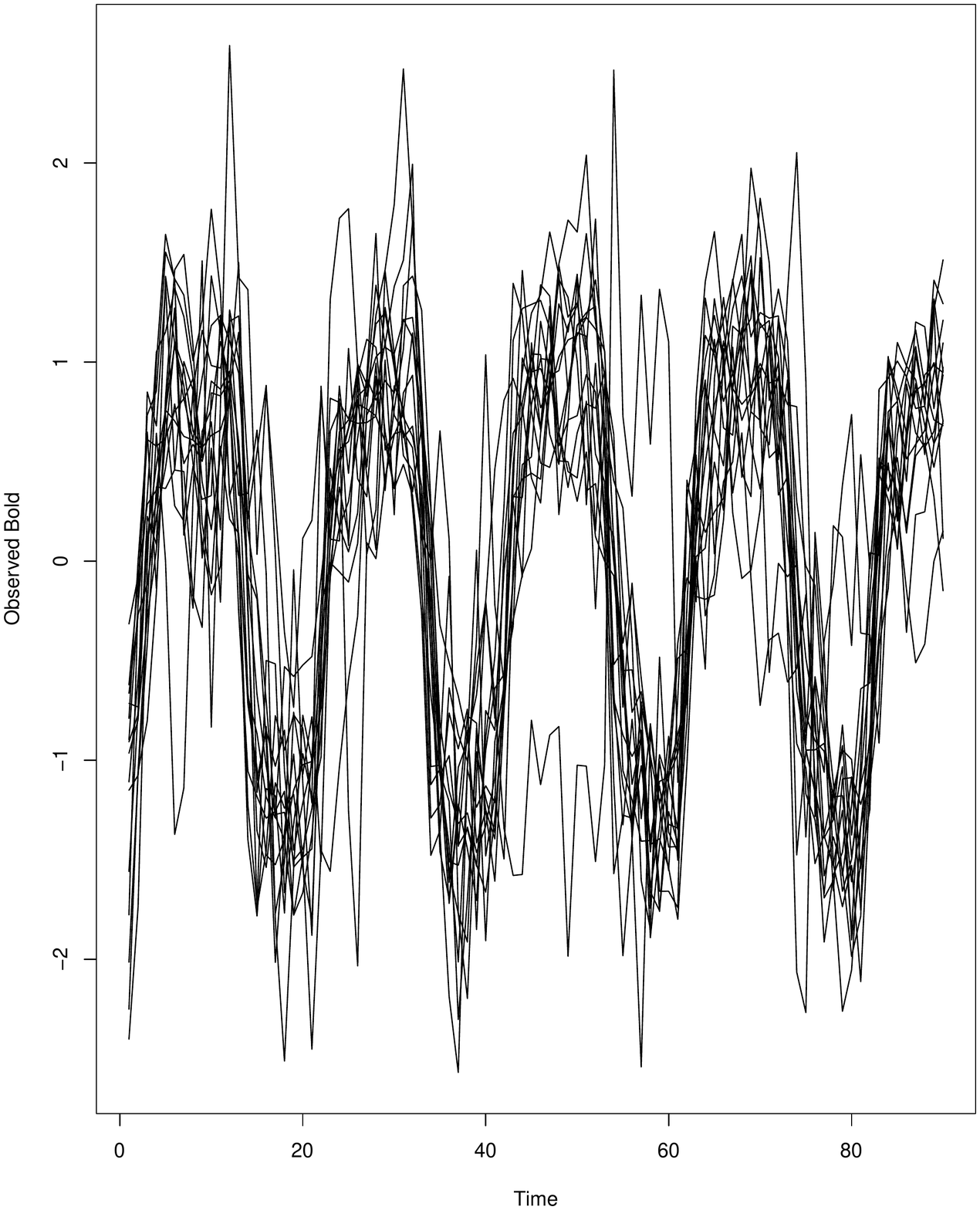}&\includegraphics[width=.35\textwidth]{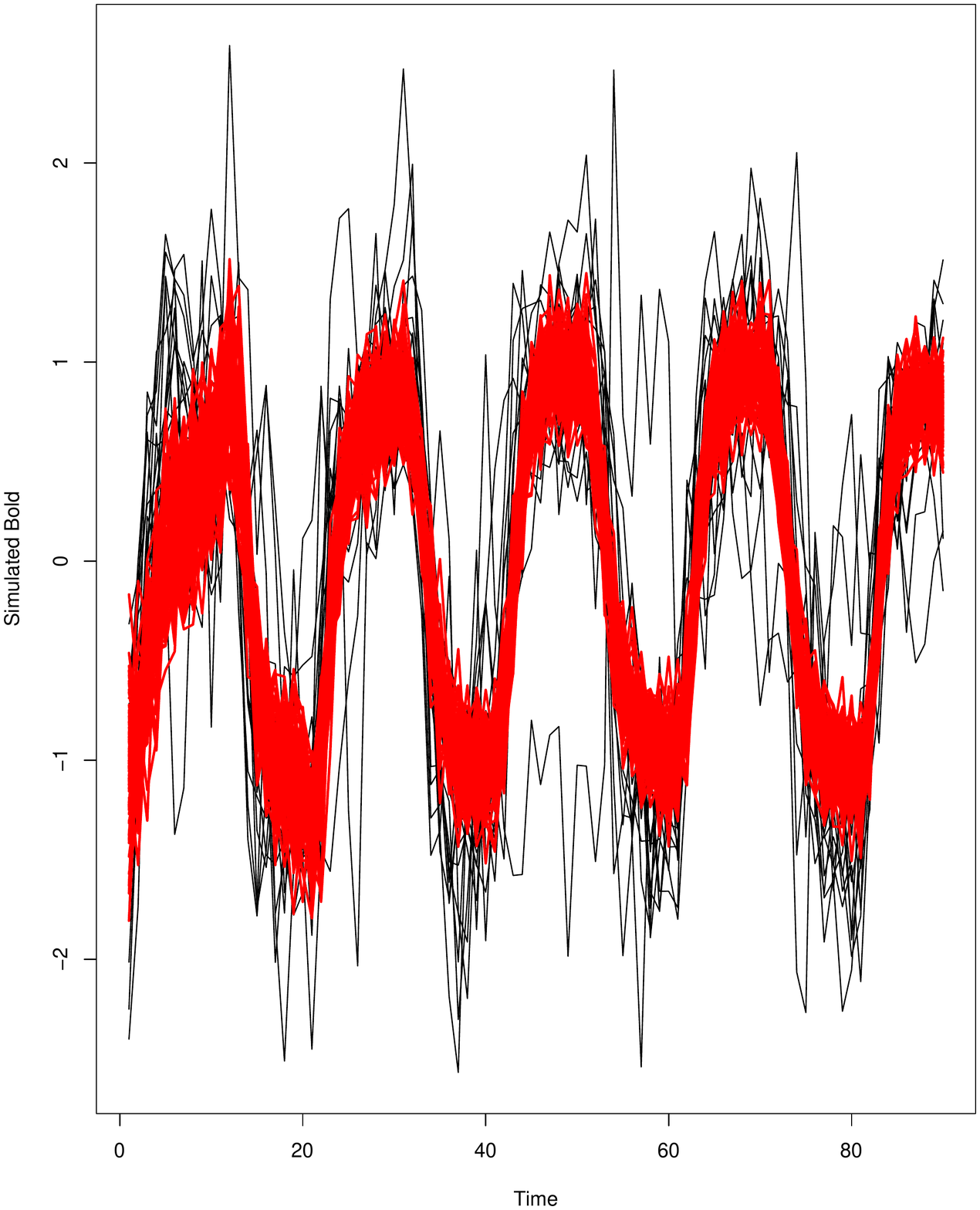}\\
\multicolumn{2}{c}{\includegraphics[width=.35\textwidth]{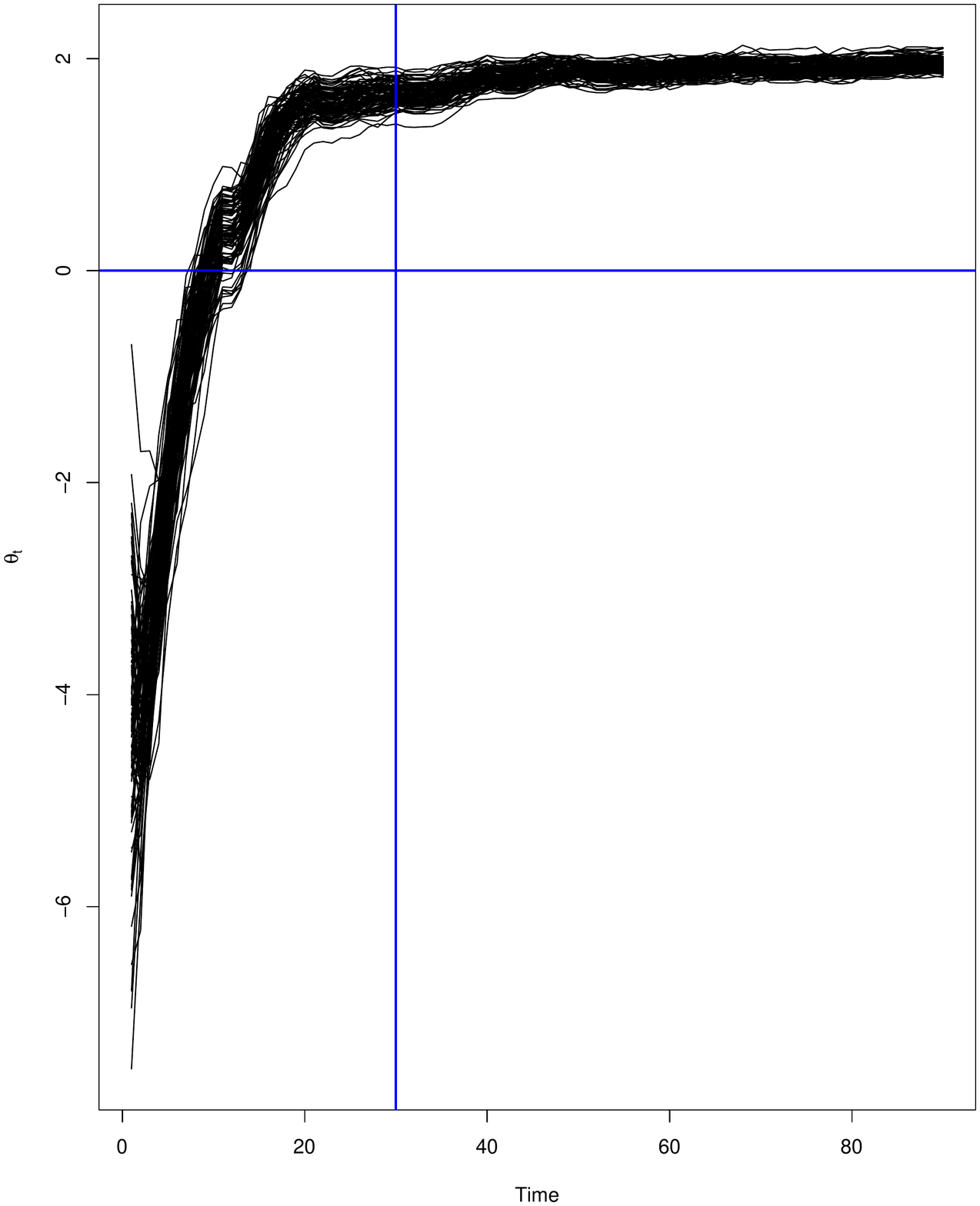}}\\
\end{tabular}
\end{center}
  \caption{
  Top left panel: Observed BOLD response for each subject (in the patients' group) in a fixed location of the visual cortex. Top right panel: Simulated BOLD response obtained with our proposed algorithm (red curves) superposed over the observed bold (black curves). Bottom panel: the on-line estimated trajectory of the parameter $\theta_t$ for the same voxel from the visual cortex.}
  \label{cap3:fig13} 
\end{figure}
\end{table}

\begin{table}[H]
\begin{figure}[H]
  \centering
\begin{center}
\begin{tabular}{cc}\\
\multicolumn{2}{c}{No match case: non-activated voxel}\\
\includegraphics[width=.35\textwidth]{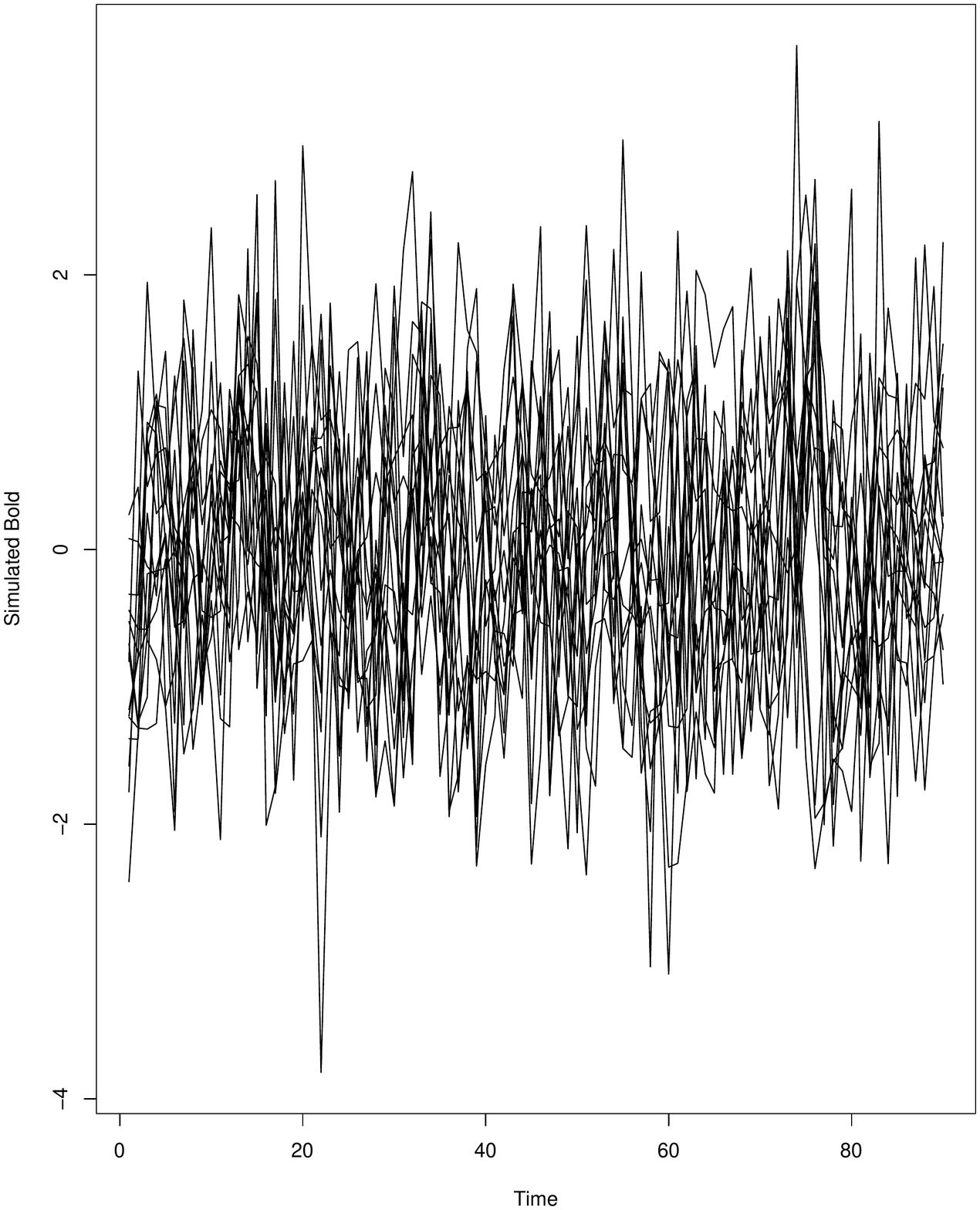}&\includegraphics[width=.35\textwidth]{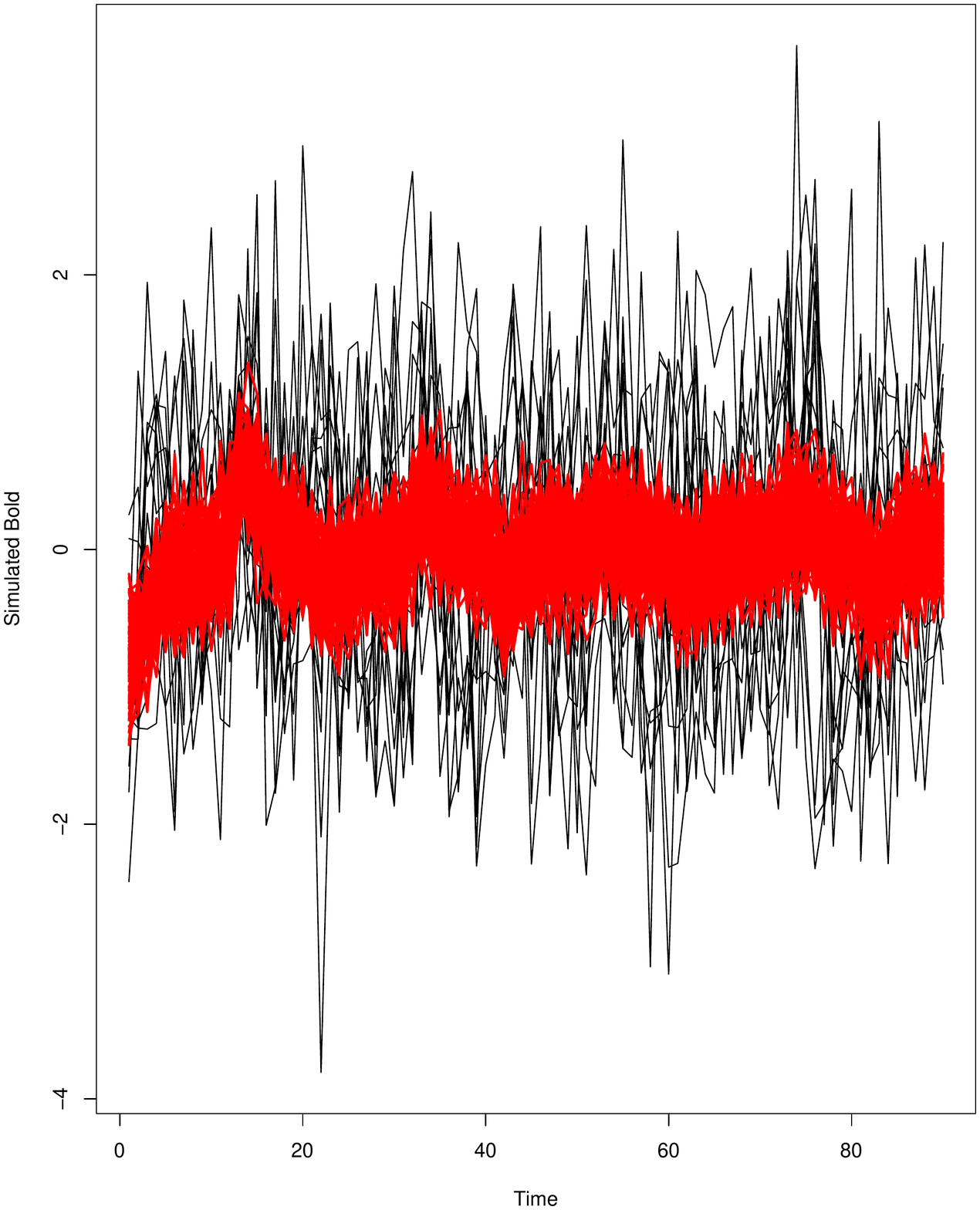}\\
\multicolumn{2}{c}{\includegraphics[width=.35\textwidth]{simulatedOnlineThera_non}}\\
\end{tabular}
\end{center}
  \caption{Top left panel: Observed Bold response for each subject (in the patients' group) in a fixed location outside the visual cortex. Top right panel: Simulated Bold response obtained with our proposed algorithm (red curves) superposed over the observed bold (black curves). Bottom panel: the on-line estimated trajectory of the parameter $\theta_t$ for the same voxel outside the visual cortex.}
  \label{cap3:fig14} 
\end{figure}
\end{table}

\subsection{Gaussian process ANOVA model}

In the previous chapter we presented the Gaussian process ANOVA model introduced by \cite{kaufman2010bayesian} as another alternative to compare batches of curves. At this moment in time, we do not have any results to be shown related to this alternative, but we expect to make an implementation of this functional ANOVA model.

\section*{Validation of the proposed method}

In order to assess the proposed method described in sections~\ref{chap3:sec1} and \ref{chap3:sec2}, we follow the same approach as in \cite{eklund2012does} and \cite{eklund2016cluster}. We will use resting-state fMRI data from healthy controls, obtained from the 1000 Functional Connectomes Project \cite{biswal2010toward}. We will create fictitious covariates related to block and event-related designs and execute individual and group voxel-wise analysis. Resting-state data should not contain systematic changes in brain activity.  Therefore, all voxels identified as active must be considered as false-positive. Thus, the assessment of the method relies on the empirical rate of false-positive activations. 

\section{Computational aspects}

An important issue in this thesis is the massive amount of data related to an fMRI array. For instance, in the applications presented above, we had 15 and 35 fMRI arrays for the patient and control groups respectively. Each of these arrays is composed of $90\times 90\times 100$ voxels and each one of these represents a temporal series of $90$ observations. Thus, in this kind of applications we could deal with arrays of dimension $90\times 90\times 100\times 90$ (or even greater) for each subject. One of the main objectives of this thesis is to build an R package with the implementation of voxel-wise group analysis.  Almost all of the source code has been written in the C programming language in order to speed up the computation time. Specifically, we use the GNU Scientific Library (\cite{gough2009gnu}), which has implemented some useful functions for linear algebra operations and statistical analysis.

        % associado ao arquivo: 'cap-conclusoes.tex'

% cabeçalho para os apêndices
\renewcommand{\chaptermark}[1]{\markboth{\MakeUppercase{\appendixname\ \thechapter}} {\MakeUppercase{#1}} }
\fancyhead[RE,LO]{}
\appendix

%\include{ape-conjuntos}      % associado ao arquivo: 'ape-conjuntos.tex'

% ---------------------------------------------------------------------------- %
% Bibliografia
\backmatter \singlespacing   % espaçamento simples
\bibliographystyle{apacite}
\bibliography{bibliografia}  % associado ao arquivo: 'bibliografia.bib'

% ---------------------------------------------------------------------------- %
% Índice remissivo
%\index{TBP|see{periodicidade região codificante}}
%\index{DSP|see{processamento digital de sinais}}
%\index{STFT|see{transformada de Fourier de tempo reduzido}}
%\index{DFT|see{transformada discreta de Fourier}}
%\index{Fourier!transformada|see{transformada de Fourier}}

%\printindex   % imprime o índice remissivo no documento 

\end{document}